\documentclass[11pt,a4paper]{article}

\usepackage[]{geometry}
\usepackage{graphicx}
\usepackage{amsmath}
\usepackage{amssymb}
\usepackage{xspace}
\usepackage{subfigure}
\usepackage{epsfig}
\usepackage{psfrag}
\usepackage{appendix}

\newcommand{\filt}[1]{#1_{\mbox{\tiny filt}}}
\newcommand{\bbbar}{b\bar{b}}
\newcommand{\bb}{\bar{b}}
\newcommand{\as}{\alpha_s}
\newcommand{\Li}{\mbox{Li}}
\newcommand{\erf}{\mbox{erf}}

\title{Non-Global Logarithms in Filtered Jet Algorithms}
\author{Mathieu Rubin\\
  LPTHE\\ UPMC Univ.\ Paris 6\\ CNRS UMR 7589\\ Paris 05, France.  }

\date{\ }

\begin{document}

\maketitle
\begin{abstract}

We analytically and numerically study the effect of perturbative gluons emission on the ``Filtering analysis'', which is part of a subjet analysis procedure proposed two years ago to possibly identify a low-mass Higgs boson decaying into $b\bb$ at the LHC. This leads us to examine the non-global structure of the resulting perturbative series in the leading single-log large-$N_c$ approximation, including all-orders numerical results, simple analytical approximations to them and comments on the structure of their series expansion. We then use these results to semi-analytically optimize the parameters of the Filtering analysis so as to suppress as much as possible the effect of underlying event and pile-up on the Higgs mass peak reconstruction while keeping the major part of the perturbative radiation from the $b\bb$ dipole.

\end{abstract}

\section{Introduction}

\begin{figure}[htbp]
  \centering
  \includegraphics[scale=0.7]{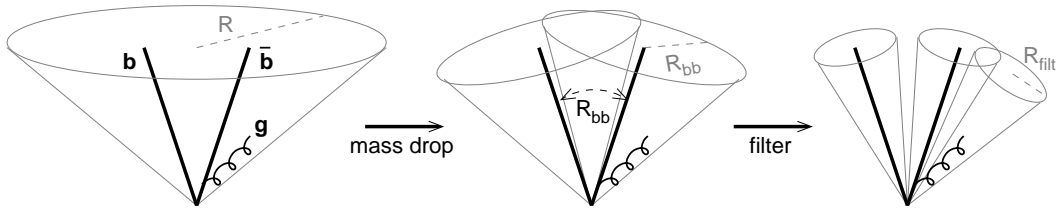}
  \caption{The Mass Drop and Filtering analysis in the procedure used to enhance the signal from a light Higgs decaying into $b\bb$ at the LHC.}
  \label{MD_and_F_analysis}
\end{figure}

In recent years, there has been a growing interest in jets studies in order to identify a boosted massive particle decaying hadronically, for instance the $W$ boson \cite{Seymour:1993mx,Butterworth:2002tt,Skiba:2007fw,Holdom:2007nw}, top quarks \cite{Almeida:2008tp,Kaplan:2008ie,Thaler:2008ju,Plehn:2009rk}, supersymmetric particles \cite{Butterworth:2007ke,Butterworth:2009qa} and heavy resonances \cite{Baur:2008uv,FileviezPerez:2008ib,Bai:2008sk} (see also \cite{Ellis:2009me} for related work on general massive jets). Some of these studies revealed themselves to be successful in looking for a boosted light Higgs boson decaying into $b\bb$ at the LHC \cite{Plehn:2009rk,MyFirstPaper,ATL-PHYS-PUB-2009-088,Kribs:2009yh}. That of \cite{MyFirstPaper,ATL-PHYS-PUB-2009-088} can be briefly summed up as follows: after having clustered the event with a radius $R$ large enough to catch the $b$ and $\bb$ from the Higgs decay into a single jet,\footnote{The value chosen was $R=1.2$.} this jet can be analysed in $2$ steps:
\begin{itemize}
  \item A Mass Drop (MD) analysis that allows one to identify the splitting responsible for the large jet mass, i.e. separate the $b$ and $\bb$ and thus measuring the angular distance $R_{bb}$ between them, while suppressing as much QCD background as possible.
  \item A Filtering analysis where one reclusters the $2$ resulting subjets with a smaller radius and takes the $3$ highest-$p_t$ subjets\footnote{The value of $3$ was found to work well in \cite{MyFirstPaper}.} obtained in order to keep the major part of the perturbative radiation while getting rid of as many underlying event (UE) and pile-up (PU) particles as possible (used also in \cite{Plehn:2009rk,Kribs:2009yh,Cacciari:2008gd}, and a variant is proposed in \cite{Krohn:2009th}).
\end{itemize}
Concerning the MD analysis, the only thing we need to know is that we end up with $2$ $b$-tagged jets, each with a radius roughly equal to $R_{bb}$. Notice that due to angular ordering \cite{Fadin:1983aw,Ermolaev:1981cm,Mueller:1981ex,Dokshitzer:1982xr,Bassetto:1984ik}, these $2$ jets should capture the major part of the perturbative radiation from the $b\bb$ dipole. The whole procedure is depicted in figure~\ref{MD_and_F_analysis} (taken from \cite{MyFirstPaper}).

In this paper, we are going to focus on the Filtering analysis. One can generalize it with respect to its original definition using $2$ parameters, $\filt{n}$ and $\filt{\eta}$ (as discussed also in \cite{Krohn:2009th}): after the MD analysis was carried out, one reclusters the $2$ resulting subjets with a radius $\filt{R} = \filt{\eta}R_{bb}$ and takes the $\filt{n}$ hardest jets obtained. Obviously, the larger the value of $\filt{\eta}$ the more perturbative radiation we keep, but also the more the UE/PU degrades the Higgs peak. The same holds for $\filt{n}$. So there is a compromise to make between losing too much perturbative radiation and being contaminated by soft particles from UE/PU. In \cite{MyFirstPaper}, the values that were found to give nice results were $\filt{n} = 3$ and $\filt{\eta} = \min(0.3/R_{bb},1/2)$. However, these values had been chosen on brief Monte-Carlo event generator study and one would like to gain a little more analytical control over them. One question would be for instance to understand even approximately how the optimal ($\filt{n}$,$\filt{\eta}$) values change when one increases the Higgs $p_{t_H}$ cut, or when the PU becomes more and more important during the high luminosity running of the LHC. Though the MD and Filtering analysis were originally designed to identify a light Higgs boson, one should be aware that similar calculations may apply in other uses of filtering, for instance to study any boosted colorless resonance decaying hadronically, including $W$ and $Z$ bosons.

The article will be devoted in large part to the study of the dependence of the perturbative radiation loss with respect to the filtering parameters. As usual in this kind of work, large logarithms arise due to soft or collinear gluon emissions, and one is forced to deal with them in order to obtain reliable results in the region where the observable is sensitive to this kind of emission. We will thus compute analytically the two first orders in the leading soft logarithmic (LL) approximation when $\filt{n}=2$ and to all-orders in the large-$N_c$ limit\footnote{$N_c=3$ denotes as usual the number of QCD colors.} when $\filt{n}=2$ or $3$ for small enough values of $\filt{\eta}$ (section~\ref{NG_structure_analytical_insights}). With these in hand, and using a program that allows one to make all-orders leading-log calculations in the large-$N_c$ approximation, we check the analytical results and examine if the small-$\filt{\eta}$ limit and/or the truncation of the LL expansion can be trusted to estimate the loss of perturbative radiation in practice (section~\ref{NG_structure_numerical_results}). Finally, in section~\ref{choice_of_the_filtering_parameters}, we will analyse the Higgs mass peak width due respectively to the loss of perturbative radiation and to the presence of UE/PU, before combining them in a simple and approximate but physically reasonable way in order to be able to conclude about the optimal parameters choices.

\section{Non-Global structure: analytical insights \label{NG_structure_analytical_insights}}

\subsection{The filtered Higgs mass: a Non-Global observable \label{the_filtered_higgs_mass_a_non_global_observable}}

 It is now very well known \cite{Catani:1991kz,Catani:1991pm,Catani:1992ua,Catani:1992jc,Catani:1998sf,Dokshitzer:1998kz,Antonelli:1999kx,Burby:1999yb,Burby:2001uz,Banfi:2000si,Dasgupta:2001sh,Banfi:2003jj,Banfi:2004nk,Banfi:2004yd,Becher:2008cf,Dissertori:2009ik} that soft or collinear gluons can give rise, in multiscale problems, to the appearance of large logarithms in the perturbative expansion of an observable, and more precisely in a region of phase space where it is sensitive to the soft or collinear divergences of QCD. In this article, the observable considered is $\Delta M = M_H-M_{\mbox{\scriptsize filtered jet}}$, where $M_{\mbox{\scriptsize filtered jet}}$ is the reconstructed Higgs-jet mass and $M_H$ is its true mass. $\Delta M$ has the property that it is $0$ when no gluon is emitted. We are interested in $\Sigma(\Delta M)$, the probability for the difference between the reconstructed and true Higgs masses to be less than a given $\Delta M$. In this case, large soft logarithms have to be resummed at all-orders to obtain a reliable description of the small $\Delta M$ distribution.

\begin{figure}[htb]
  \centering
  \subfigure[]{
    \includegraphics[scale=0.4]{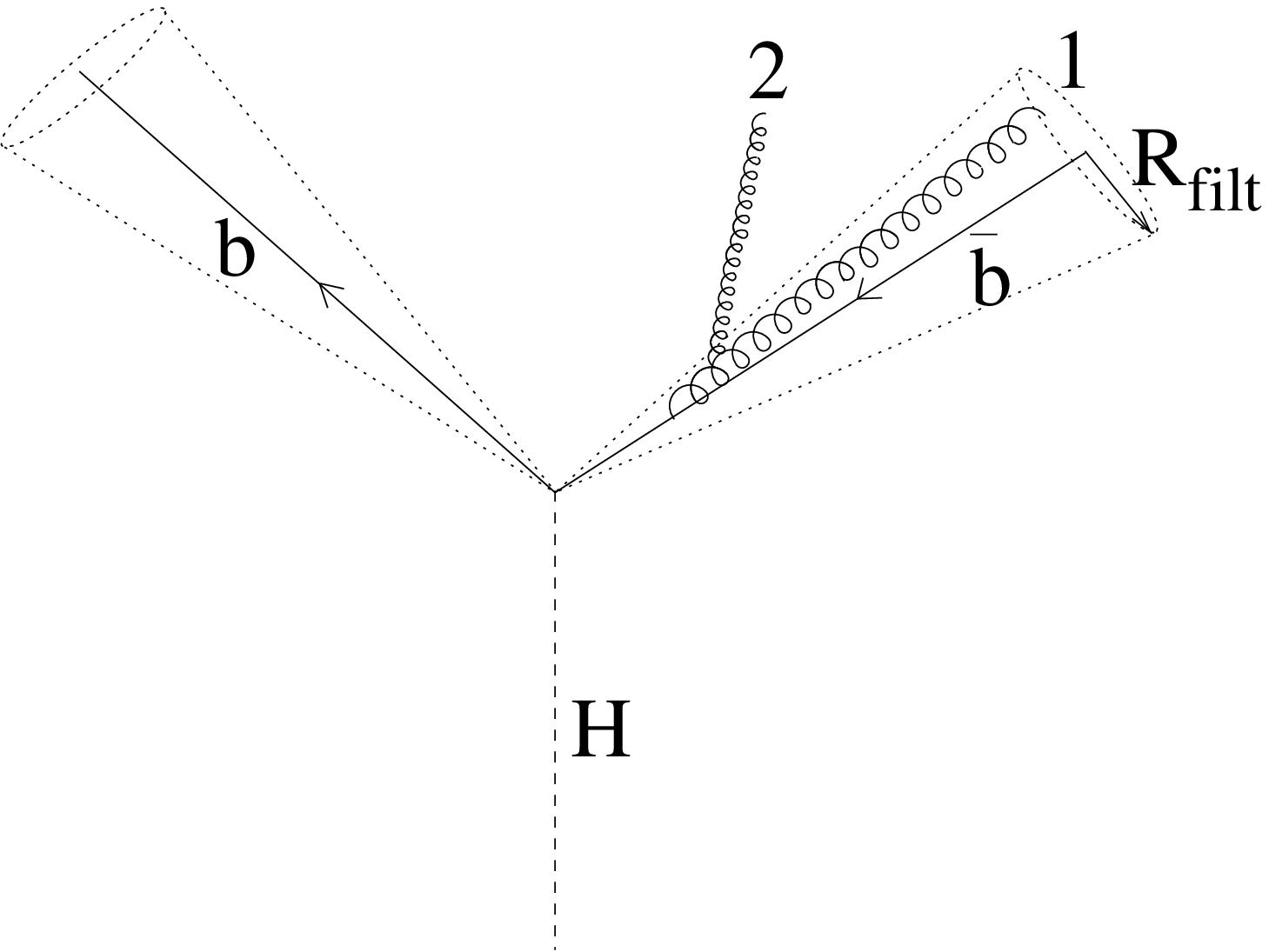}
    \label{NG_nfilt2}
  }~
  \subfigure[]{
    \includegraphics[scale=0.4]{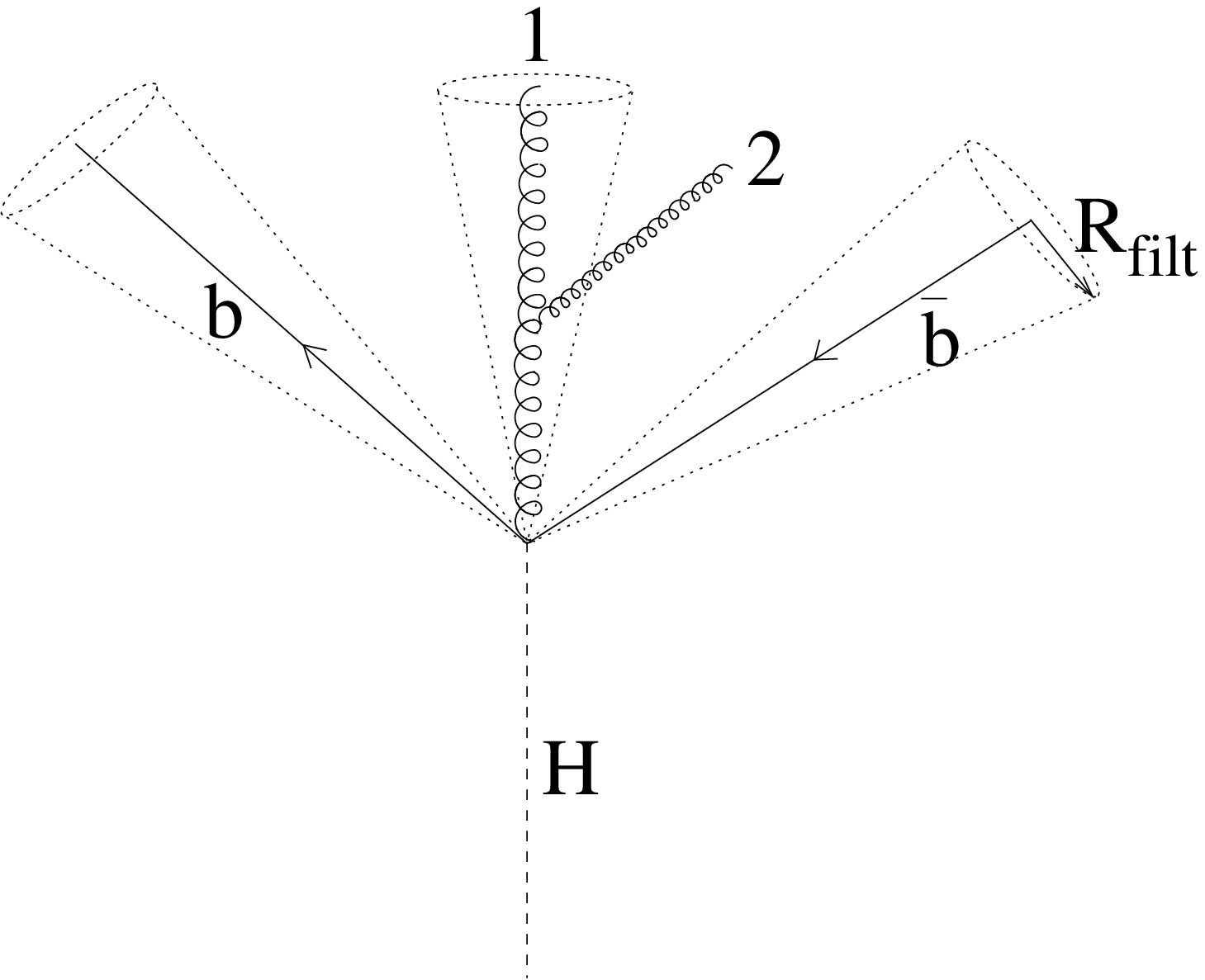}
    \label{NG_nfilt3}
  }
  \caption{Configurations leading to non-global logarithms when \subref{NG_nfilt2} $\filt{n}=2$ and \subref{NG_nfilt3} $\filt{n}=3$. In each case, the hardest gluon $1$, which is inside the filtered jet region, emits a softer gluon $2$ outside the filtered jet region.}
  \label{non-global_configuration}
\end{figure}

For this observable, soft gluons emissions lead to powers of $\ln\frac{M_H}{\Delta M}$, whereas collinear gluons emissions leads to powers of $\ln\frac{R_{bb}}{\filt{R}}$. In this study, gluons are strongly ordered in energy (the first emitted gluon being the most energetic one, and so on), and we aim to control the $\left(\as\ln\frac{M_H}{\Delta M}\right)^k$ series, in a region where
\begin{equation}
  \ln\frac{M_H}{\Delta M}\gg \ln\frac{R_{bb}}{\filt{R}}\,.
\end{equation}
Therefore, at leading-log accuracy, one has to resum terms like 
\begin{equation}
  I_k(\Delta M) = f_k\left(\frac{R_{bb}}{\filt{R}}\right)\left(\alpha_s\ln\frac{M_H}{\Delta M}\right)^k\,,
\end{equation}
where all the $f_k$ are functions to be computed. We thus disregard all the subleading terms, i.e. those suppressed by at least one power of $\ln\frac{M_H}{\Delta M}$. Unfortunately, such a calculation is highly non-trivial due to the fact that the observable is {\it non-global}. This property, first studied in \cite{Dasgupta:2001sh}, means that it is sensitive to radiation in only a part of the phase space. In the case of $\Delta M$, only emissions of gluons outside the filtered jets region contribute to the observable (cf figure~\ref{non-global_configuration}).
As a consequence of this property, one must consider soft gluons emissions not just from the $\bbbar$ dipole (usually called primary emissions, the only ones that would be present in QED) but also from the whole ensemble of already emitted gluons \cite{Dasgupta:2001sh,Dasgupta:2002bw}. As the number of gluons is increased, the geometry and the color structure of all these gluons become rapidly too complex to perform an analytical calculation. Therefore, to deal with this, one is forced to apply numerical Monte-Carlo calculations that can easily take care of the geometry. But the colour structure remains prohibitive, and one must usually also resort to the large-$N_c$ approximation in order to go beyond the $2$ first orders in perturbation theory~\cite{Dasgupta:2001sh,Dasgupta:2002bw,Delenda:2006nf,Appleby:2002ke} (though some authors have derived some analytical results in special cases \cite{Banfi:2002hw,Hatta:2009nd} and others have examined contributions beyond the leading large-$N_c$ approximation \cite{Weigert:2003mm,Forshaw:2006fk}).

However, before considering numerical calculations, some results can be derived analytically at $2^{nd}$ order for $\filt{n}=2$ (where $f_1$ and $f_2$ are computed exactly) and $\filt{n}=3$ (where only the leading behaviour of the $f_k$ in $\ln\frac{R_{bb}}{\filt{R}}$ and $N_c$ is looked for).

\subsection{Some results for $\filt{n}=2$ \label{some_results_for_nf2}}

Perturbatively, one can write $\Sigma(\Delta M)$ as 
\begin{equation}
  \Sigma(\Delta M) = 1+\sum_{k=1}^{\infty}I_k(\Delta M)\,,
  \label{perturbative_expansion_of_Sigma_DeltaM}
\end{equation}
where $I_k(\Delta M)$ is the ${\cal O}\left(\alpha_s^k\right)$ contribution to the observable. To simplify the calculation, $\Sigma(\Delta M)$ will be computed using the anti-$k_t$ algorithm \cite{Cacciari:2008gp}, even if the numerical study will be done using the $C/A$ algorithm \cite{Dokshitzer:1997in,Wobisch:1998wt} to be in accordance with the choice in \cite{MyFirstPaper}. However, the anti-$k_t$ algorithm is enough to catch the dominant behaviour of the leading-log series, in the sense that it does not affect the leading large collinear logarithm in the function $f_k$ at small $\filt{R}$:
\begin{equation}
  f_k\left(\frac{R_{bb}}{\filt{R}}\right) = a_k\ln^k\left(\frac{R_{bb}}{\filt{R}}\right) + {\cal O}\left(\ln^{k-1}\left(\frac{R_{bb}}{\filt{R}}\right)\right)\,,\label{function_f_k}
\end{equation}
i.e. $a_k$ is unchanged when moving from $C/A$ to anti-$k_t$.\footnote{When $\filt{R}\sim \frac{R_{bb}}{2}$, the discarding of the ${\cal O}\left(\ln^{k-1}\left(\frac{R_{bb}}{\filt{R}}\right)\right)$ terms is not a priori justified, but fig.~\ref{comparison_with_theory}, which compares numerical results obtained using $C/A$ with analytical estimates using anti-$k_t$, supports the dominance of the leading collinear logarithms.} This jet algorithm gives simpler results because the gluons outside the filtered jet region tend not to cluster with the ones inside. It is this property which ensures that the hardest jets in an event are generally perfect cones, as particles usually cluster with the hardest ones in their neighbourhood first \cite{Cacciari:2008gp}.

As a first step, primary emissions are considered, defined to be those one would obtain if gluons were only emitted from the $\bbbar$ dipole (as for photons in QED).

\subsubsection{Primary emissions}

\begin{figure}[htbp]
  \begin{center}
    \includegraphics[scale=0.32]{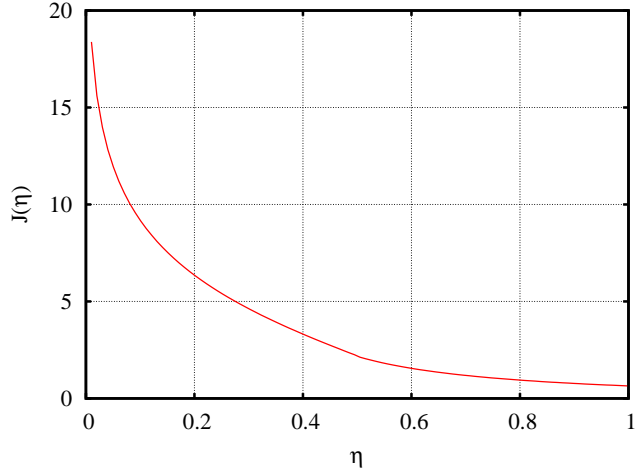}
  \end{center}
  \vspace{-0.8cm}
  \caption{The coefficient $J(\eta)$ in front of the primary soft logarithm $\ln\frac{M_H}{\Delta M}$.}
  \label{J_eta}
\end{figure}

Due to the use of the anti-$k_t$ algorithm, the result of the primary emissions can be easily shown to exponentiate, as will be roughly seen in the next section with the ${\cal O}\left(\alpha_s^2\right)$ analysis. Here, we just review the very well known result that the contribution to $\Sigma(\Delta M)$ from primary emissions, denoted $\Sigma^{(P)}(\Delta M)$, can be written as:\footnote{The superfix $(P)$ serves as a reminder that only primary emissions are being accounted for.}
\begin{equation}
  \Sigma^{(P)}(\Delta M) = e^{I_1(\Delta M)}\,,
  \label{primary_exponentiation}
\end{equation}
with:
\begin{equation}
  I_1(\Delta M) = \int \frac{d^3\vec{k}_1}{(2\pi)^32|\vec{k}_1|}M(k_1)\left(\Theta\left(\vec{k}_1\in J_{b\bb}\right) + \Theta\left(\vec{k}_1\notin J_{b\bb}\right)\Theta\left(\Delta M - \Delta M(\vec{k}_1)\right) - 1\right)\,.\label{definition_of_I_1}
\end{equation}
$M(k_1)$ is the matrix element squared for emitting one soft gluon from the $\bbbar$ dipole (the $b$ quark is taken to be massless):
\begin{equation}
  M(k_1) = 4\pi\alpha_sC_F\frac{2(p_b.p_{\bar{b}})}{(p_b.k_1)(k_1.p_{\bar{b}})}\,.
\end{equation}
Concerning the notations, $\Theta\left(\vec{k}_1\in J_{b\bb}\right)$ equals $1$ when gluon $1$ is emitted inside the jet regions around $b$ and $\bar{b}$, denoted by $J_{b\bb}$ (and is $0$ otherwise), which, for $\filt{R}<R_{bb}$, is just $2$ cones of radius $\filt{R}$ centered on $b$ and $\bar{b}$ (figure~\ref{NG_nfilt2}). Then, concerning the expression in brackets in eq.~(\ref{definition_of_I_1}), we separate the $2$ different regions where the gluon can be: either inside or outside the filtered Higgs jet. The first term $\Theta(\vec{k}_1\in J_{b\bb})$ means that the gluon does not contribute to the observable (as it is kept in the Higgs jet, the reconstructed Higgs mass is the true Higgs mass: $\Delta M(k_1)=0$). If the gluon is outside the filtered jet region (second term), then it does contribute to the observable:
\begin{equation}
  \Delta M(k)\sim k_t\frac{M_H}{p_{t_H}}\,,
\end{equation}
up to prefactors that can be neglected in the leading-log approximation, see appendix~\ref{app:analytical_considerations}. Finally, the $-1$ stands for the virtual corrections, for which there's obviously no loss of mass for the Higgs, and whose matrix element is just the opposite of the soft real one.\footnote{Even if the result seems obvious here, this way of doing the calculation can be easily generalised to higher orders and other kinds of jet algorithms.} One thus obtains:
\begin{equation}
  I_1(\Delta M) = -\int_{\vec{k}_1\notin J_{b\bb}}\frac{d^3\vec{k}_1}{(2\pi)^32|\vec{k}_1|}M(k_1)\Theta\left(\Delta M(k_1) - \Delta M\right)\,.
\label{PrimaryIntegral}
\end{equation}
The computation of this integral in the boosted regime, where $p_{tH}\gg M_H$, or equivalently $R_{bb}\ll 1$, is done in appendix~\ref{app:analytical_considerations}. From now on, we will essentially use $\filt{\eta} = \filt{R}/R_{bb}$ instead of $\filt{R}$ and we define $n\equiv \filt{n}$ and $\eta\equiv\filt{\eta}$ for more clarity in mathematical formulae. In order to keep in mind that it depends on the $2$ parameters of the Filtering analysis, the distribution $\Sigma(\Delta M)$ is renamed $\Sigma^{(n)}(\eta,\Delta M)$. What we obtain at fixed coupling is the following:\footnote{To obtain the result at running coupling, one simply makes the replacement (see eqs.~(\ref{t_running_coupling},\ref{t_fixed_coupling}) later in the article):
\begin{equation*}
\as\ln\frac{M_H}{\Delta M} \rightarrow  \frac{1}{2\beta_0}\ln\left(\frac{1}{1-2\beta_0\as(M_H)\ln\frac{M_H}{\Delta M}}\right)\,.
\end{equation*}}
\begin{equation}
  \Sigma^{(2),(P)}(\eta,\Delta M) = e^{-\frac{\alpha_sC_F}{\pi}J(\eta)\ln\frac{M_H}{\Delta M}}\,,
\label{exponentiated_primary_result}
\end{equation}
with
\begin{align}
  J(\eta) & = 2\ln\left(\frac{1-\eta^2}{\eta^2}\right) \hspace{5.02cm} \mbox{ if } \eta < \frac12 \label{primary_emission_result_eta_small}\,,\\
  & = \frac{8}{\pi}\int_{\eta}^{+\infty}\frac{du}{u(u^2-1)}\arctan\left(\frac{u-1}{u+1}\sqrt{\frac{2u+1}{2u-1}}\right) \mbox{ if } \frac12 < \eta < 1\,. \label{primary_emission_result_eta_large}
\end{align}
We give the value of $J(1)$, a quantity that is important to discuss some aspects of the results obtained in the following sections:
\begin{equation}
  J(1)\simeq 0.646\,. 
\end{equation}
Notice that the case $\eta>1$ will not be used in this study, but is mentioned in appendix~\ref{app:analytical_considerations}. The function $J(\eta)$ is plotted in figure~\ref{J_eta}.

Two remarks can be made:
\begin{enumerate}
  \item The result does not depend on the energy fraction $z$ of the Higgs splitting into $\bbbar$.
  \item When $\eta\ll 1$, another large logarithm $\ln\frac{1}{\eta}$ appears due to collinear enhancement.
\end{enumerate}

\subsubsection{Non-global contributions}

Now, we turn to the ${\cal O}\left(\alpha_s^2\right)$ term, and more precisely to the contribution of the non-global terms that have to be added to the primary logarithms computed in the previous section. That corresponds to the analysis of $I_2(\Delta M)$ in the perturbative expansion of $\Sigma(\Delta M)$ from eq.~(\ref{perturbative_expansion_of_Sigma_DeltaM}). The matrix element squared for $2$ real gluons emission is expressed as \cite{Bassetto:1984ik,Dasgupta:2001sh,Fiorani:1988by,Dokshitzer:1992ip}:
\begin{align}
  M(k_1\mbox{ real},k_2\mbox{ real}) & = (4\pi\alpha_s)^2(W_1+W_2)\,,\\
  \mbox{with } W_1 & = 4C_F^2\frac{(p_b.p_{\bar{b}})}{(p_b.k_1)(k_1.p_{\bar{b}})}\frac{(p_b.p_{\bar{b}})}{(p_b.k_2)(k_2.p_{\bar{b}})}\,, \\
  W_2 & = 2C_FC_A\frac{(p_b.p_{\bar{b}})}{(p_b.k_1)(k_1.p_{\bar{b}})}\left(\frac{(p_b.k_1)}{(p_b.k_2)(k_2.k_1)}+\frac{(p_{\bar{b}}.k_1)}{(p_{\bar{b}}.k_2)(k_2.k_1)}-\frac{(p_b.p_{\bar{b}})}{(p_b.k_2)(k_2.p_{\bar{b}})}\right)\,.
\end{align}
This expression is valid when there is a strong energy ordering between the two real gluons $1$ and $2$, either $E_1\gg E_2$ or $E_2\gg E_1$ (the formula is completely symmetric under the interchange $k_1\leftrightarrow k_2$). For the cases with one or both gluons being virtual, the following matrix elements are obtained, valid only when $E_1\gg E_2$~\cite{Dokshitzer:1992ip}:
\begin{align}
  M(k_1\mbox{ real},k_2\mbox{ virt}) & = -(4\pi\alpha_s)^2(W_1+W_2)\,, \nonumber\\
  M(k_1\mbox{ virt},k_2\mbox{ real}) & = -(4\pi\alpha_s)^2W_1\,, \nonumber \\
  M(k_1\mbox{ virt},k_2\mbox{ virt}) & = (4\pi\alpha_s)^2W_1\,.
\end{align}
Using these properties, separating the $4$ phase space regions depending on whether the gluons are inside or outside the filtered jet region in the same way as was done for $I_1$, and defining $dk$ as:
\begin{equation}
  dk = \frac{d^3\vec{k}}{(2\pi)^32|\vec{k}|}\,,
\end{equation}
we can then write $I_2$ in the following form:
\begin{align}
  I_2 = & \int dk_1dk_2 (4\pi\alpha_s)^2\Theta(E_1-E_2) \big\{ \nonumber \\
  & \quad\hspace{0.08cm} \Theta(k_1\in J_{b\bb})\Theta(k_2\in J_{b\bb})\big((W_1+W_2) - (W_1+W_2) - W_1 +W_1\big)\nonumber \\
  & {} + \Theta(k_1\in J_{b\bb})\Theta(k_2\notin J_{b\bb})\big((W_1+W_2)\Theta(\Delta M - \Delta M(k_2))-(W_1+W_2)-W_1\Theta(\Delta M - \Delta M(k_2))+W_1\big)\nonumber \\
  & {} + \Theta(k_1\notin J_{b\bb})\Theta(k_2\in J_{b\bb})\big((W_1+W_2)\Theta(\Delta M - \Delta M(k_1))-(W_1+W_2)\Theta(\Delta M - \Delta M(k_1))-W_1+W_1 \big)\nonumber \\
  & {} + \Theta(k_1\notin J_{b\bb})\Theta(k_2\notin J_{b\bb})\big((W_1+W_2)\Theta(\Delta M - \Delta M(k_1,k_2)) - (W_1+W_2)\Theta(\Delta M - \Delta M(k_1))\nonumber\\
  & {} \hspace{4.5cm} -W_1\Theta(\Delta M - \Delta M(k_2))+W_1 \big) \big\}\,.
  \label{SystematicApproach}
\end{align}
For each phase space region, the $4$ terms $(k_1,k_2)=$ (real,real) $-$ (real,virt) $-$ (virt,real) $+$ (virt,virt) are considered. The strong energy ordering $E_1\gg E_2$ implies that $\Delta M(k_1,k_2) = \Delta M(k_1)$, and one immediately gets:
\begin{align}
  I_2 = & \quad  \int dk_1dk_2 (4\pi\alpha_s)^2\Theta(E_1-E_2)\Theta(k_1\notin J_{b\bb})\Theta(k_2\notin J_{b\bb})W_1\Theta(\Delta M(k_2) - \Delta M) \nonumber\\
  & {} - \int dk_1dk_2 (4\pi\alpha_s)^2\Theta(E_1-E_2)\Theta(k_1\in J_{b\bb})\Theta(k_2\notin J_{b\bb})W_2\Theta(\Delta M(k_2) - \Delta M)\,, \nonumber\\
 = & \quad I_2^{(P)}(\Delta M) + I_2^{(NG)}(\Delta M)\,, \label{DefinitionOfI_2}
\end{align}
where $I_2^{(P)}(\Delta M)$ corresponds to the first integral containing the function $W_1$ whereas $I_2^{(NG)}(\Delta M)$ corresponds to the second integral with the function $W_2$. $I_2^{(P)}$ is just the second order contribution to the primary emissions, already computed above. To be convinced, one can notice that $(4\pi\alpha_s)^2W_1$  can be expressed as the product of $2$ one-gluon matrix elements $M(k_1)M(k_2)$ and, when $E_1\gg E_2$, 
\begin{equation}
\Theta\left(\Delta M(k_2)-\Delta M\right) = \Theta\left(\Delta M(k_2)-\Delta M\right)\Theta\left(\Delta M(k_1)-\Delta M\right)\,,
\end{equation}
if $k_1$ and $k_2$ belong to the same phase space region. Therefore $I_2^{(P)}$ can be written in a more symmetric way:
\begin{align}
  I_2^{(P)}(\Delta M) & = \frac12\int dk_1dk_2 (4\pi\alpha_s)^2\Theta(k_1\notin J_{b\bb})\Theta(k_2\notin J_{b\bb})W_1\Theta(\Delta M(k_1) - \Delta M)\Theta(\Delta M(k_2) - \Delta M)\,, \nonumber \\
  & = \frac12\left(\int dk \Theta(k\notin J_{b\bb})M(k)\Theta(\Delta M(k) - \Delta M)\right)^2\,, \nonumber \\
  & = \frac12 \left(I_1(\Delta M)\right)^2\,,
\end{align}
so that it corresponds to the second order perturbative expansion of the result eq.~(\ref{primary_exponentiation}), obtained with primary emissions only.

The important term for this section is the one containing $W_2$, denoted by $I_2^{(NG)}$. As mentioned in section~\ref{the_filtered_higgs_mass_a_non_global_observable}, it receives a non-zero contribution when the hardest gluon $1$ is emitted inside the filtered jet region whereas the softest gluon $2$ is emitted outside. For the opposite configuration, there is an exact cancellation between gluon $2$ being real and virtual. Here again the computation of $I_2^{(NG)}$ is postponed to appendix~\ref{app:analytical_considerations}, giving directly what will help to interpret some results later. $S_2$ is defined such that
\begin{equation}
   I_2^{(NG)}(\eta,\Delta M) = \frac{1}{2}C_FC_A\left(\frac{\alpha_s}{\pi}\ln\left(\frac{M_H}{\Delta M}\right)\right)^2S_2(\eta)\,,
\end{equation}
where we explicitly introduce the dependence on $\eta$ and we factorize out the soft divergence, still revealed in the large logarithm $\ln\frac{M_H}{\Delta M}$. When $\eta<1/2$, the result for $S_2$ can be written as:
\begin{align}
  S_2(\eta) & = -\frac{\pi^2}{3}+8\int_0^1\frac{du_1}{u_1}\int_0^1\frac{du_2}{u_2}\left(\frac{1}{\sqrt{(1-\eta^2(u_1^2+u_2^2))^2-4\eta^4u_1^2u_2^2}}-\frac{1}{1-\eta^2(u_1^2+u_2^2)}\right) \,,\nonumber \\
  & = -\frac{\pi^2}{3} + 4\eta^4 + 12\eta^6 + {\cal O}\left(\eta^8\right)\,.
  \label{S_2_computed}
\end{align}
The important point to notice in this result is the absence of collinear logarithms, which would appear as $\ln\frac{1}{\eta}$, contrary to the primary emission case (eq.~(\ref{primary_emission_result_eta_small})). So that the primary emissions dominate for this observable, at least for $\eta$ sufficiently small.

As mentioned in previous studies \cite{Dasgupta:2001sh,Dokshitzer_et_al}, one notices the presence of ``$\pi^2$ terms'' in non-global results at second order.

\subsection{Some results for $\filt{n}=3$ \label{Some_results_for_nfilt_3}}

The goal in this section is to have an estimate of the analytical behavior in the large $N_c$ limit of $\Sigma^{(n)}(\eta,\Delta M)$ for $n=3$, which is the probability of having no second gluon emission leading to a $\Delta M'$ greater than $\Delta M$. Notice that, contrary to the previous part where we obtained the function $\Sigma^{(2)}$, only the leading behavior in $L=\ln\frac{1}{\eta}$ and $N_c$ will be derived, so that in this context $\Sigma^{(2)}$ can be simply written:\footnote{This results simply from the combination of equations~(\ref{exponentiated_primary_result}) and~(\ref{primary_emission_result_eta_small}) with $\eta \ll 1$, and $C_F=\frac{N_c}{2}$ in the large $N_c$ limit.}
\begin{equation}
  \Sigma^{(2)}(L,t) = e^{-4N_cLt} \label{Sigma2LargeLLimit}
\end{equation}
where for further convenience we introduce the parameter $t = \frac{\alpha_s}{2\pi}\ln\frac{M_H}{\Delta M}$ and we change the arguments of $\Sigma$ which becomes now a function of $L$ and $t$. In this formula, $2L = 2\ln\frac{R_{bb}}{\filt{R}}$ can be interpreted as the ``logarithmic size'' of the $\bbbar$ dipole, i.e. the allowed phase space in rapidity for an emission from this dipole (in its center of mass frame) outside the jet region. The parameter $t$ means that this emission cannot occur with a $t'$ between $0$ and $t$.

\begin{figure}[htbp]
  \begin{psfrags}
    \psfrag{S}{\large $\Sigma^{(3)}(L,t)\hspace{0.3cm} \simeq$}
    \psfrag{A}{$2L$}
    \psfrag{C}{$2l$}
    \psfrag{I}{\huge $\int$}
    \psfrag{J}{$dl dt'$}
    \psfrag{D}{$g(t')$}
  \begin{center}
    \includegraphics[scale=0.55]{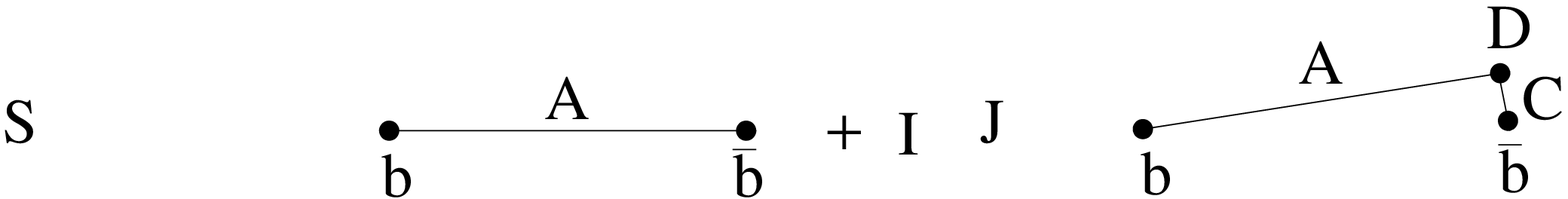}
  \end{center}
  \end{psfrags}
  \caption{How to compute the leading behavior of $\Sigma^{(3)}(L,t)$ from $\Sigma^{(2)}(L,t)$ when $L\gg 1$ and $N_c\gg 1$. In the second term, $t'$ is the gluon's emission scale.}
\label{PictureOfSigma3}
\end{figure}

Now we turn to $\Sigma^{(3)}(L,t)$. To have no second gluon emission in $[0,t]$, either there is no first gluon emission in $[0,t]$ outside the jet region (which corresponds to $\Sigma^{(2)}(L,t)$), or there is such an emission but the new dipole configuration is prohibited from emitting a second gluon in $[0,t]$ outside the jet region. This is depicted in figure~\ref{PictureOfSigma3}.
As the calculation is done in the large-$N_c$ limit, after the emission of a first gluon, the second one cannot be emitted from the $\bbbar$ dipole, but only from the $bg$ and $\bb g$ ones. Fig.~\ref{PictureOfSigma3} can be translated mathematically as:
\begin{equation}
  \Sigma^{(3)}(L,t) \simeq \Sigma^{(2)}(L,t) + \int_0^{t}dt'4N_c\Sigma^{(2)}(L,t')\int_0^Ldl\hspace{0.1cm}\Sigma^{(2)}(L,t-t')\Sigma^{(2)}(l,t-t')\,. \label{Sigma3_not_yet_integrated}
\end{equation}
Notice that $L_{bg}$, the logarithmic size of the $bg$ dipole in figure~\ref{PictureOfSigma3}, does not depend on $l$ in the leading collinear log approximation.\footnote{One can easily show the following relation:
\begin{equation*}
  L_{bg} = 2L + {\cal O}\left(e^{l-L}\right)\,.
\end{equation*}
If we introduce the neglected component of $L_{bg}$ in the $\Sigma^{(3)}$ calculation eq.~(\ref{Sigma3_not_yet_integrated}), then we have to compute an integral of the form
\begin{equation*}
  \int_0^Ldl\, \frac{1-e^{(l+{\cal O}(e^{l-L}))t}}{l+{\cal O}(e^{l-L})}\,.
\end{equation*}
Expanding the exponential and keeping the term of order $k$ gives
\begin{equation*}
  \int_0^Ldl\, \left(l+{\cal O}(e^{l-L})\right)^{k-1}t^k = \frac{(Lt)^k}{k}+{\cal O}(L^{k-2}t^k)\,.
\end{equation*}
The leading ${\cal O}\left((Lt)^k\right)$ term is already taken into account in eq.~(\ref{Sigma3LargeLLimit}). Therefore, including the $l$ dependent component of $L_{bg}$ gives rise to terms of the form $N_c^kL^{k-2}t^k$ at order $k$, suppressed by $2$ powers of $L$ with respect to the leading one.} In this expression, $4N_cL\Sigma^{(2)}(L,t')dt'$ is the probability not to emit the first gluon in $[0,t']$ and to emit it only at $t\in [t',t'+dt']$. The remaining part $\frac{1}{L}\int_0^Ldl\,\Sigma^{(2)}(L,t-t')\Sigma^{(2)}(l,t-t')$ is the probability to emit no second gluon from the $bg$ and $\bb g$ dipoles in $[t',t]$. Using eq.~(\ref{Sigma2LargeLLimit}) for $\Sigma^{(2)}$, $\Sigma^{(3)}$ is then given by:
\begin{equation}
  \Sigma^{(3)}(L,t) \simeq e^{-4N_cLt}\left(1+\int_0^{4N_cLt}dt'\frac{1-e^{-t'}}{t'}\right)\,.
  \label{Sigma3LargeLLimit}
\end{equation}
 Two limits can be considered:
\begin{equation}
  \Sigma^{(3)}(L,t) \simeq \left\{\begin{array}{ll}
        1-\frac{3}{4}(4N_cLt)^2+{\cal O}\left((4N_cLt)^3+N_ct\right) & \mbox{ if } 4N_cLt\ll 1\,,\\
        & \\
        e^{-4N_cLt}\left(\ln\left(4N_cLt\right)+{\cal O}(1)\right) & \mbox{ if } 4N_cLt\gg 1\,. \end{array} \right.
    \label{Sigma3allTLimit}
\end{equation}
The limit $4N_cLt\ll 1$ reveals two important aspects:
\begin{enumerate}
  \item One can notice the absence of the ${\cal O}(Lt)$ term, which is indeed the goal of the filtering analysis as it was presented in its original version \cite{MyFirstPaper}: it is intended to catch the major part of the ${\cal O}(\as)$ perturbative radiation. It cannot catch {\it all} the ${\cal O}(\as)$ contribution because a hard gluon emitted at an angle $\theta>R_{bb}$ from the $b$ and $\bb$ escapes the filtering process as it is rejected by the Mass Drop analysis. Therefore, when $4N_cLt\ll 1$, the expansion eq.~(\ref{Sigma3allTLimit}) misses a term ${\cal O}(N_ct)$, but this is legitimate in a leading collinear log estimate. Notice that the missing term is simply $-J(1)N_ct$ where $J(\eta)$ was given in eq.~(\ref{primary_emission_result_eta_large}).
  \item it shows that the purely non-global result for $n=3$ contains large collinear logarithms $L$, contrary to the case $n=2$ (eq.~(\ref{S_2_computed})). Indeed, the primary result for $n=3$ at second order can be proven to behave as\footnote{In fact, one can show the following general estimate for the primary emissions in the leading soft and collinear approximations: $$\Sigma^{(n)}(L,t) = e^{-8C_FLt}\sum\limits_{k=0}^{n-2}\frac{(8C_FLt)^k}{k!}\,.$$} $-32C_F^2(Lt)^2$ at order $\alpha_s^2$, so that the $S_2$ term for $n=3$ should be equivalent to $-8C_FC_A(Lt)^2$ at large $L$.
\end{enumerate}

Having understood some analytical features of the Filtering analysis, we now examine what can be learnt from a numerical calculation of the reconstructed Higgs mass observable.

\section{Non-Global structure: numerical results \label{NG_structure_numerical_results}}

In all that follows $t$ is defined so as to gather all the information about the soft logarithms in a running coupling framework:
\begin{align}
  t & = \frac{1}{2\pi}\int_0^{p_{t_H}}\frac{dk_t}{k_t}\alpha_s\left(k_t\frac{M_H}{p_{t_H}}\right)\Theta\left(\Delta M(k)-\Delta M\right)\,,\nonumber\\
  & = \frac{1}{2\pi}\int_{p_{t_H}\frac{\Delta M}{M_H}}^{p_{t_H}}\frac{dk_t}{k_t}\alpha_s\left(k_t\frac{M_H}{p_{t_H}}\right)\,,\nonumber\\
  & = \frac{1}{4\pi\beta_0}\ln\left(\frac{1}{1-2\beta_0\alpha_s(M_H)\ln\frac{M_H}{\Delta M}}\right)\,, \label{t_running_coupling}
\end{align}
where the last equality holds at the one-loop level and $\beta_0=\frac{11C_A-2n_f}{12\pi}$. The argument of $\alpha_s$ was taken as the gluon's transverse momentum with respect to the Higgs boson direction, of order $k_t\frac{M_H}{p_{t_H}}$, $k_t$ being its transverse momentum with respect to the beam. In the case of a fixed coupling constant $\alpha_s$, the definition for $t$ here coincides with that of section~\ref{Some_results_for_nfilt_3}:
\begin{equation}
  t = \frac{\alpha_s}{2\pi}\ln\frac{M_H}{\Delta M}\,.
  \label{t_fixed_coupling}
\end{equation}
But from now on, and unless stated otherwise, $t$ is given in the running coupling framework, eq.~(\ref{t_running_coupling}), and the function $\Sigma(\eta,\Delta M)$ is rewritten as $\Sigma(\eta,t)$.

To get an idea of the range of values covered by $t$, table~\ref{some_particular_t_values} presents a few $t$ values corresponding to a given $\Delta M$ for a Higgs mass of $115$ GeV ($\alpha_s(M_H)=0.114$). It reveals that the physical values for $t$ are below $0.15$.

\begin{table}[htb]
  \centering
  \begin{tabular}{|c|c|c|c|c|c|c|c|}
    \hline
    $\Delta M$ (GeV) & 1 & 2 & 5 & 10 & 20 & 50 & 115 \\
    \hline
    $t$ & 0.141 & 0.108 & 0.075 & 0.054 & 0.036 & 0.016 & 0 \\
    \hline
  \end{tabular}
  \caption{Correspondence between $\Delta M$ and $t$ for some particular values.}
  \label{some_particular_t_values}
\end{table}

\vspace{0.2cm}

To numerically investigate non-global observables, two approaches can be followed: 
\begin{itemize}
  \item an all-orders approach where one resums the leading-logs at all-orders in the large-$N_c$ limit, the output being the function $\Sigma(t)$, i.e. the probability that the loss of perturbative emission results in a Higgs mass in the range $[M_H-\Delta M(t),M_H]$, with
    \begin{equation}
      \Delta M(t) = M_He^{-\frac{1}{2\beta_0\as}\left(1-e^{-4\pi\beta_0t}\right)}\,,
    \end{equation}
    simply obtained by inverting the relation eq.~(\ref{t_running_coupling}).
  \item a fixed-order approach where the first few coefficients from the expansion of $\Sigma(t)$ are computed in the large-$N_c$ limit. More precisely, if $\Sigma(t) = \sum\limits_{k=0}^{\infty}\frac{c_k}{k!}\left(N_ct\right)^k$, then the program returns the first few coefficients $c_k$.
\end{itemize}

From a numerical point of view, the way to write an all-orders program was explained in \cite{Dasgupta:2001sh}. On the other side, a result at fixed-order may be obtained by developping a systematic approach like the one presented at second order in eq.~(\ref{SystematicApproach}). For the filtered Higgs jet mass observable, we used the Fastjet package \cite{Cacciari:2005hq} to perform the clustering (and mass-drop $+$ filtering) with the $C/A$ algorithm, consistently with the choice made in \cite{MyFirstPaper}.

As the all-orders program gives immediately what we are looking for, which is $\Sigma(t)$, we will use it (section~\ref{study_of_the_Higgs_perturbative_width}) to compute the perturbative Higgs width. But in order to check it and be confident with the results obtained, we compare them with the previous analytical estimates and see how well the perturbative leading log series fits them. This leads us to study the behaviour of the higher order terms and to gain a better understanding of the convergence and structure of the non-global series. Though treated in more details in appendix~\ref{app:convergence_of_the_non-global_series}, the main points are mentioned in this section.

\subsection{Comparison with analytics \label{comparison_with_analytics}}

\begin{figure}[htb]
  \centering
  \subfigure[]{
      \includegraphics[scale=0.275]{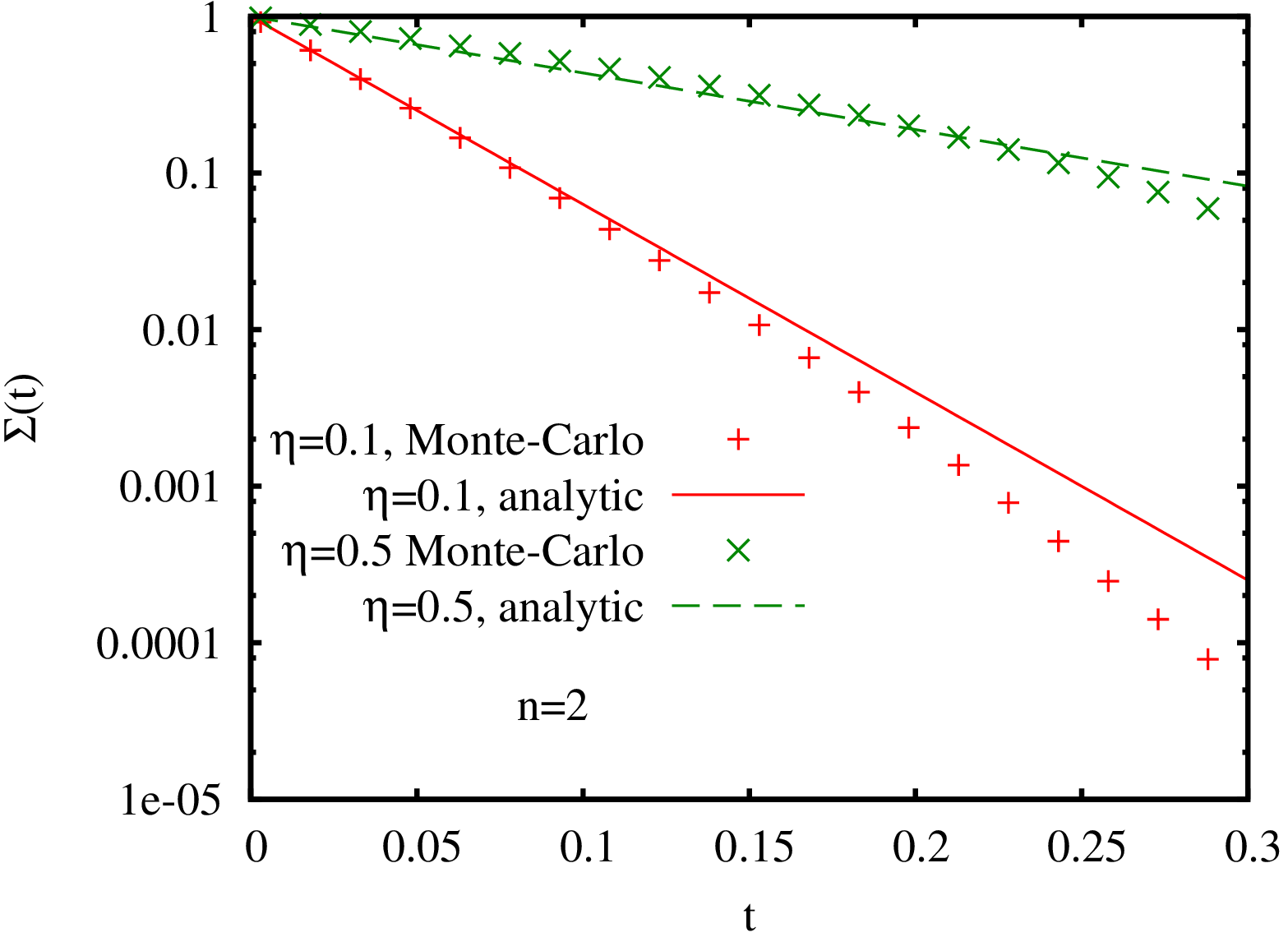}
      \label{ao_vs_an_nf2}
    }~
    \subfigure[]{
      \includegraphics[scale=0.275]{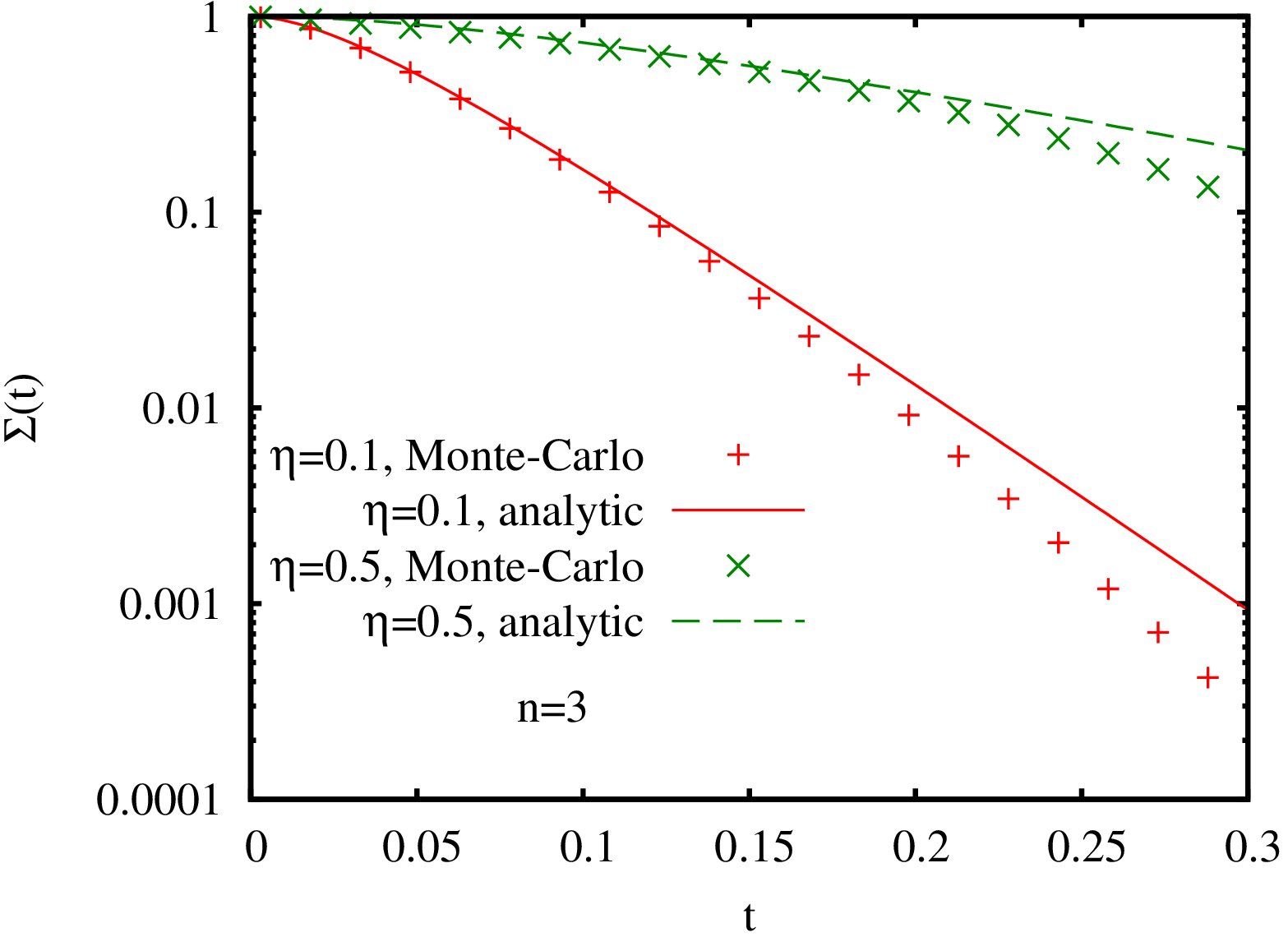}
      \label{ao_vs_an_nf3}
    }
    \caption{Comparison between numerical all-orders result (obtained using $C/A$ algorithm) and leading collinear logarithm estimate of $\Sigma(t)$ derived with anti-$k_t$ for \subref{ao_vs_an_nf2} $n=2$ and \subref{ao_vs_an_nf3} $n=3$ for $2$ values of $\eta$: $0.1$ and $0.5$.}
  \label{comparison_with_theory}
\end{figure}

Using the all-orders Monte-Carlo program, a comparison between the all-orders numerical curves obtained using the $C/A$ algorithm and their corresponding analytical estimates obtained previously with anti-$k_t$ in eqs.~(\ref{Sigma2LargeLLimit},\ref{Sigma3LargeLLimit}) can be done. The results are presented in figure~\ref{comparison_with_theory} and show good agreement, at least in the region of physical $t$ values. 

Notice that the slight discrepancy between analytical estimations and numerics starts to occur at $t>0.1$, which is at the edge of the physical region (cf table~\ref{some_particular_t_values}), beyond which $\Delta M$ would be below the perturbative scale of around $1$ GeV. This agreement manifests that:
\begin{itemize}
  \item In the physical region, the leading terms in $(\alpha_sLt)^k$, with $L=\ln\frac{1}{\eta}$ seem to completely dominate and we do not need to compute the subleading corrections.
  \item One can use these analytical expressions to get an accurate estimate of the reconstructed Higgs peak width.
\end{itemize}

\subsection{Comparison with fixed-order results \label{comparison_with_fixed_order_results}}

The structure of the non-global series at fixed-order is now examined so as to independently cross-check the all-orders program and to understand if the perturbative leading-log series can be usefully truncated.

\begin{figure}[htb]
  \centering
  \subfigure[]{
    \includegraphics[scale=0.275]{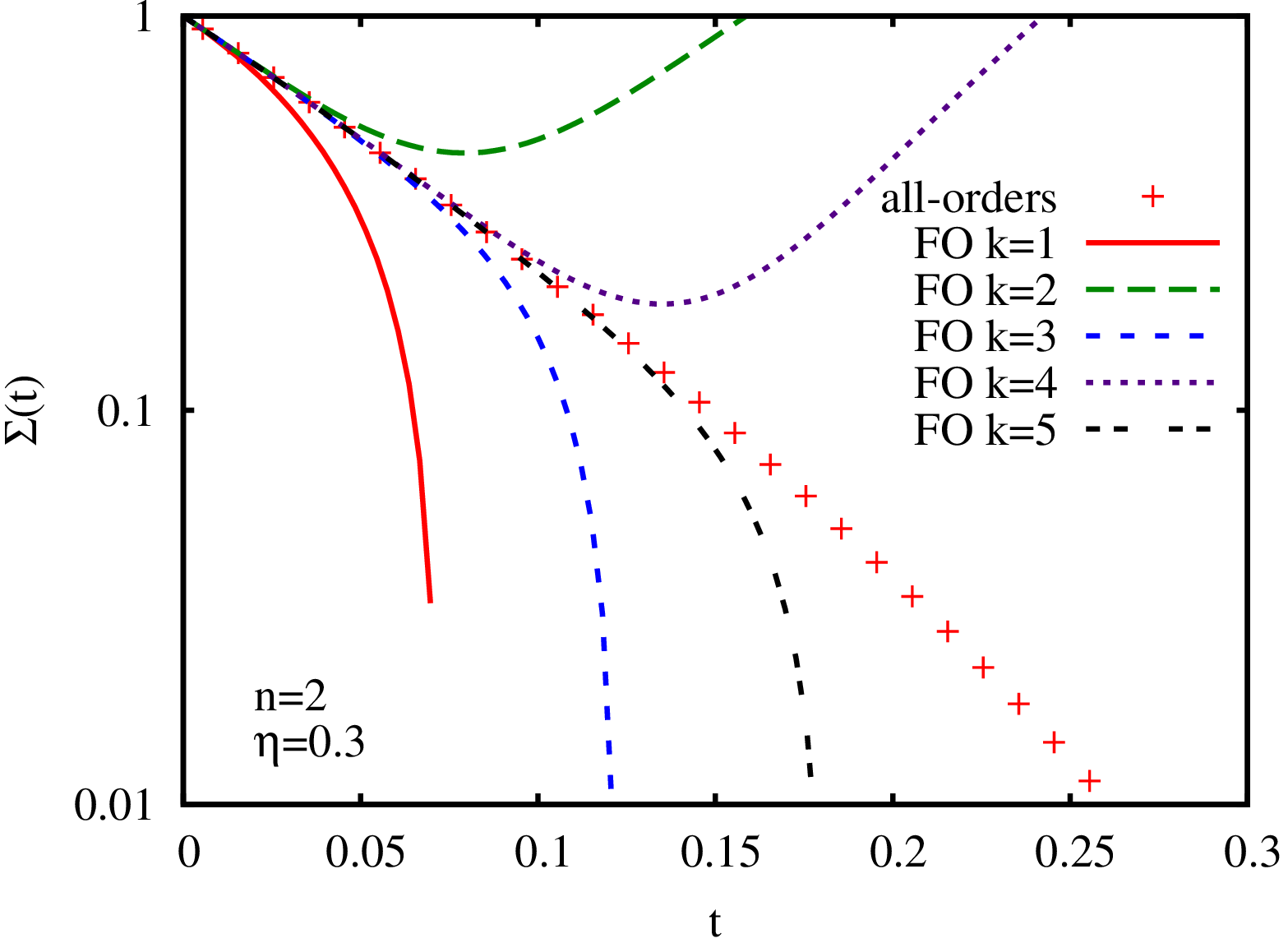}
    \label{ao_vs_fo_nf2_eta0.3}
  }~
  \subfigure[]{
    \includegraphics[scale=0.275]{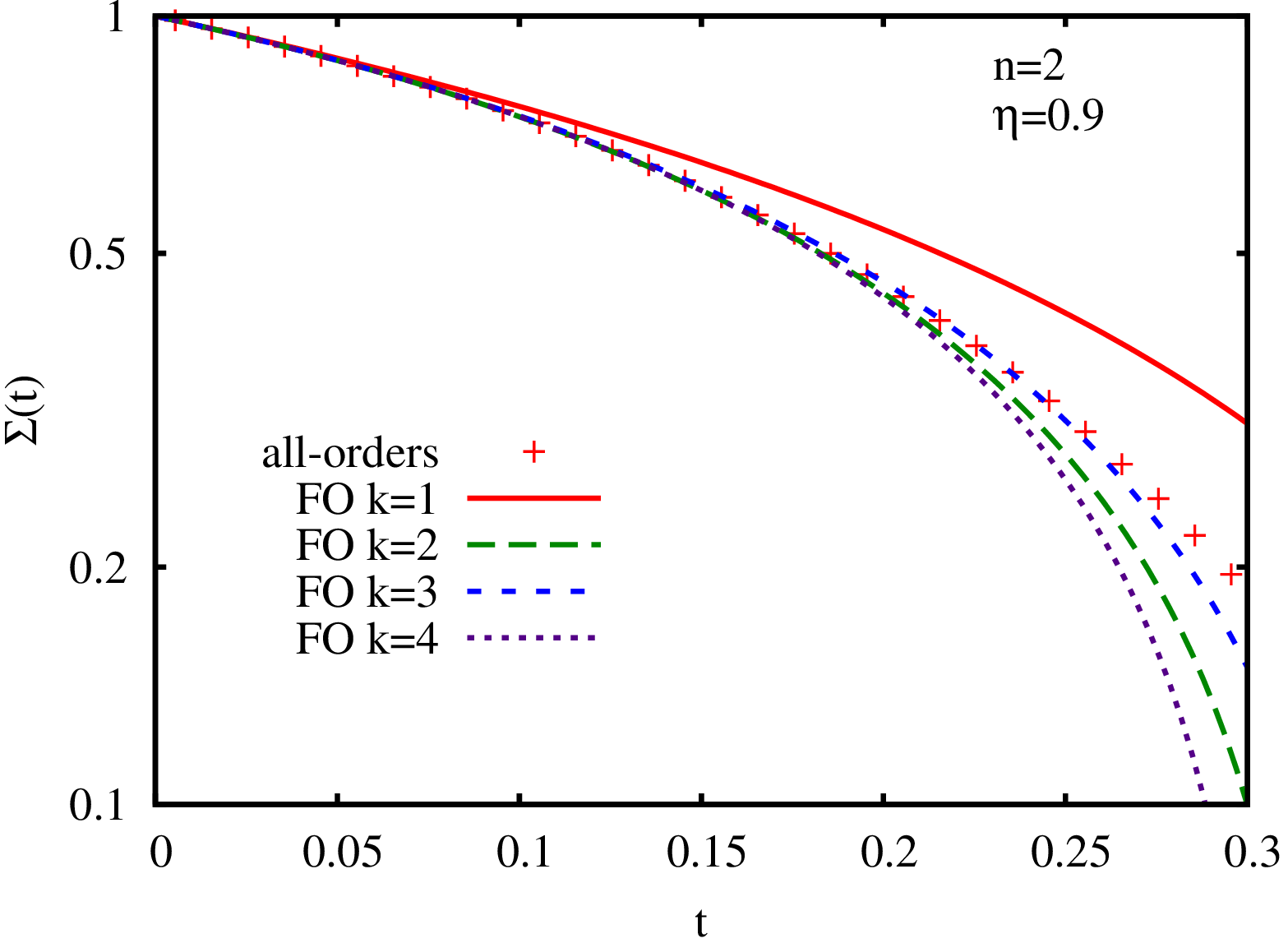}
    \label{ao_vs_fo_nf2_eta0.9}
  }
  \caption{Comparison between fixed-order (FO) expansion and all-orders result when $n=2$ for \subref{ao_vs_fo_nf2_eta0.3} $\eta=0.3$ and \subref{ao_vs_fo_nf2_eta0.9}  $\eta=0.9$. Both fixed-order and all-orders results were obtained using the $C/A$ algorithm.}
  \label{ao_vs_fo_nf2}
\end{figure}

As an example, figure~\ref{ao_vs_fo_nf2} compares the all-orders result to the fixed-order ones up to $\alpha_s^5$ for $n=2$ and two different values of $\eta$ (only the coefficients with an uncertainty of at most a few percent are plotted\footnote{This uncertainty obviously increases with the perturbative order, but also with $\eta$ because at small $\eta$, the coefficients are sensitive to the large logarithm $\ln\frac{1}{\eta}$, which is easy to compute.}). The curves are represented up to $t=0.3$, which is far beyond the physical region but is instructive to study the convergence of the series.

The left plot for $\eta=0.3$ shows a nice convergence of the perturbative series eq.~(\ref{perturbative_expansion_of_Sigma_DeltaM}), as the $t$ range for which the all-orders and fixed-order curves coincide grows with $k$. However, the second plot for $\eta=0.9$ gives an unexpected result: the fourth order diverges with respect to the third one, in the sense that the point of disagreement is shifted to smaller $t$. The question arises whether this divergence will remain at higher orders. To answer it, one needs to go further in perturbation theory. In appendix~\ref{app:convergence_of_the_non-global_series}, a parallel is made between the filtered Higgs jet observable and the slice observable, studied for instance in \cite{Dasgupta:2002bw}, for which, due to computationnal speed, it is possible to obtain reliable coefficients up to order $6$. The same effect is observed and is even enhanced at orders $5$ and $6$. Therefore, it seems that the fixed-order information cannot be safely used in general: one has to be aware that the leading-log large-$N_c$ non-global series may be divergent for any value of $t$.

\section{Choice of the filtering parameters \label{choice_of_the_filtering_parameters}}

In the previous sections we examined the structure and convergence of the perturbative leading-log series, analytically and numerically. We could then cross-check the analytical expressions and the fixed-order approach with the all-orders program, which we are going to use throughout this part.

We would like to decide how one should choose the filtering parameters ($n$,$\eta$) depending on the level of UE and PU as well as the $p_t$ of the Higgs boson. Here, we do not claim to make an exact and complete analysis, but we want to obtain some estimates. First, we consider the width of the Higgs mass distribution separately in presence of perturbative radiation (using the all-orders results) and UE/PU (using a simple model for it). Then, we try to minimize the Higgs width in presence of both of these effects. Finally, we will estimate hadronisation corrections.

In all this part, we set the Higgs mass $M_H$ at $115$ GeV, as in \cite{MyFirstPaper}.

\subsection{Study of the Higgs perturbative width \label{study_of_the_Higgs_perturbative_width}}

As we could see in the previous sections, even without considering additional particles from UE/PU, $\Delta M \equiv M_H-M_{\mbox{\tiny filtered jet}}\neq 0$ because of the loss of perturbative radiation. The Higgs boson thus acquires a perturbative width, denoted $\delta M_{PT}$. At first sight, knowing the distributions $\Sigma^{(n)}(\eta,\Delta M)$, one might simply define it as:
\begin{equation}
  \delta M_{PT} = 2\sqrt{\langle \Delta M^2\rangle - \langle \Delta M\rangle^2}\,,\label{simple_definition_for_delta_M}
\end{equation}
as we do for gaussian distributions for instance. Unfortunately, if we simply take $n=2$ as an example and if we consider the primary emission result eq.~(\ref{exponentiated_primary_result}), we can deduce the following distribution for $\Delta M$:
\begin{equation}
  \frac{d\Sigma^{(2)}(\eta,\Delta M)}{d\Delta M} = \frac{\as C(\eta)}{M_H^{\as C(\eta)}}\frac{1}{\Delta M^{1-\as C(\eta)}}\,,\label{differential_Sigma_2}
\end{equation}
with
\begin{equation}
  C(\eta) = \frac{C_F J(\eta)}{\pi}\,.
\end{equation}
Computing $\langle \Delta M \rangle$ and $\langle \Delta M^2 \rangle$ implies dealing with integrals of the form
\begin{align}
  \int_0^{M_H}\frac{d\Delta M}{\Delta M^{1-\as C(\eta)}}\Delta M & = \int_0^{M_H} d\Delta M \, \Delta M^{\as C(\eta)}\,,\\
  \int_0^{M_H}\frac{d\Delta M}{\Delta M^{1-\as C(\eta)}}\Delta M^2 & = \int_0^{M_H} d\Delta M \, \Delta M^{1+\as C(\eta)}\,.
\end{align}
Such integrals give a large importance to the $\Delta M \sim M_H/2$ region, where there should be very few events, and do not describe what happens in the neighbourhood of the peak near $\Delta M = 0$. Therefore, the definition eq.~(\ref{simple_definition_for_delta_M}) does not seem adequate for the perturbative width. That's why we shall adopt another definition, adapted from \cite{Buttar:2008jx}. The Higgs perturbative width is defined as the size $\delta M_{PT}$ for which a given fraction $f$ of events satisfy $0<\Delta M<\delta M_{PT}$. Using the all-orders function previously computed, this is equivalent to solving the equation $\Sigma^{(n)}(\eta,\Delta M) = f$. This leads to the width function $\delta M_{PT}(n,\eta,f)$.
\begin{figure}[hbt]
  \centering
  \subfigure[]{
    \includegraphics[scale=0.275]{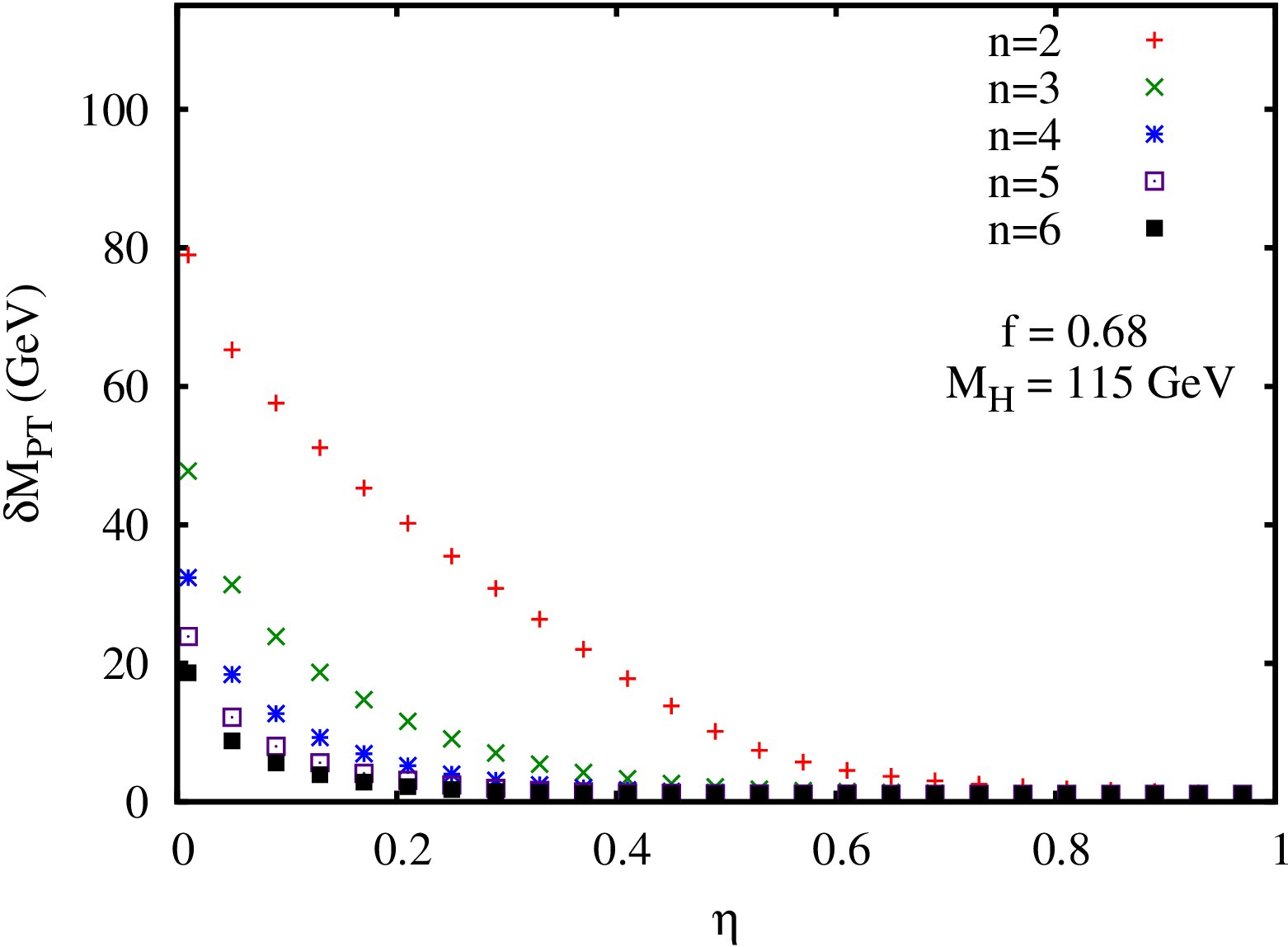}
    \label{Higgs_PT_width_GeV_f0.68}
  }~
  \subfigure[]{
    \includegraphics[scale=0.275]{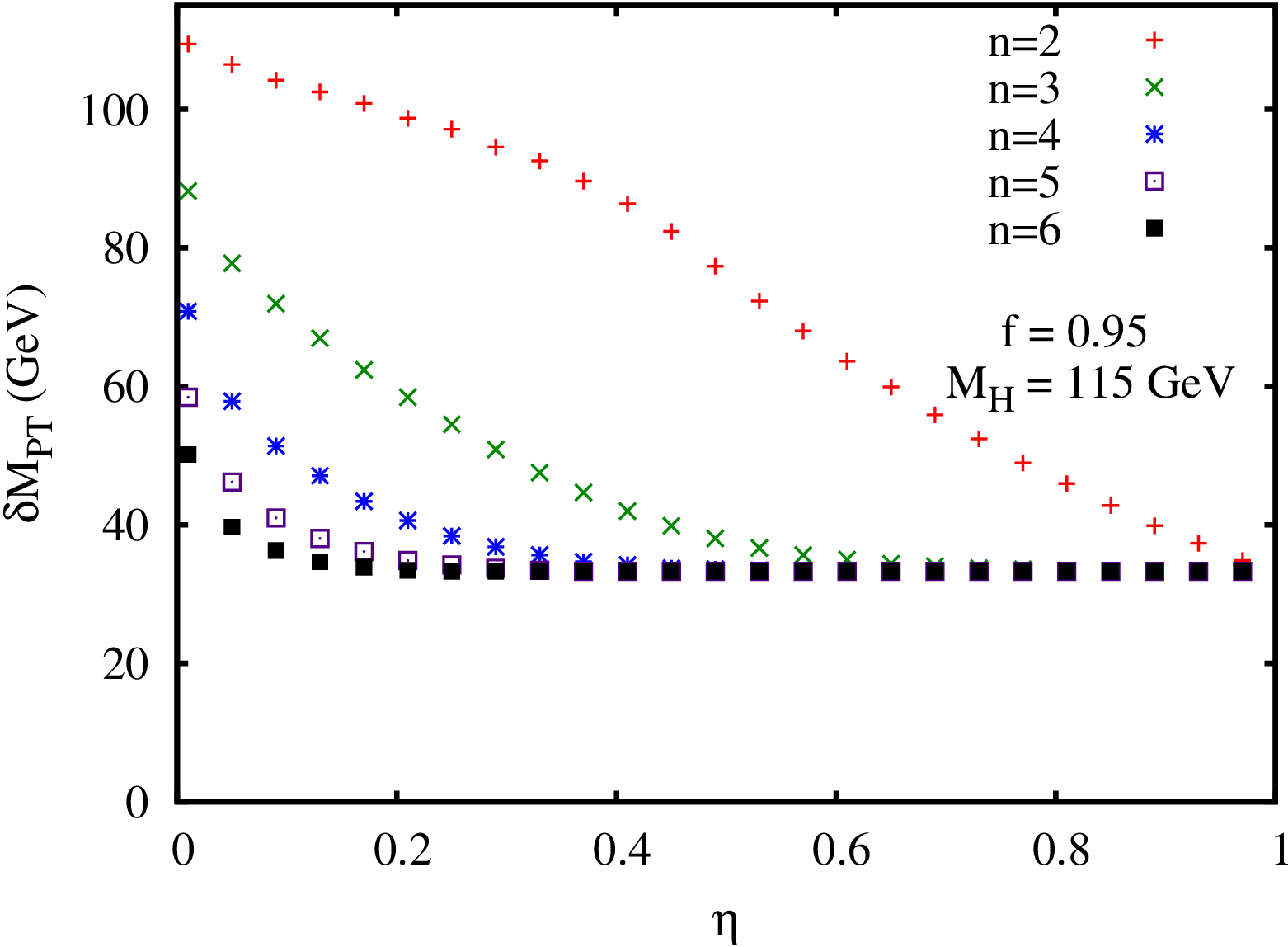}
    \label{Higgs_PT_width_GeV_f0.95}
  }
  \caption{Perturbative width of the Higgs boson (in GeV) as a function of $\eta$ for several values of $n$ when \subref{Higgs_PT_width_GeV_f0.68} $f=0.68$ and \subref{Higgs_PT_width_GeV_f0.95} $f=0.95$.}
  \label{Higgs_PT_width_GeV}
\end{figure}

Fig.~\ref{Higgs_PT_width_GeV} shows $\delta M_{PT}$ as a function of $\eta$ for $n=2\mathellipsis 6$. When $\Delta M\sim 50$ GeV (i.e. $\sim M_H/2$), one should be aware that soft approximation loses sense and results on these plots should no longer be taken seriously. We chose the values $f=0.68$ and $f=0.95$, corresponding respectively to $2\sigma$ and $4\sigma$ for gaussian distributions, to show that the Higgs mass perturbative distribution is not gaussian (otherwise, going from $2\sigma$ to $4\sigma$ would have multiplied the width by a factor of $2$, see also eq.~(\ref{differential_Sigma_2})). One important thing to notice is a kind of ``saturation'' effect that one observes for $\eta$ close enough to $1$ for every fraction $f$. It manifests itself as a flat curve at a value $\delta M_{PT} = \delta M_{sat}(f)$, independent of $n$. For instance, $\delta M_{sat}(f=0.68) \simeq 1$ GeV and $\delta M_{sat}(f=0.95) \simeq 33$ GeV. This can be understood simply by considering that when the radius of the filtering is large enough, say $\eta>\eta_{sat}(n)$, it captures (almost) all the particles resulting from the Mass Drop analysis, i.e. all those that are within angular distance $R_{bb}$ from $b$ or $\bb$, but it still fails to capture particles outside the Mass Drop region.\footnote{The probability to emit a gluon outside the MD region in $[0,t]$ is roughly given by $1-e^{-J(1)N_ct}$.} Of course, the larger $n$, the smaller $\eta_{sat}(n)$ as we keep more jets. This saturation property is equivalent to saying that all the functions $\Sigma^{(n)}(\eta,\Delta M)$ become independent of $n$ and $\eta$ when $\eta>\eta_{sat}(n)$.

For the rest of this analysis we keep the value $f=0.68$, even if it is not clear which value should be chosen, and more generally what should be the relevant definition of the Higgs perturbative width. However, we will mention in section~\ref{variations_of_the_results_with_z_and_f} what happens if we vary $f$ between $f=0.5$ and $f=0.8$, so as to obtain a measure of the uncertainty of the calculations.

\begin{figure}[htb]
  \centering
  \subfigure[]{
    \includegraphics[scale=0.275]{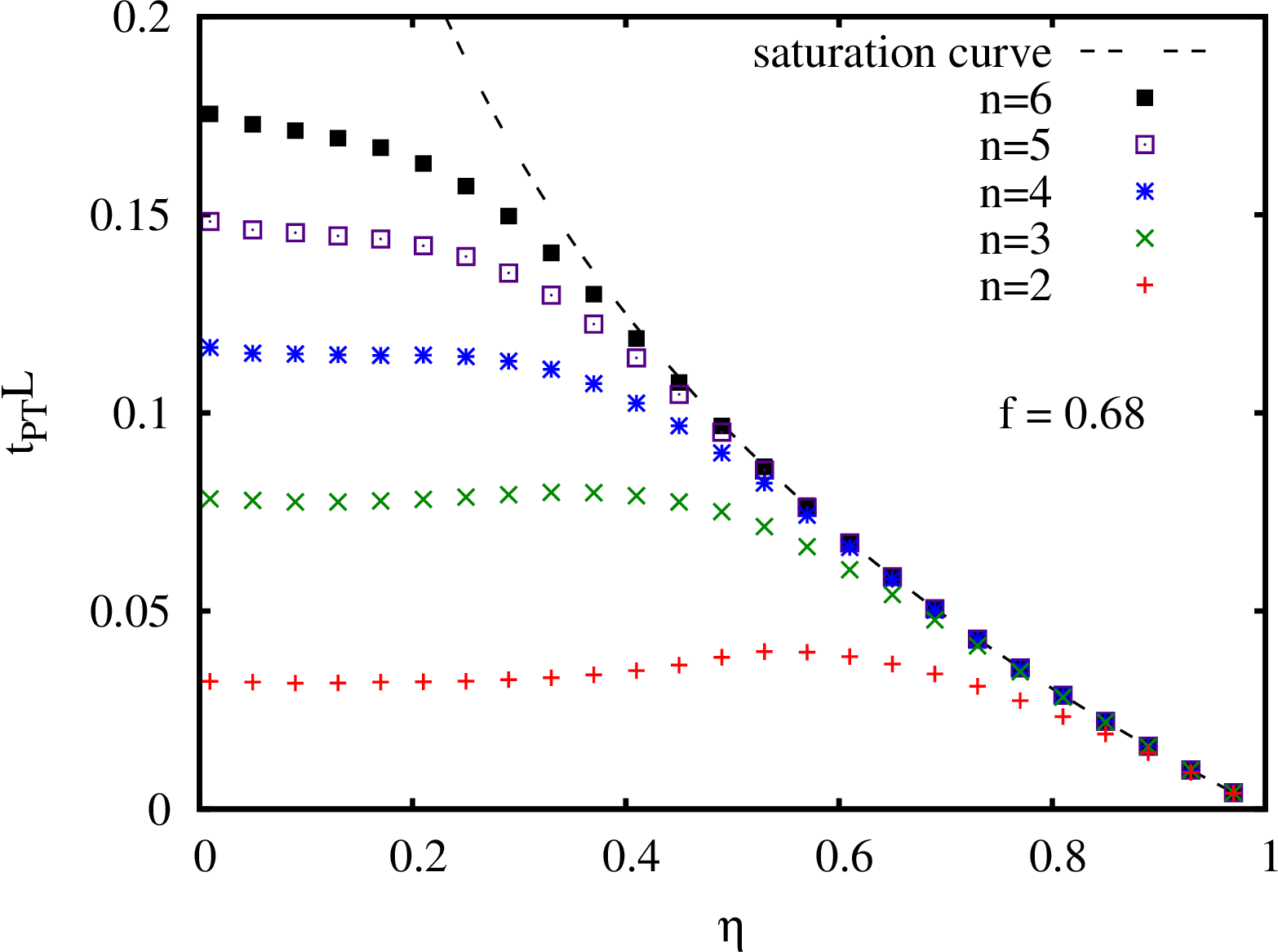}
    \label{Higgs_PT_width_t}
  }~
  \subfigure[]{
    \includegraphics[scale=0.275]{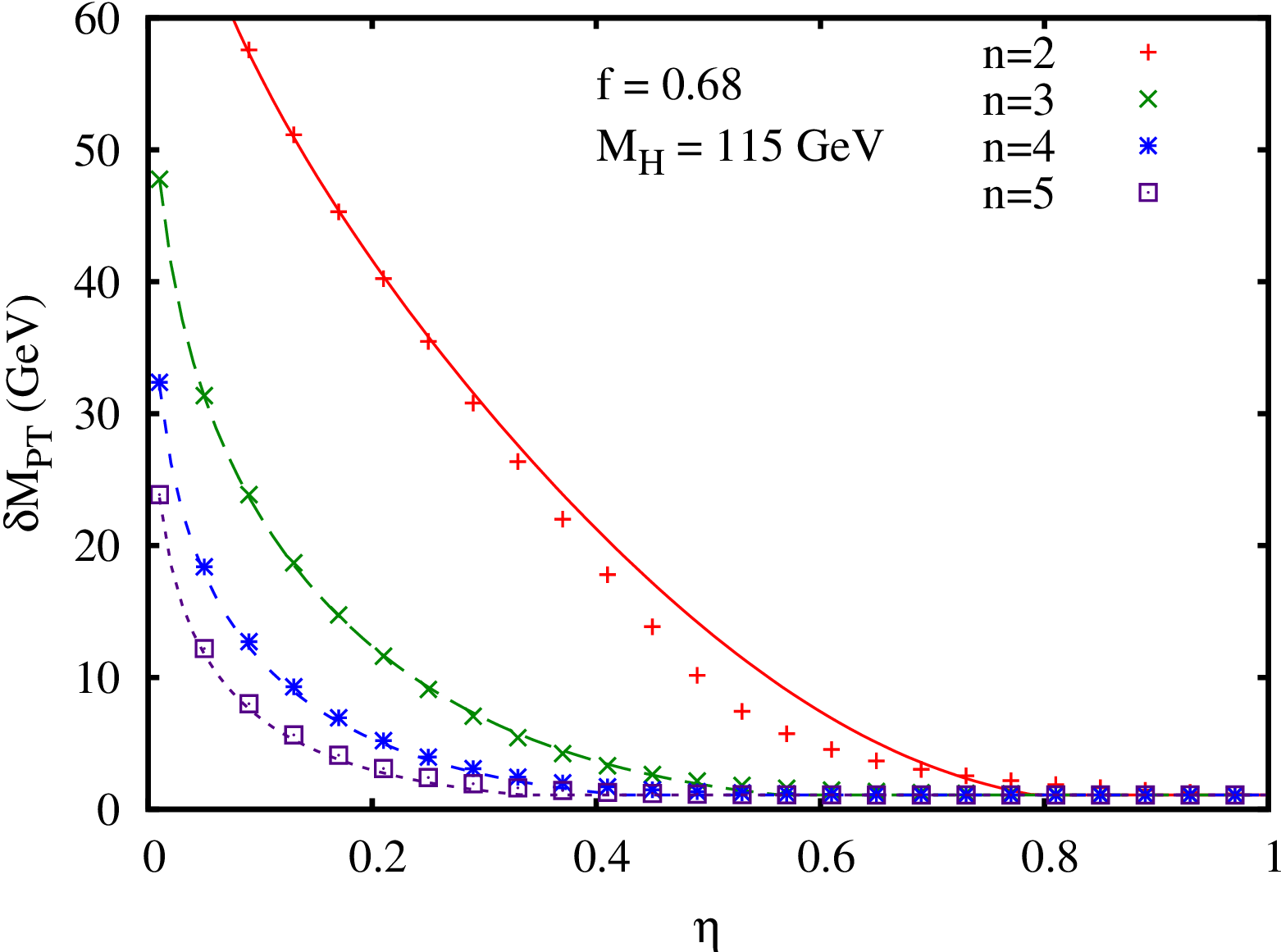}
    \label{Higgs_PT_width_dM_f0.68}
  }
  \caption{\subref{Higgs_PT_width_t} $t_{PT}L$ as a function of $\eta$ for $f=0.68$ and different values of $n$. The saturation curve simply comes from the fact that all widths saturate to the same constant $\delta M_{sat}$ for $\eta$ large enough, and its equation is therefore given by $t_{sat}L$ (with $t_{sat}(f=0.68)\simeq 0.136$). \subref{Higgs_PT_width_dM_f0.68} $\delta M_{PT}$ as a function of $\eta$ for $f=0.68$ and different values of $n$ (curves with points). For each $n$ is also represented the corresponding approximate width (lines) given by eq.~(\ref{parametrisation_of_PT_width}).}
  \label{Higgs_PT_width_t_dM}
\end{figure}

The curves in figure~\ref{Higgs_PT_width_GeV} only give us an overview of the scales involved in the Higgs boson width. But one can go a little further. At small $\eta$, we should get a large collinear enhancement revealing itself as a large logarithm $L=\ln\frac{1}{\eta}$ multiplying $t$. The perturbative expansion is thus a series in $\left(N_cLt\right)^k$. As a direct consequence, at small $\eta$, the all-orders function $\Sigma^{(n)}(\eta,t)$ can be written as a function of a single variable $\Sigma^{(n)}(N_cLt)$. Solving the ``width equation''
\begin{equation}
  \Sigma^{(n)}(N_cLt) = f\,,
\end{equation}
gives
\begin{equation}
  t_{PT} = \frac{C_{PT}(n,f)}{L}\,,
\end{equation}
where $t_{PT}$ is simply related to $\delta M_{PT}$ by
\begin{equation}
   t_{PT} = \frac{1}{4\pi\beta_0}\ln\left(\frac{1}{1-2\beta_0\as(M_H)\ln\frac{M_H}{\delta M_{PT}}}\right)\,,
   \label{Definition_of_t_PT}
\end{equation}
and where $C_{PT}(n,f)$ is a function, independent of $\eta$, which increases with $n$ and decreases when $f$ increases. This is confirmed by figure~\ref{Higgs_PT_width_t} which shows that $t_{PT}L$ is indeed independent of $\eta$ as long as $\eta$ and $n$ are not too large.

As an example, for $n=2$, let us take the simple result $\Sigma^{(2)}(L,t) = e^{-4N_cLt}$ from eq.~(\ref{Sigma2LargeLLimit}) in the small $\eta$ limit. It was shown in section~\ref{comparison_with_analytics} that this result is very close to the all-orders one in the physical $t$ region. Solving $\Sigma^{(2)}(L,t)=f$ immediately implies
\begin{equation}
  C_{PT}(2,f) = \frac{\ln\frac{1}{f}}{4N_c}\,, \label{C_PT_2_f}
\end{equation}
which, for $f=0.68$, gives $C_{PT} \simeq 0.032$ in accordance with figure~\ref{Higgs_PT_width_t}.

One observes that $t_{PT}L$ is not strictly speaking a constant for higher $n$ values. This may be due to the saturation effects discussed above. Indeed, even at large $L$, the perturbative expansion is not only a function of $Lt$ but also of $t$ for the lowest orders, as mentioned at the end of section~\ref{Some_results_for_nfilt_3}:
\begin{equation}
  \Sigma^{(n)}(L,t) = 1+\sum_{k=1}^{n-2}a_kt^k+\sum_{k=n-1}^{+\infty}\left(a_k(Lt)^k+{\cal O}\left(L^{k-1}t^k\right)\right)\,,
\end{equation}
If we only had QED like emissions, i.e. primary ones, with the use of the anti-$k_t$ jet algorithm, we would obtain $a_k = \frac{(-J(1)N_c)^k}{k!}$ for $k\leq n-2$, where $J(\eta)$ was derived in section~\ref{some_results_for_nf2}. As $n$ increases, the term $a_1t$ becomes more and more important with respect to $a_{n-1}(Lt)^{n-1}$, leading to larger and larger deviations from the simple law $t_{PT}L$ $=$ constant. However, until $n=5$, assuming $t_{PT}L$ is a constant at small $\eta$ seems a good approximation. Therefore, using eq.~(\ref{Definition_of_t_PT}), one can model the Higgs perturbative width in the following form:
\begin{equation}
  \delta M_{PT}(n,L,f) = \left\{\begin{array}{ll}
      M_He^{-\frac{1}{2\beta_0\as}\left(1-e^{-4\pi\beta_0\frac{C_{PT}(n,f)}{L}}\right)} & \mbox{ if } \eta<\eta_{sat}(n,f)\,,\\
      & \\
      \delta M_{sat}(f) & \mbox{ if } \eta>\eta_{sat}(n,f)\,.
      \end{array}\right.
\label{parametrisation_of_PT_width}
\end{equation}
$\eta_{sat}(n,f)$ is given by the intersection between the curve $t_{PT}=C_{PT}/L$ and $t_{PT}=t_{sat}$. Therefore:
\begin{equation}
  \eta_{sat}(n,f) = e^{-\frac{C_{PT}(n,f)}{t_{sat}}}\,.\label{eta_sat}
\end{equation}
Table~\ref{Cpt_and_eta_sat_values} shows $C_{PT}$ and $\eta_{sat}$ for $f=0.68$ and different $n$ values.
\begin{table}
  \centering
  \begin{tabular}{|c|c|c|c|c|}
    \hline
    $n$ & 2 & 3 & 4 & 5\\
    \hline
    $C_{PT}$ & 0.032 & 0.078 & 0.117 & 0.149\\
    \hline
    $\eta_{sat}$ & 0.79 & 0.56 & 0.42 & 0.34\\
    \hline
  \end{tabular}
  \caption{$C_{PT}$ and $\eta_{sat}$ as a function of $n$ when $f=0.68$.}
  \label{Cpt_and_eta_sat_values}
\end{table}
Figure~\ref{Higgs_PT_width_dM_f0.68} shows the curves corresponding to the parametrisation eq.~(\ref{parametrisation_of_PT_width}). We can see that it works rather well for all values of $n$ except $n=2$ in the region $\eta\sim 0.4-0.6$. This can be improved using the relation $J(\eta)t_{PT} = $ constant, which works better for $n=2$ because it is exact for primary emissions with anti-$k_t$. But implementing it would not change the main conclusions presented in sections~\ref{sec:higgs-width}-\ref{hadronisation_corrections}. Therefore, for the sake of simplicity, we will not use it here: we keep eq.~(\ref{parametrisation_of_PT_width}) as the expression for $\delta M_{PT}$ for the rest of this study.

Of course, were it only for the perturbative radiation, it would be nicer to choose $\eta\ge\eta_{sat}$ in order to catch as many gluons as possible, leading to $\delta M_{PT}\rightarrow \delta M_{sat}$. But we also have to take into account Initial State Radiation (ISR) from the incoming $q\bar{q}$ pair and non-perturbative effects like PU and UE that can spoil our Higgs neighbourhood, thus increasing the jet mass.

For the purpose of this article we will only add UE and PU to the Final-State Radiation (FSR) effect studied above. We will thus ignore ISR, this partially for a question of simplicity of the analysis, but also because the results of work such as \cite{Cacciari:2008gd,Dasgupta:2007wa} suggest that for LHC processes whose hard scales are few hundred GeV, the crucial interplay is that between FSR and UE/PU. This is evident in the preference for small $R$ values in dijet mass reconstructions in those references, where ISR is not playing a major role. Similarly we believe that the optimal values of $\eta$ that we will determine here will have limited impact from ISR, though we shall not check this explicitly.

\subsection{Study of the Higgs width due to underlying event and pile-up}

For this simple analysis, which does not aim to give precise numbers but only an estimate of the influence of the UE/PU on the mass of the Higgs jet when we vary for instance $\eta$, $n$, or when the Higgs boson becomes more and more boosted, we model the UE and PU as soft particles uniformly distributed in the $(y,\phi)$ plane \cite{Cacciari:2008gn,Cacciari:2009dp}, and with transverse momentum per unit area denoted by $\rho$. In order to get this estimate, we consider the simple case of a symmetric ($z=1/2$) Higgs decay along the $x$ axis. In the limit $M_H\ll p_{t_H}$, the Higgs momentum $p_H$ is given by: 
\begin{equation}
  p_H = \left(p_{t_H}+\frac{M_H^2}{2p_{t_H}},p_{t_H},0,0\right)\,.
\end{equation}
The UE/PU momentum, denoted $p_{UE}$,\footnote{For brevity, we define $p_{UE}$ to be the sum of the UE and/or PU particles' momentum but without referencing the PU dependence, which will always be implicit.} is simply the sum of all the UE/PU particles $g$ belonging to the filtered jet $J$. Still in the limit $M_H\ll p_{t_H}$, we recall the following formula \cite{MyFirstPaper}:
\begin{equation}
  R_{bb}\simeq\frac{1}{\sqrt{z(1-z)}}\frac{M_H}{p_{t_H}}\,.\label{Rbb_boosted_limit}
\end{equation}
Throughout this section, we will apply it with $z=1/2$. We can now write $\Delta M  =  M_{\mbox{\scriptsize filtered jet}} - M_H$ as:
\begin{align}
  \Delta M & = \frac{1}{2M_H}\left((p_H+p_{UE})^2-M_H^2\right)\,, \nonumber\\
  & \simeq \frac{1}{M_H}\sum_{g\in J}p_{t_g}p_{t_H}\left(\frac{\theta_{gH}^2}{2}+\frac{M_H^2}{2p_{t_H}^2}\right)\,,\nonumber\\
  & \simeq \frac{M_H}{p_{t_H}}\sum_{g\in J}p_{t_g}\,.\label{average_delta_M_UE}
\end{align}
In the last line we used the approximation:
\begin{equation}
  \theta_{gH}\sim \theta_{bH} = \frac{R_{bb}}{2}\,, \label{theta_gh_approx}
\end{equation}
which comes from the fact that the UE and PU particles tend to cluster around the perturbative radiation, which is usually close to the $b$ and $\bb$ because of the collinear logarithmic divergence of QCD. As all the filtered UE/PU particles flow approximately in the same direction, the remaining sum is just the total transverse momentum of the UE which, by definition of $\rho$, is equal to $\rho A$, $A$ being the total area of the filtered jets.\footnote{in the active sense, see \cite{Cacciari:2008gn}.} We thus obtain
\begin{equation}
  \Delta M \simeq \frac{\rho A M_H}{p_{t_H}}\,,
  \label{AverageDeltaM}
\end{equation}
with
\begin{equation}
  \langle A\rangle \simeq n\pi\eta^2R_{bb}^2\,, \label{A_average_for_C_A}
\end{equation}
for the $C/A$ jet algorithm, taking into account the anomalous dimension that comes from the fact that there should be some perturbative radiation in the jets (cf figure~$14$ in \cite{Cacciari:2008gn}). Notice that eq.~(\ref{A_average_for_C_A}) is only true if all the jets do not overlap, so usually when $\eta$ is small enough. But this is sufficient for the purpose of our study, and we shall use this formula in all the following calculations. The correction eq.~(\ref{AverageDeltaM}) for $\Delta M$ only induces a shift towards higher masses of the Higgs mass peak. However, there are $3$ sources of fluctuations that give a width to this Higgs peak:
\begin{enumerate}
  \item $\rho$ is not strictly uniform in the $(y,\phi)$ plane in a given event.
  \item $\rho$ is not the same from one event to the next.
  \item The jets' area fluctuates.
\end{enumerate}
Following \cite{Cacciari:2008gn}, we can write the total UE/PU transverse momentum contributing to the Higgs $p_t$ as
\begin{equation}
  p_{t_{UE}} = \rho A \pm \left(\sqrt{A}\sigma + A\delta\rho + \rho\Sigma \right)\,,
\end{equation}
where
\begin{align}
  \sigma & = \sqrt{\langle \rho^2\rangle -\langle \rho\rangle ^2} \quad \mbox{ with $\langle ...\rangle $ a spatial average in a given event}\,, \\
  \delta\rho & = \sqrt{\langle \rho^2\rangle -\langle \rho\rangle ^2} \quad \mbox{ with $\langle ...\rangle $ an average over events}\,, \\
  \Sigma & = \sqrt{\langle A^2\rangle -\langle A\rangle ^2} \quad \mbox{ with $\langle A\rangle $ the average over events of the filtered jets' area}\,.
\end{align}
For pure UE events, i.e. without PU, these terms can be estimated \cite{Cacciari:2008gn,Cacciari:2009dp}:
\begin{align}
  \rho_{UE} & \simeq 2-3 \mbox{ GeV/area} \,, \label{rho_UE}\\
  \sigma_{UE} & \simeq 0.6\rho_{UE} \,, \label{sigma_UE}\\
  \delta\rho_{UE} & \simeq 0.8\rho_{UE} \,, \label{delta_rho_UE}\\
  \Sigma & \simeq 0.26\sqrt{n}\pi\eta^2R_{bb}^2 \,. \label {Sigma_UE}
\end{align}
Though $\rho_{UE}$ seems to be around $2$ GeV/area, the tuning used in \cite{MyFirstPaper} was closer to $3$ GeV/area, the value that we choose here. In presence of PU, i.e. when there is more than $1$ $pp$ collision per bunch crossing at the LHC (thus leading to the emission of other soft particles), $\rho$, $\sigma$ and $\delta\rho$ have to be modified. We define $N_{PU}$ to be the number of $pp$ collisions in a bunch crossing except the one at the origin of the hard interaction. We use a simple model to write the parameters of the UE/PU as:
\begin{align}
  \rho & \simeq \left(1+\frac{N_{PU}}{4}\right)\rho_{UE} \,, \label{rho_PU}\\
  \sigma & \simeq \sqrt{1+\frac{N_{PU}}{4}}\sigma_{UE} \,, \label{sigma_PU}\\
  \delta\rho & \simeq \sqrt{1+\frac{N_{PU}}{4}}\delta\rho_{UE} \,. \label{delta_rho_PU}
\end{align}
Some comments are needed: since $\rho$ measures the level of noise, it should grow like $N_{PU}$. In the expression $1+N_{PU}/4$, the $1$ corresponds to the $pp$ collision that leads to the UE and to the hard interaction, whereas the $N_{PU}/4$ term simply corresponds to the other $pp$ interactions and could be derived from the numbers given in \cite{Cacciari:2007fd}. The intra and inter events fluctuations of $\rho$ are modelled as growing like $\sqrt{\rho}$: we thus just give $\sigma$ and $\delta\rho$ the factor $\sqrt{1+N_{PU}/4}$, though further studies might be of value to parametrize these terms in a more adequate manner. Notice that the value given for $\delta\rho$ ignores the fluctuations in the number of PU events from one bunch crossing to the next, but this is beyond the accuracy of our model here. At high luminosity at LHC, $N_{PU}$ is expected to be $\sim 20$, which implies $\rho\sim10-20$ GeV \cite{Cacciari:2007fd,Sjostrand:2000wi,Sjostrand:2003wg}.

Assuming gaussian distributions for these three kinds of fluctuations, one can deduce the Higgs width due to the presence of UE/PU,\footnote{Here again, for brevity, we define $\delta M_{UE}$ to be the Higgs width in presence of UE and/or PU without referencing the PU dependence. Actually it serves only to distinguish the width due to UE/PU from the perturbative width $\delta M_{PT}$.} $\delta M_{UE} = 2\sqrt{\langle \Delta M^2\rangle  - \langle \Delta M\rangle ^2}$:
\begin{equation}
  \delta M_{UE} = 2\sqrt{A\sigma^2+A^2\delta\rho^2+\rho^2\Sigma^2}\frac{M_H}{p_{t_H}}\,.\label{WidthUE}
\end{equation}
For a gaussian peak, defining a $2\sigma$ width means that we keep roughly $68\%$ of the events around the average, which is in correspondence with the value $f=0.68$ chosen for the perturbative calculation.

We now have all the important results in hand to consider both UE/PU and FSR simultaneously.

\subsection{Study of the Higgs width in presence of both UE/PU and perturbative radiation}
\label{sec:higgs-width}

\begin{figure}[htb]
  \centering
  \includegraphics[scale=0.275]{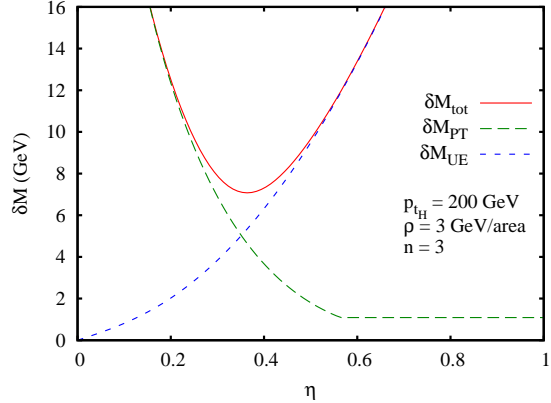}
  \caption{The Higgs width due to UE/PU and loss of perturbative radiation, combined as if the $2$ distributions were gaussians, i.e. $\delta M_{tot} = \sqrt{\delta M_{PT}^2+\delta M_{UE}^2}$ when $n=3$. In this case, $\eta_{opt}$, though slightly larger, is approximately given by the intersection of the $2$ curves, at least as long as $\eta$ is not in the saturation region.}
  \label{Higgs_total_width}
\end{figure}

The purpose of this part is to give an estimate of how one should choose the couple of filtering parameters ($n$,$\eta$). For that, one has to convolute the effects of UE/PU and perturbative radiation and compute the resulting reconstructed Higgs peak width, and then minimize it with respect to the filtering parameters. This is highly non trivial to do analytically and we leave it for future work. The simple choice made here is to say that, for a given $n$, the optimal $\eta$, denoted $\eta_{opt}$, is the one for which the two widths are equal. This is obviously not true in general, but seems reasonable to obtain an estimate (figure~\ref{Higgs_total_width}) and to understand how $\eta_{opt}$ changes when we vary $p_{t_H}$ and $N_{PU}$. Notice that, using this method, we have to impose $\eta_{opt}<\eta_{sat}$ where $\eta_{sat}$ is the saturation point (eq.~(\ref{eta_sat})), because beyond $\eta_{sat}$, increasing $\eta$ makes $\delta M_{UE}$ larger without decreasing $\delta M_{PT}$, thus solving the equation $\delta M_{PT}=\delta M_{UE}$ has no sense in this region. Finally, we numerically minimize $\sqrt{\delta M_{PT}^2+\delta M_{UE}^2}$, calculated at $\eta=\eta_{opt}(n)$, with respect to $n$ in order to find $n_{opt}$.

\begin{figure}[htb]
  \centering
  \subfigure[]{
    \includegraphics[scale=0.275]{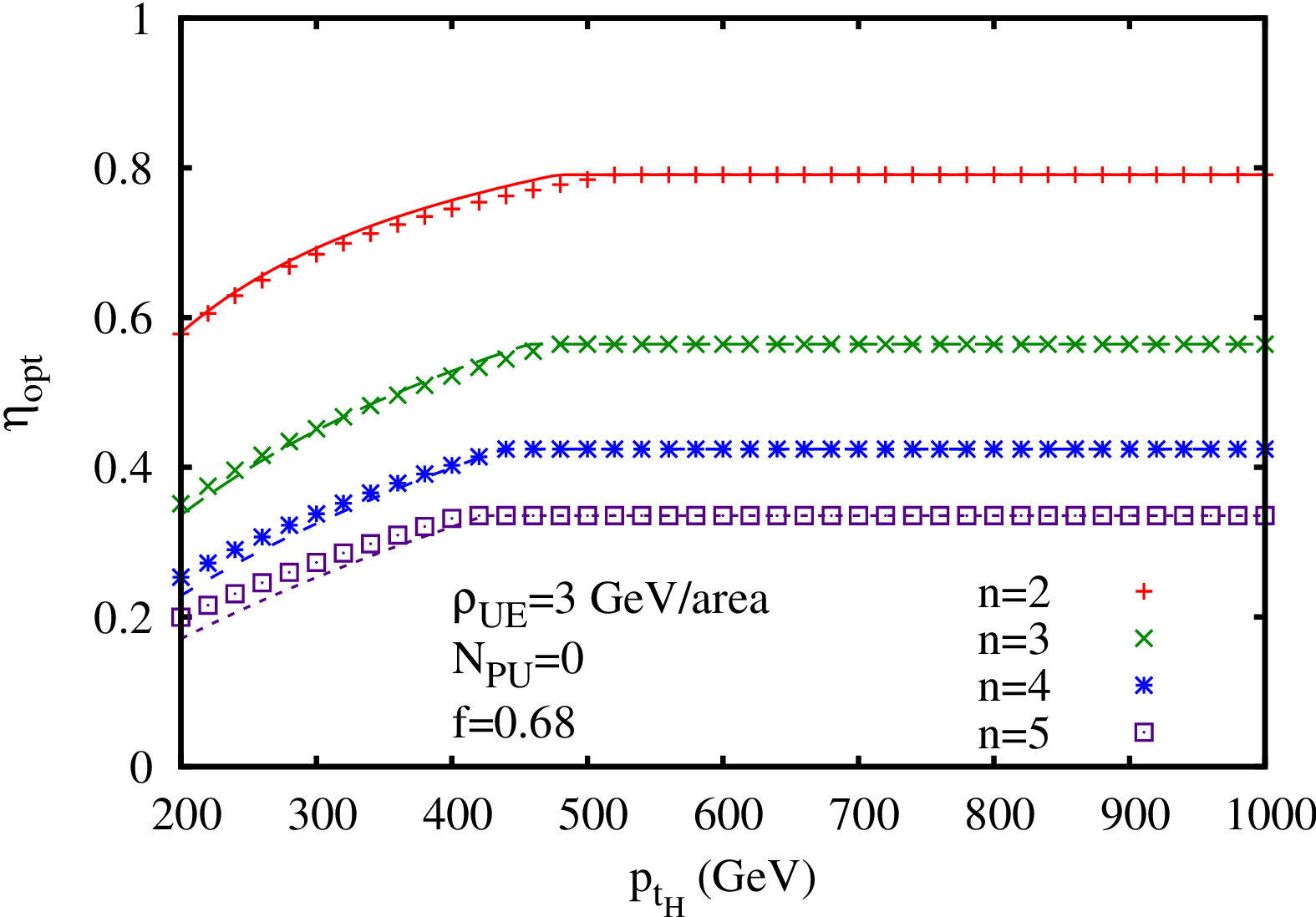}
    \label{solution_for_L_ptH_curve}
  }~
  \subfigure[]{
    \includegraphics[scale=0.275]{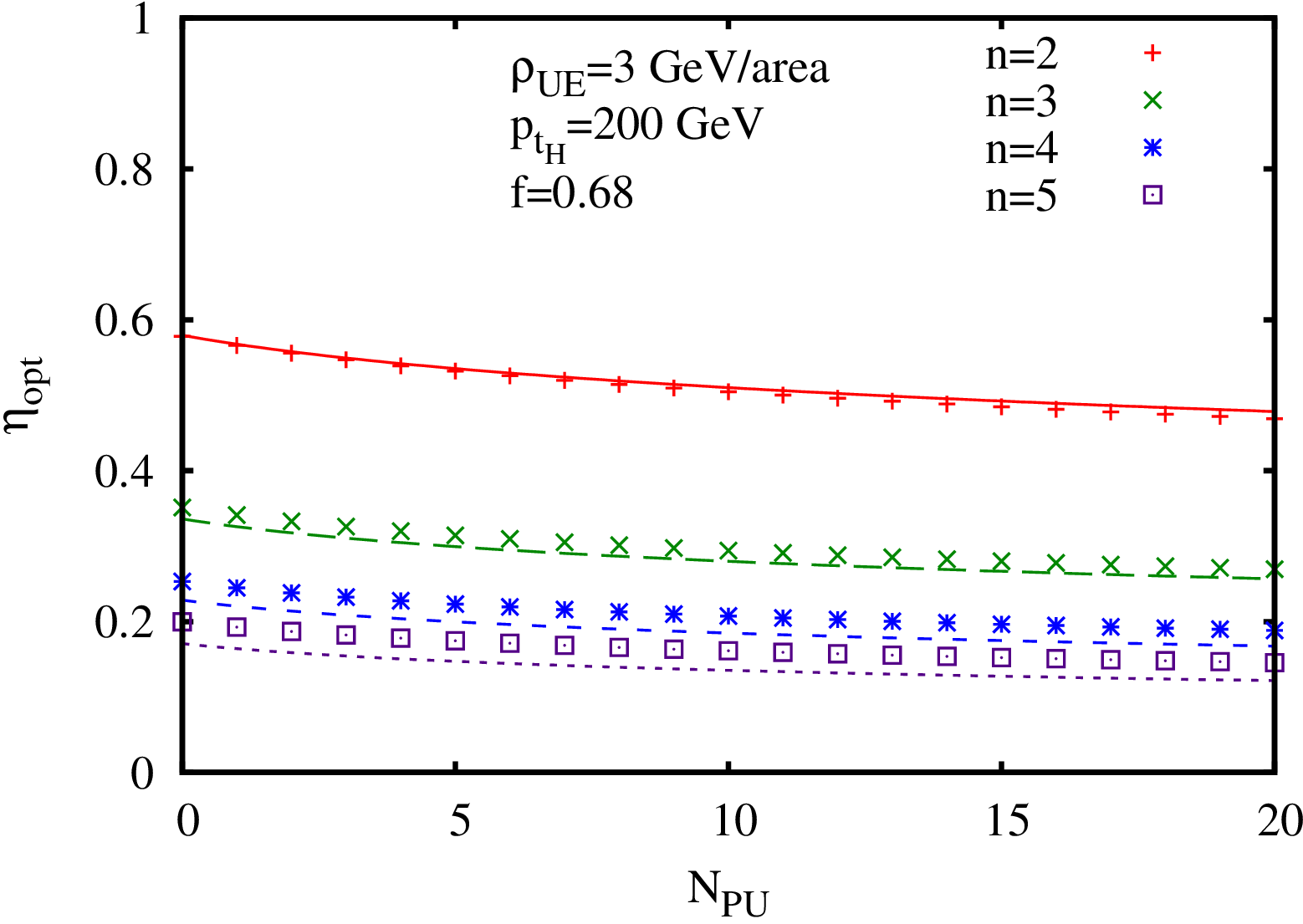}
    \label{solution_for_L_NPU_curve}
  }
  \caption{The numerical solutions (points) of eq.~(\ref{exact_equation_for_L}) shown for different values of $n$: \subref{solution_for_L_ptH_curve} as a function of $p_{t_H}$ when $N_{PU}=0$, and \subref{solution_for_L_NPU_curve} as a function of $N_{PU}$ when $p_{t_H}=200$ GeV. We also show the corresponding approximate analytical solutions (lines) derived in eq.~(\ref{eta_approximate_solution}).}
  \label{solutions_of_the_equation_for_L}
\end{figure}

First, we would like to understand how $\eta_{opt}$ evolves with respect to the physical parameters. The equality $\delta M_{PT}=\delta M_{UE}$ gives an equation in $L=\ln\frac{1}{\eta}$:
\begin{equation}
   M_He^{-\frac{1}{2\beta_0\as}\left(1-e^{-4\pi\beta_0\frac{C_{PT}}{L}}\right)} = 2\sqrt{c_{\sigma}^2e^{-2L}+c_{\delta\rho}^2e^{-4L}+c_{\Sigma}^2e^{-4L}}\rho_{UE}\frac{M_H}{p_{t_H}}\,, \label{exact_equation_for_L}
\end{equation}
where the coefficients $c_{\sigma}$, $c_{\delta\rho}$ and $c_{\Sigma}$ can be easily calculated using eqs.~(\ref{A_average_for_C_A},\ref{rho_UE}-\ref{delta_rho_PU},\ref{WidthUE}):
\begin{align}
  c_{\sigma}(n,N_{PU},R_{bb}) & \simeq 0.6\sqrt{\pi}\sqrt{n}R_{bb}\sqrt{1+\frac{N_{PU}}{4}}\,,\label{c_sigma_with_PU}\\
  c_{\delta\rho}(n,N_{PU},R_{bb}) & \simeq 0.8\pi n R_{bb}^2\sqrt{1+\frac{N_{PU}}{4}}\,,\label{c_delta_rho_with_PU}\\
  c_{\Sigma}(n,N_{PU},R_{bb}) & \simeq 0.26\pi\sqrt{n}R_{bb}^2\left(1+\frac{N_{PU}}{4}\right)\,.\label{c_Sigma_with_PU}
\end{align}
If the solution of eq.~(\ref{exact_equation_for_L}) for a given $n$ is found to be above $\eta_{sat}(n,f)$, then $\eta_{opt}=\eta_{sat}(n,f)$ in order to take the saturation of $\delta M_{PT}$ into account. We start by solving this equation numerically. In figure~\ref{solutions_of_the_equation_for_L} we show $\eta_{opt}$ as a function of $p_{t_H}$ and $N_{PU}$ for different values of $n$. As it should, $\eta_{opt}$ increases with $p_{t_H}$ at fixed $N_{PU}$. Indeed, if $p_{t_H}$ grows at fixed $\eta$, $R_{bb}$ decreases and so does the effect of UE/PU, whereas the perturbative radiation is kept fixed (no dependence on $R_{bb}$). Notice also, for $n=3$, that the values obtained for $\eta_{opt}$ are roughly consistent with the choice in \cite{MyFirstPaper} where we had $\eta = \min(0.3/R_{bb},1/2)$. The saturation comes into effect at relatively low $p_{t_H}$, around $400-500$ GeV. Above this value, the total width is small and hadronisation corrections start to become relevant, so that the results presented on these plots become not very reliable. However, for $p_t>\sim 500$ GeV and $\eta>\eta_{sat}$, the Higgs width due to perturbative radiation and UE/PU vary slowly with $\eta$ and we thus believe that the precise value chosen for $\eta$ is not so important: one can take any value above $\eta_{sat}$ without changing the result too much. The decrease of $\eta_{opt}$ with $N_{PU}$ seems to be weaker than one might have expected {\it a priori}. However, in fig.~\ref{Higgs_total_width}, we can see that the negative slope of the perturbative width is very large, and therefore increasing the noise from PU will not change too much the $\eta_{opt}$ value.

\begin{figure}[bht]
  \centering
  \subfigure[]{
    \includegraphics[scale=0.275]{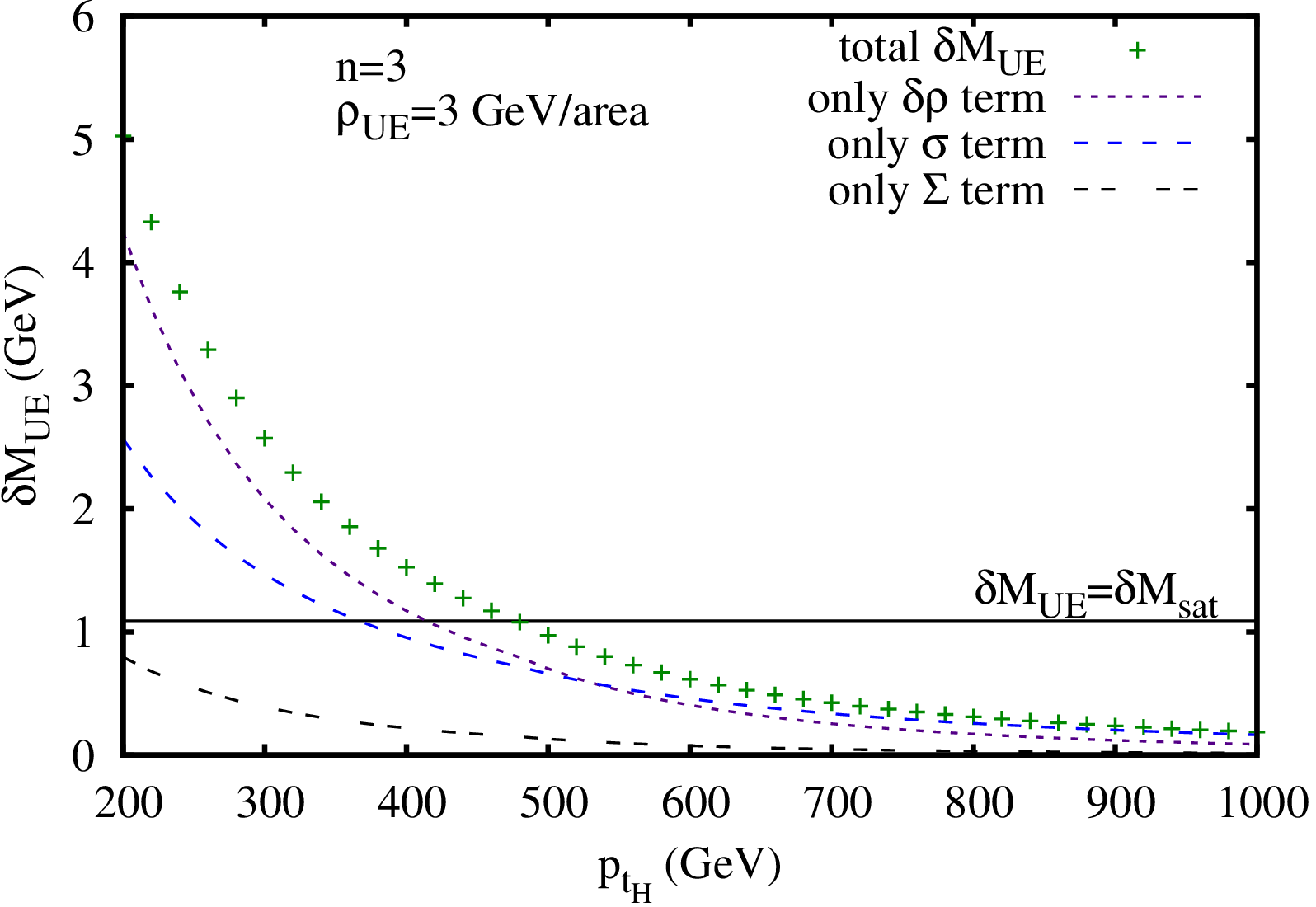}
    \label{dMUE_wrt_ptH}
  }~
  \subfigure[]{
    \includegraphics[scale=0.275]{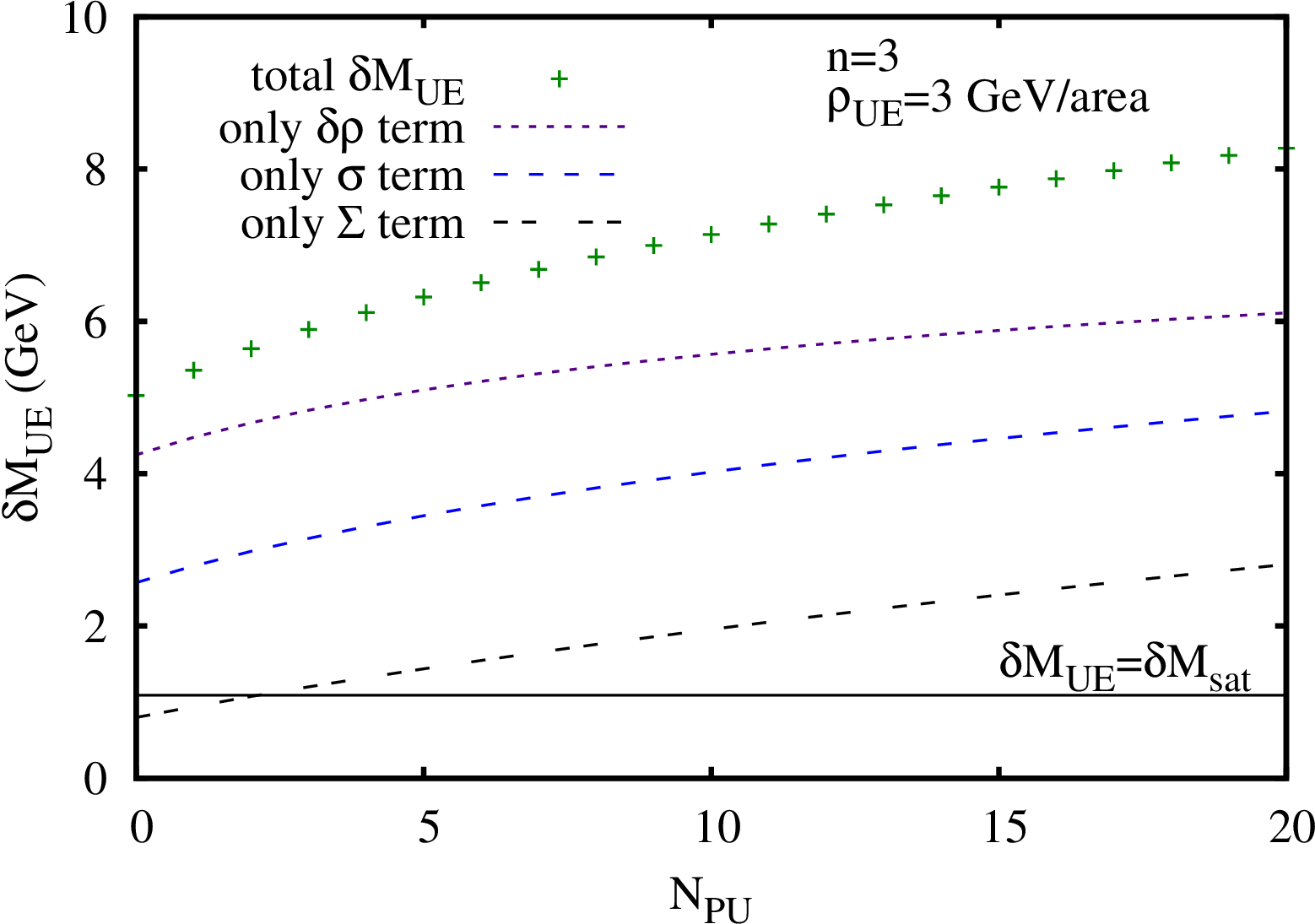}
    \label{dMUE_wrt_NPU}
  }
  \caption{$\delta M_{UE}$ computed at $\eta = \eta_{opt}$ with respect to \subref{dMUE_wrt_ptH} $p_{t_H}$ when $N_{PU}=0$ and \subref{dMUE_wrt_NPU} $N_{PU}$ when $p_{t_H}=200$ GeV. On these plots is also represented the contribution to $\delta M_{UE}$ of each term separately. When the UE/PU width falls below the saturation line $\delta M_{UE}=\delta M_{sat}$, then $\eta_{opt}=\eta_{sat}$.}
  \label{which_UE_term_is_dominant}
\end{figure}

It would be interesting to understand analytically the evolution of $\eta_{opt}$ with respect to the physical parameters $p_{t_H}$ and $N_{PU}$. Unfortunately, eq.~(\ref{exact_equation_for_L}) cannot be easily dealt with. That's why we have to make an approximation: in this equation, one of the $3$ terms under the square root may be dominant when $\eta = \eta_{opt}$. At first sight, one would expect that at low $\eta_{opt}$, the $c_{\sigma}^2e^{-2L}$ term, which scales like $\eta^2$, should be the largest, whereas at large $N_{PU}$, it should be the $c_{\Sigma}^2e^{-4L}$ term that is the largest one as it scales like $N_{PU}^2$. But figure~\ref{which_UE_term_is_dominant} for $n=3$ reveals that the $c_{\delta\rho}$ term surprisingly brings the largest contribution to $\delta M_{UE}$ for physical values of the parameters (the same holds for other values of $n$). 
Therefore, to simplify things a little, one can consider eq.~(\ref{exact_equation_for_L}) and put $c_{\sigma}=c_{\Sigma}=0$. However, to be more general, and to consider the possible situation where one of the other terms might be dominant,\footnote{The subtraction procedure proposed in \cite{Cacciari:2007fd} seems to eliminate most of the fluctuations from the $c_{\delta\rho}$ and $c_{\Sigma}$ terms, so that the remaining $c_{\sigma}$ term would be dominant in this case.} we rewrite eq.~(\ref{exact_equation_for_L}) in the following approximate form:
\begin{equation}
  M_He^{-\frac{1}{2\beta_0\as}\left(1-e^{-4\pi\beta_0\frac{C_{PT}}{L}}\right)} = C_{UE}\rho_{UE}e^{-pL}R_{bb}^p\frac{M_H}{p_{t_H}}\,,\label{approximate_equation_for_L_1}
\end{equation}
where $p=1$ if the $c_{\sigma}$ term dominates and $p=2$ otherwise. Moreover:
\begin{equation}
  C_{UE}(n,N_{PU}) = \left\{\begin{array}{ll}
      1.2\sqrt{\pi}\sqrt{n}\sqrt{1+\frac{N_{PU}}{4}}\,, & \mbox{ if the $c_{\sigma}$ term is dominant,}\\
      1.6\pi n \sqrt{1+\frac{N_{PU}}{4}}\,, & \mbox{ if the $c_{\delta\rho}$ term is dominant,}\\
      0.52\pi\sqrt{n}\left(1+\frac{N_{PU}}{4}\right)\,, & \mbox{ if the $c_{\Sigma}$ term is dominant.}\end{array}\right.
  \label{definition_of_C_UE}
\end{equation}
Eq.~(\ref{approximate_equation_for_L_1}) can be written in a slightly different way:
\begin{equation}
  \frac{B_{PT}}{L} = \ln\left(\frac{1}{B_{UE}-2\beta_0\as pL}\right)\,,\label{approximate_equation_for_L_2}
\end{equation}
with:
\begin{align}
  B_{PT} & = 4\pi\beta_0C_{PT}\,, \label{value_of_B_PT}\\
  B_{UE} & = 1-2\beta_0\as\ln\left(\frac{p_{t_H}}{C_{UE}\rho_{UE}R_{bb}^p}\right)\,.
\end{align}
Despite its simpler form, eq.~(\ref{approximate_equation_for_L_2}) for $L$ cannot be solved analytically. Here comes the second approximation, which is to make a perturbative expansion:
\begin{equation}
  \frac{B_{PT}}{L} = \ln\frac{1}{B_{UE}}+\frac{2\beta_0\as p}{B_{UE}}L+{\cal O}\left((\as L)^2\right)\,.
\end{equation}
Neglecting the ${\cal O}\left((\as L)^2\right)$ term, the resulting quadratic equation immediately implies
\begin{equation}
  L_{opt} = \frac{-B_{UE}\ln\frac{1}{B_{UE}}+\sqrt{B_{UE}^2\ln^2\frac{1}{B_{UE}}+8\beta_0\as pB_{UE}B_{PT}}}{4\beta_0\as p}\,.\label{solution_for_L_opt}
\end{equation}
Taking into account the saturation effect, $\eta_{opt}$ is then given by:
\begin{equation}
  \eta_{opt} = \left\{\begin{array}{ll}
       e^{-L_{opt}}\,, & \mbox{ if } L_{opt}>-\ln\eta_{sat}\,,\\
       \eta_{sat}\,, & \mbox{ otherwise}\,. \end{array}\right.
       \label{eta_approximate_solution}
\end{equation}
We used this expression with $C_{UE}$ corresponding to the $\delta\rho$ term in eq.~(\ref{definition_of_C_UE}) and $p=2$ in order to plot the approximate solutions in figure~\ref{solutions_of_the_equation_for_L}. This reveals that the above relation for $\eta_{opt}$ (eq.~(\ref{eta_approximate_solution})) works rather well, within a few $\%$.

As a second step, we would like to find the optimal $n$, denoted $n_{opt}$. This also should depend on the way UE/PU and perturbative radiation are combined. However, as a simple approximation, one can combine them as if they were both gaussian distributions. Therefore, one should minimize
\begin{equation}
  \delta M_{tot}(n) = \sqrt{\delta M_{PT}^2(n)+\delta M_{UE}^2(n)}\,,
\end{equation}
computed at $\eta = \eta_{opt}(n)$ for a given $p_{t_H}$ and $N_{PU}$. 
\begin{figure}[htb]
  \centering
  \subfigure[]{
    \includegraphics[scale=0.275]{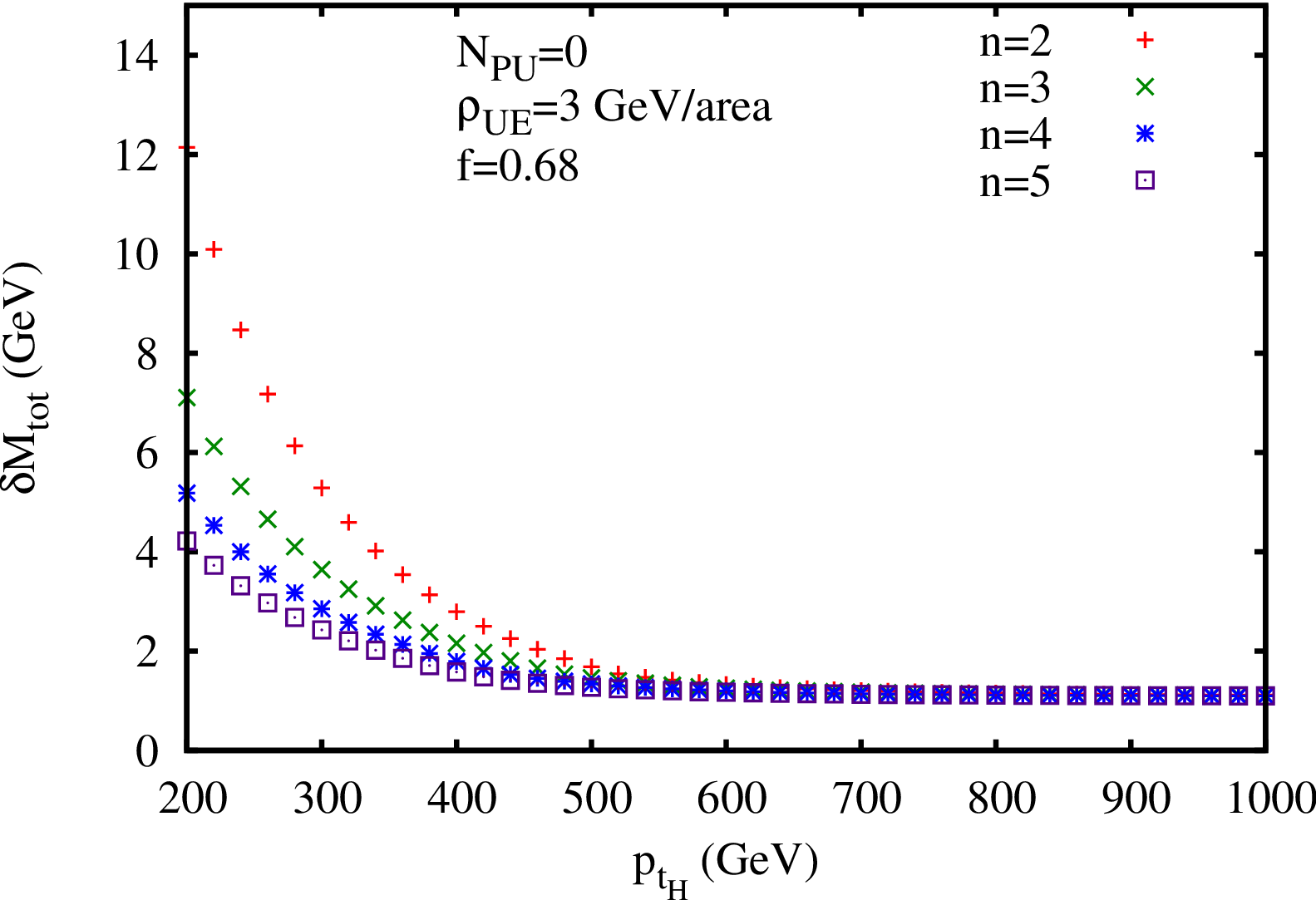}
    \label{dMtot_wrt_ptH}
  }~
  \subfigure[]{
    \includegraphics[scale=0.275]{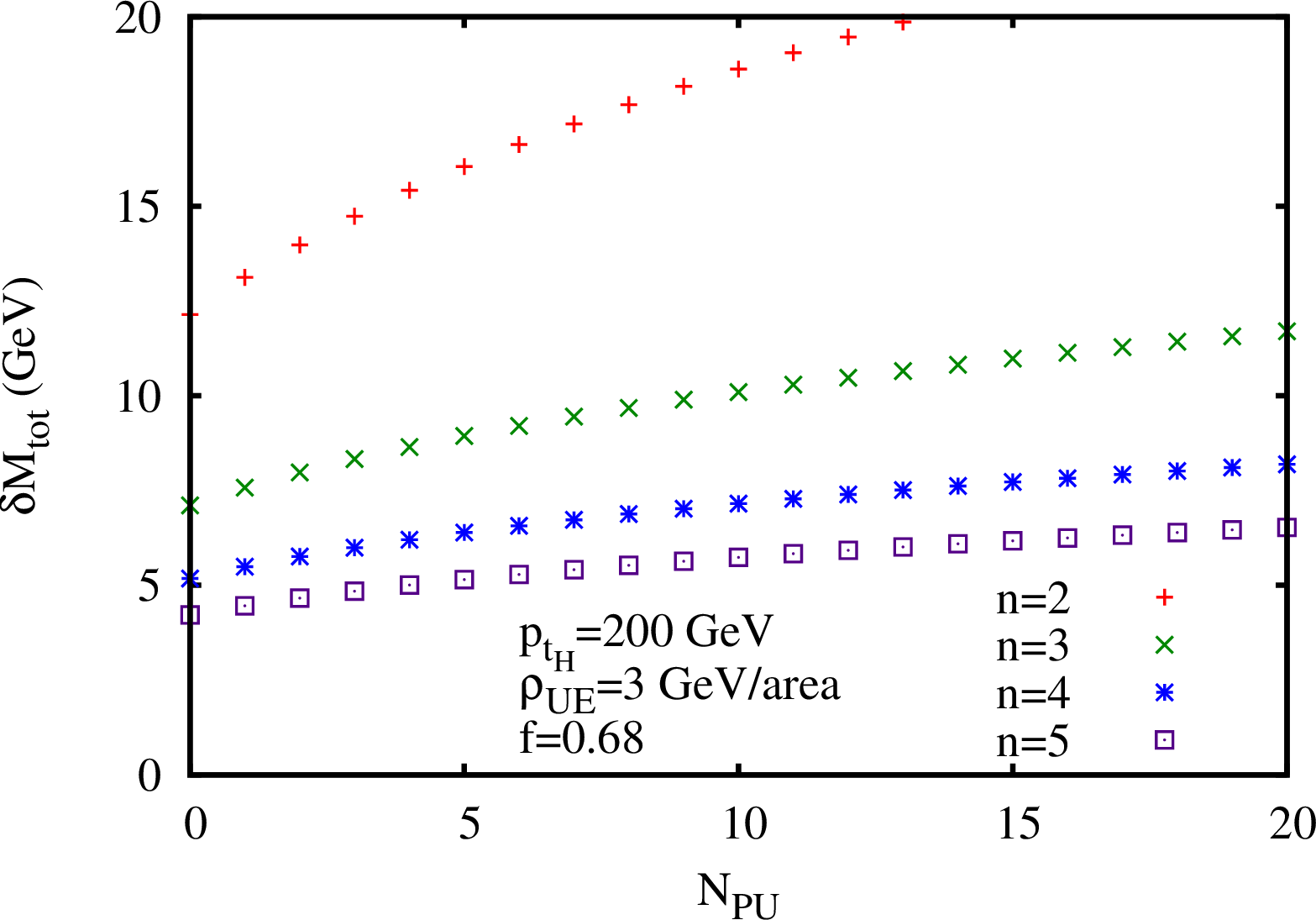}
    \label{dMtot_wrt_NPU}
  }
  \caption{$\delta M_{tot}$ computed at $\eta = \eta_{opt}$ as a function of \subref{dMtot_wrt_ptH} $p_{t_H}$ and \subref{dMtot_wrt_NPU} $N_{PU}$ for different values of $n$.}
  \label{solution_for_n_opt}
\end{figure}

The results are plotted in figure~\ref{solution_for_n_opt}. We can notice that the larger $n$, the narrower the peak, and thus the better the result. However, one should keep in mind that when $n$ increases, the optimal $\filt{R}=\eta R_{bb}$ becomes small, and we have to deal with hadronisation corrections that grow as $1/\filt{R}$ \cite{Dasgupta:2007wa} as well as detector resolution and granularity $\delta\eta\times\delta\phi = 0.1\times0.1$ that both start to have an important impact on the reconstructed Higgs width, and thus degrade the results presented here. In section~\ref{hadronisation_corrections} we will examine what happens when we include a very rough estimate for hadronisation corrections. However, at first sight, it seems that one should definitely not take $n=2$. The value $n=3$ chosen in \cite{MyFirstPaper} is good, but it may be possible to do better with $n=4$. Beyond this value, the optimal $\filt{R}$ falls below $\sim$ $0.2$ (cf figure~\ref{solutions_of_the_equation_for_L}), which is too small for this study to be fully reliable, as we shall see in section~\ref{hadronisation_corrections}. 

\subsection{Variations of the results with $z$ and $f$ \label{variations_of_the_results_with_z_and_f}}

Until now, we have only presented some results for $f=0.68$ and $z=1/2$, $z$ being defined as
\begin{equation}
  z=\min\left(\frac{E_b}{E_H},\frac{E_{\bb}}{E_H}\right)\,,
\end{equation}
with $E_i$ the energy of particle $i$ in the Higgs splitting into $b\bb$. What happens if we change these values?

Let us start with $z$. Though the Higgs splitting into $b\bb$ is more often symmetric than in QCD events (and this is what was used in \cite{MyFirstPaper} to distinguish it from pure QCD splittings), it still has a distribution in $z$ that is uniform in the range:
\begin{equation}
  \frac12\left(1-\frac{1}{\sqrt{1+\frac{M_H^2}{p_{t_H}^2}}}\right) < z <\frac12\,,\label{kinematical_limit_for_z}
\end{equation}
which follows from simple kinematics in the limit $m_b=0$. But in order to reduce the large QCD background, one usually cuts on small $z$, so that 
\begin{equation}
  z_{cut}<z<\frac12\,, \label{z_acceptance}
\end{equation}
with $z_{cut}\sim 0.1$. As an example, assume the $b$ quark carries the fraction $z$ of the Higgs splitting. In such a case, $b$ and $\bb$ are not equidistant from the Higgs direction: they are respectively at an angular distance $(1-z)R_{bb}$ and $zR_{bb}$ from $H$ (see for instance figure~\ref{variables} in appendix~\ref{app:analytical_considerations}). Therefore, as UE/PU particles tend to cluster around the perturbative radiation, eq.~(\ref{theta_gh_approx}) has to be modified:
\begin{equation}
  \theta_{gH}\sim zR_{bb} \mbox{ or } \theta_{gH}\sim (1-z)R_{bb}\,,
\end{equation}
for a given UE/PU particle $g$ in the filtered jet. This leads to the modification of eq.~(\ref{average_delta_M_UE}) according to $g$ being relatively close to $b$ (region called ``$J_1$'') or $\bb$ (region called ``$J_2$''):
\begin{align}
  \Delta M & \simeq \frac{p_{t_H}}{M_H}\left(\sum_{g\in J_1}p_{t_g}\left(\frac{(1-z)^2R_{bb}^2}{2}+\frac{M_H^2}{2p_{t_H}^2}\right)+\sum_{g\in J_2}p_{t_g}\left(\frac{z^2R_{bb}^2}{2}+\frac{M_H^2}{2p_{t_H}^2}\right)\right)\,,\nonumber\\
  & \simeq \frac{M_H}{2p_{t_H}}\left(\frac1z\sum_{g\in J_1}p_{t_g}+\frac{1}{1-z}\sum_{g\in J_2}p_{t_g}\right)\,,\nonumber\\
  & = \frac{M_H}{2p_{t_H}}\left(\frac1z\rho A_1+\frac{1}{1-z}\rho A_2\right)\,. \label{average_DeltaM_for_any_z_and_n_2}
\end{align}
In this calculation we used eq.~(\ref{Rbb_boosted_limit}). To compute the dependence of the fluctuations on $z$, we take the simplest case $n=2$. For the $\sigma$ and $\Sigma$ fluctuations, the terms $\rho A_1$ and $\rho A_2$ vary independently, leading to the following contribution to $\delta M_{UE}$:\footnote{the factor of $4$ comes from the fact that we compute the width at $2\sigma$.}
\begin{equation}
  \delta M_{UE,\sigma,\Sigma}^2=4\left(\frac{M_H\rho_{UE}}{2p_{t_H}}\right)^2\left(\frac{1}{z^2}\delta_{1,\sigma,\Sigma}^2+\frac{1}{(1-z)^2}\delta_{2,\sigma,\Sigma}^2\right)\,, \label{z_dependence_n_2_for_s_and_drho}
\end{equation}
where
\begin{equation}
  \delta_{1,\sigma,\Sigma}^2 = \delta_{2,\sigma,\Sigma}^2 = c_{\sigma}^2e^{-2L}+c_{\Sigma}^2e^{-4L}\,,\label{delta_sigma_Sigma}
\end{equation}
with $c_{\sigma}$ and $c_{\Sigma}$ given by eqs.~(\ref{c_sigma_with_PU},\ref{c_Sigma_with_PU}) for $n=1$. Concerning the $\delta\rho$ fluctuations, the $2$ terms $\rho A_1$ and $\rho A_2$ vary the same way from one event to another. Therefore, if it were only for the $\delta\rho$ term, we would write $\rho A_1 = \rho A_2$ leading to:
\begin{equation}
  \delta M_{UE,\delta\rho}^2 = 4\left(\frac{M_H\rho_{UE}}{2p_{t_H}}\right)^2\frac{1}{z^2(1-z)^2}\delta_{\delta\rho}^2\,,
\end{equation}
where
\begin{equation}
  \delta_{\delta\rho}^2 = c_{\delta\rho}^2e^{-4L}\,,\label{delta_delta_rho}
\end{equation}
with $c_{\delta\rho}$ given by eq.~(\ref{c_delta_rho_with_PU}) for $n=1$. Adding all these contributions,
\begin{equation}
  \delta M_{UE}^2 = \delta M_{UE,\sigma,\Sigma}^2 + \delta M_{UE,\delta\rho}^2\,,
\end{equation}
this apparently leads to an enhancement of the width by a factor of $1/z$. But we have to take into account that the coefficients $c_{\delta\rho}$ and $c_{\Sigma}$ also contain a factor $R_{bb}^2$ (eqs.~(\ref{c_sigma_with_PU},\ref{c_delta_rho_with_PU})) leading to another factor $1/z$, and thus an enhancement $1/z^2$ at small $z$.\footnote{This is valid when $p_{t_H}>\frac{1}{\sqrt{z(1-z)}}\frac{M_H}{R_0}$, with $R_0$ the radius of the initial clustering of the event, in order for the $b$ and $\bb$ to be clustered together. For lower $p_{t_H}$, there is a kinematic cut on $z$ and the enhancement is less strong.} Therefore, we can conclude that the effect of $z\neq 1/2$ is to broaden the reconstructed Higgs peak. Such a factor may partly explain the width of $\sim$ $14$ GeV that was observed in \cite{Rubin:2009ft}, to be compared with the various widths found in the previous subsection (see for instance figure~\ref{solution_for_n_opt}), and should also lead to decreasing $\eta_{opt}$. This is illustrated in fig.~\ref{eta_optimal_and_dMtot_for_z_0.2}, which was obtained with the results derived in appendix~\ref{app:some_analytical_results_for_the_dependence_on_z_and_f}, where we carry out the above analysis for a general $n$.
\begin{figure}[hbt]
  \centering
  \subfigure[]{
    \includegraphics[scale=0.275]{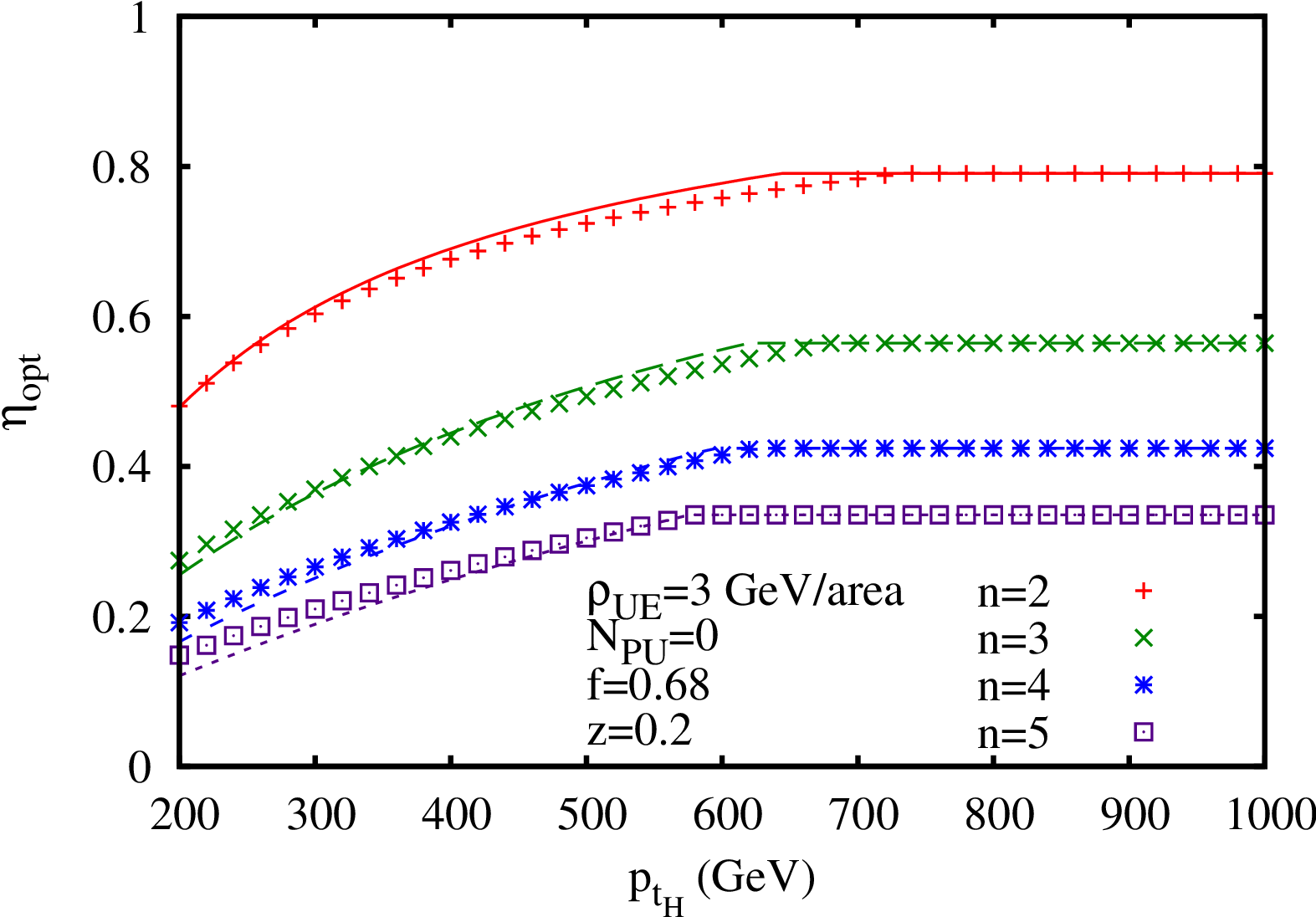}
    \label{eta_optimal_ptH_z0.2}
  }~
  \subfigure[]{
    \includegraphics[scale=0.275]{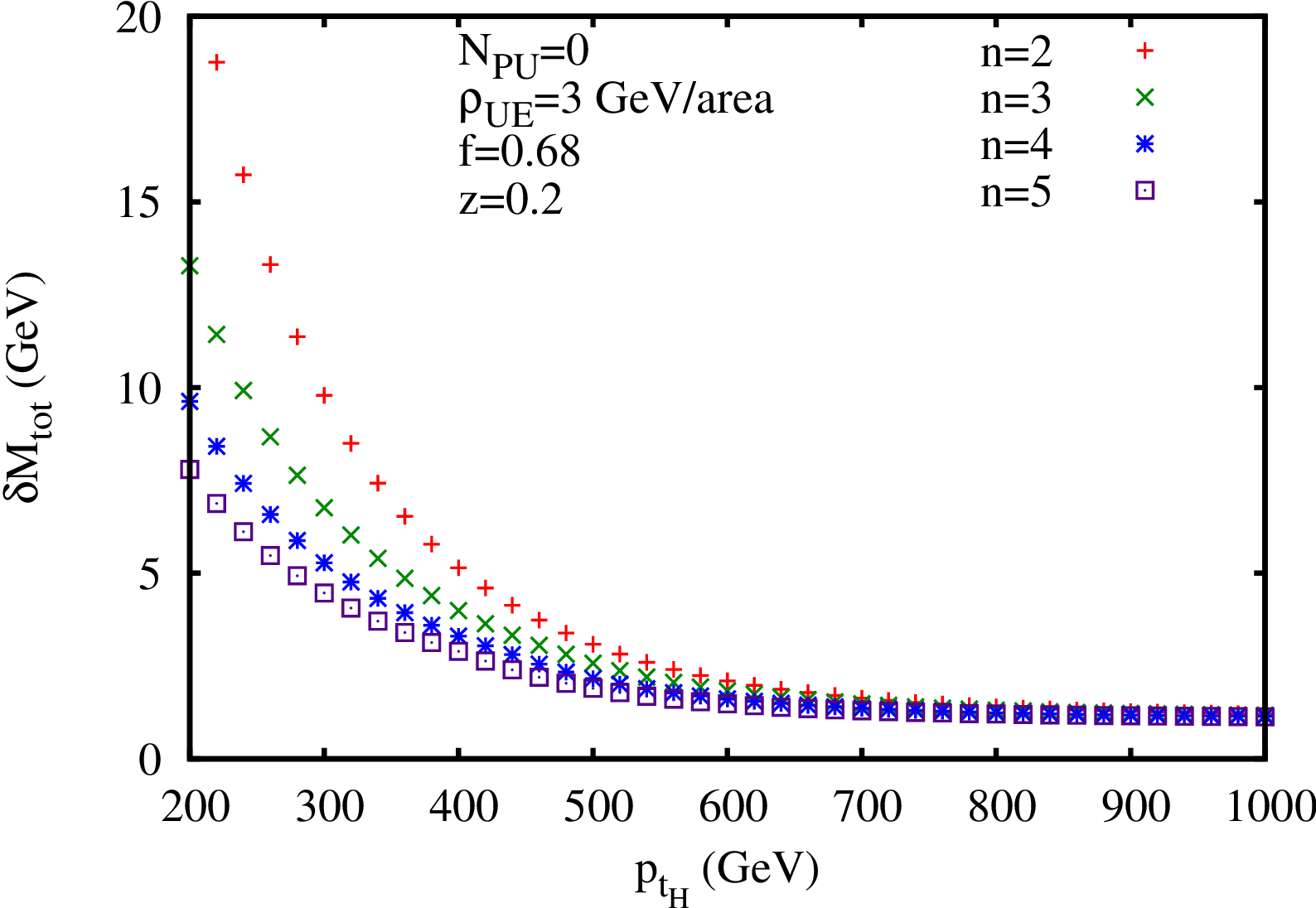}
    \label{dMtot_ptH_z0.2}
  }
  \caption{\subref{eta_optimal_ptH_z0.2} $\eta_{opt}$ as a function of $p_{t_H}$ when $f=0.68$ and $z=0.2$ for different values of $n$. The points correspond to the numerical determination of $\eta_{opt}$, found solving eq.~(\ref{exact_equation_for_L}) whose $z$ dependence is derived in appendix~\ref{app:some_analytical_results_for_the_dependence_on_z_and_f}, whereas the curves correspond to its approximate analytical solutions. \subref{dMtot_ptH_z0.2} $\delta M_{tot}=\sqrt{\delta M_{PT}^2+\delta M_{UE}^2}$ computed at $\eta=\eta_{opt}$ as a function of $p_{t_H}$ for $f=0.68$ and $z=0.2$.}
  \label{eta_optimal_and_dMtot_for_z_0.2}
\end{figure}

Now, we turn to the $f$ value. As we explained in section~\ref{study_of_the_Higgs_perturbative_width}, the choice $f=0.68$ was made to correspond to a $2\sigma$ gaussian width, as we did for $\delta M_{UE}$, which is somewhat arbitrary. We would like to estimate how the results change when $f$ is modified. We thus also consider a range of values for $f$ between $0.5$ and $0.8$. In this case the $C_{PT}(n,f)$ constants caracterizing $\delta M_{PT}$ are changed (see for instance eq.~(\ref{C_PT_2_f})), and $\delta M_{UE}$ is also changed, i.e. eqs.~(\ref{WidthUE},\ref{exact_equation_for_L},\ref{definition_of_C_UE}) have to be slightly modified:
\begin{align}
  \delta M_{UE} & = 2\sqrt2\,\erf^{-1}(f)\,\sqrt{A\sigma^2+A^2\delta\rho^2+\rho^2\Sigma^2}\frac{M_H}{p_{t_H}}\,,\\
  & = 2\sqrt2\,\erf^{-1}(f)\,\sqrt{c_{\sigma}^2e^{-2L}+c_{\delta\rho}^2e^{-4L}+c_{\Sigma}^2e^{-4L}}\rho_{UE}\frac{M_H}{p_{t_H}}\,,
\end{align}
where $\erf(x)$ is the usual error function:
\begin{equation}
  \erf(x) = \frac{2}{\sqrt{\pi}}\int_0^{x}e^{-u^2}du\,.
\end{equation}
Notice that the constants $c_{\sigma}$, $c_{\delta\rho}$ and $c_{\Sigma}$ are left unchanged with this convention. However $C_{UE}$ becomes:
\begin{equation}
  C_{UE}(n,f,N_{PU}) = \left\{\begin{array}{ll}
      2\sqrt2\,\erf^{-1}(f)\,0.6\sqrt{\pi}\sqrt{n}\sqrt{1+\frac{N_{PU}}{4}}\,, & \mbox{ if the $c_{\sigma}$ term is dominant,}\\
      2\sqrt2\,\erf^{-1}(f)\,0.8\pi n \sqrt{1+\frac{N_{PU}}{4}}\,, & \mbox{ if the $c_{\delta\rho}$ term is dominant,}\\
      2\sqrt2\,\erf^{-1}(f)\,0.26\pi\sqrt{n}\left(1+\frac{N_{PU}}{4}\right)\,, & \mbox{ if the $c_{\Sigma}$ term is dominant.} \end{array}\right.\label{C_UE_wrt_f}
\end{equation}
\begin{figure}[bth]
  \centering
  \subfigure[]{
    \includegraphics[scale=0.275]{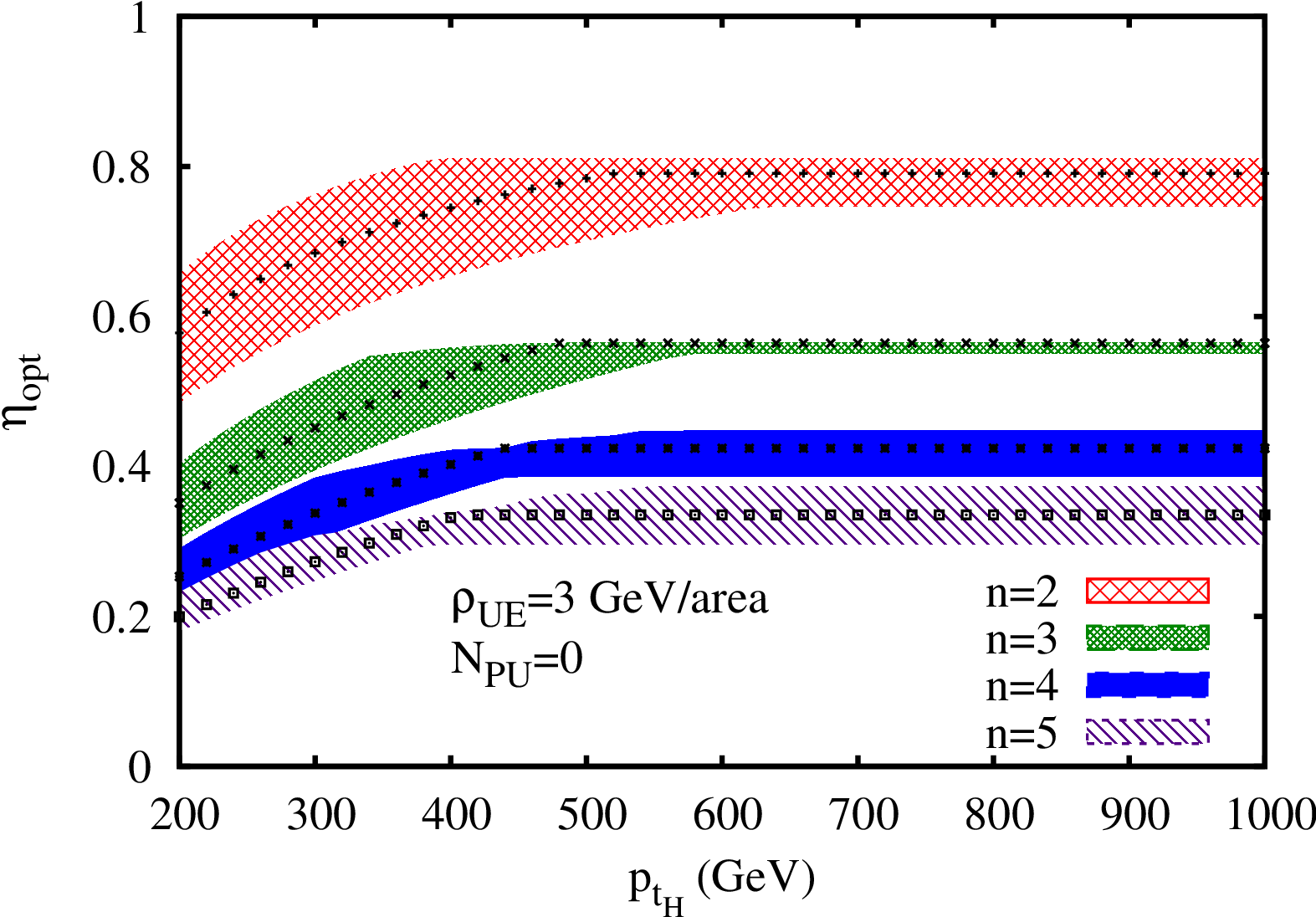}
    \label{eta_optimal_ptH_with_bands}
  }~
  \subfigure[]{
    \includegraphics[scale=0.275]{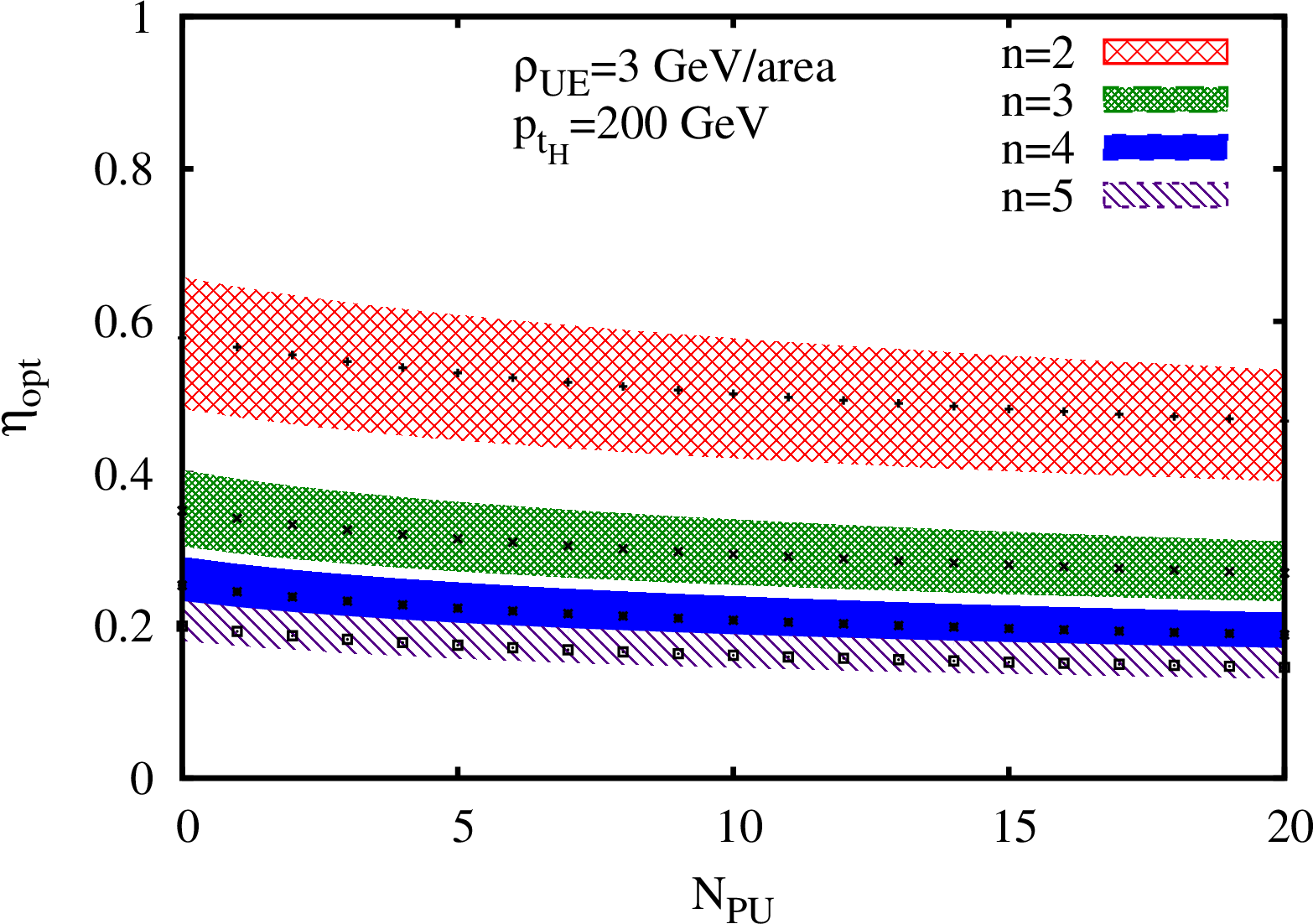}
    \label{eta_optimal_NPU_with_bands}
  }
  \caption{Uncertainty on $\eta_{opt}$  when $f$ varies from $0.5$ to $0.8$, for different values of $n$ as a function of \subref{eta_optimal_ptH_with_bands} $p_{t_H}$ when $N_{PU}=0$ and \subref{eta_optimal_NPU_with_bands} $N_{PU}$ when $p_{t_H}=200$ GeV. The results for $f=0.68$ are also plotted as a reference.}
  \label{eta_optimal_with_bands}
\end{figure}
The bands corresponding to the uncertainties on $\eta_{opt}$ that we obtain including these modifications are presented in figure~\ref{eta_optimal_with_bands}. The uncertainty that we get, $\sim 20-30\%$, is not larger than the precision of the whole study of this paper, which limits itself to a large-$N_c$ leading-log calculation. Notice that the variation with $N_{PU}$ remains small.

One finally observes that $\eta_{sat}(n,f)$ is almost independent of $f$ for $n=3$. In appendix~\ref{app:some_analytical_results_for_the_dependence_on_z_and_f}, we will show that it can be approximately written as:
\begin{equation}
  \eta_{sat} \simeq e^{-0.58}\left(1+0.044\left(f-\frac12\right)+{\cal O}\left(\left(f-\frac12\right)^2\right)\right)\,.\label{analytical_eta_sat_for_nf3}
\end{equation}
Because of the small coefficient of its first order correction, $\eta_{sat}=e^{-0.58}$ is a good approximation within less than $1\%$ for a large range of $f$ values. But this seems to be a coincidence with no deep physical reason.

\subsection{Hadronisation corrections\label{hadronisation_corrections}}

\begin{figure}[htb]
  \centering
  \subfigure[]{
    \includegraphics[scale=0.275]{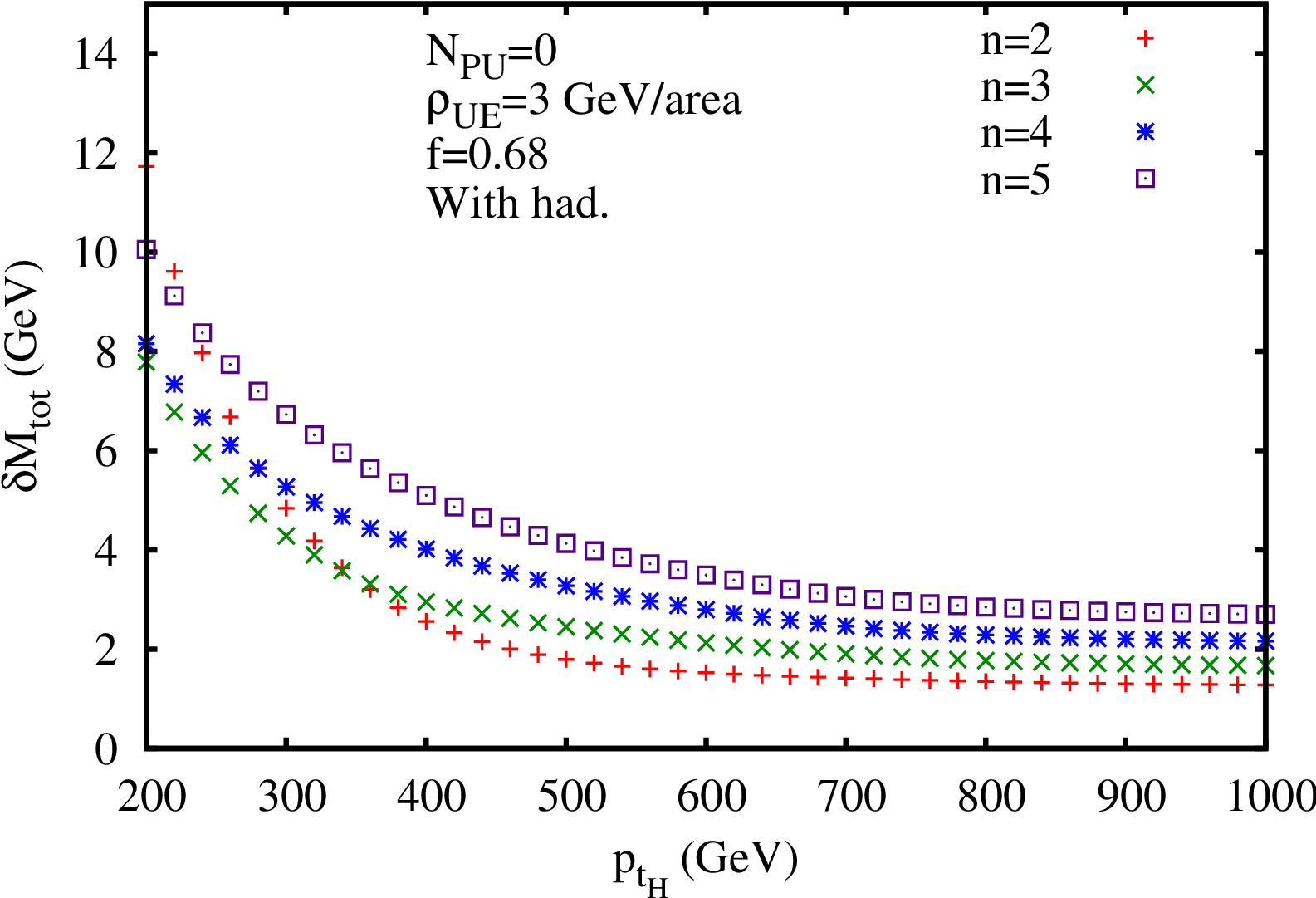}
    \label{dMtot_wrt_ptH_whad}
  }~
  \subfigure[]{
    \includegraphics[scale=0.275]{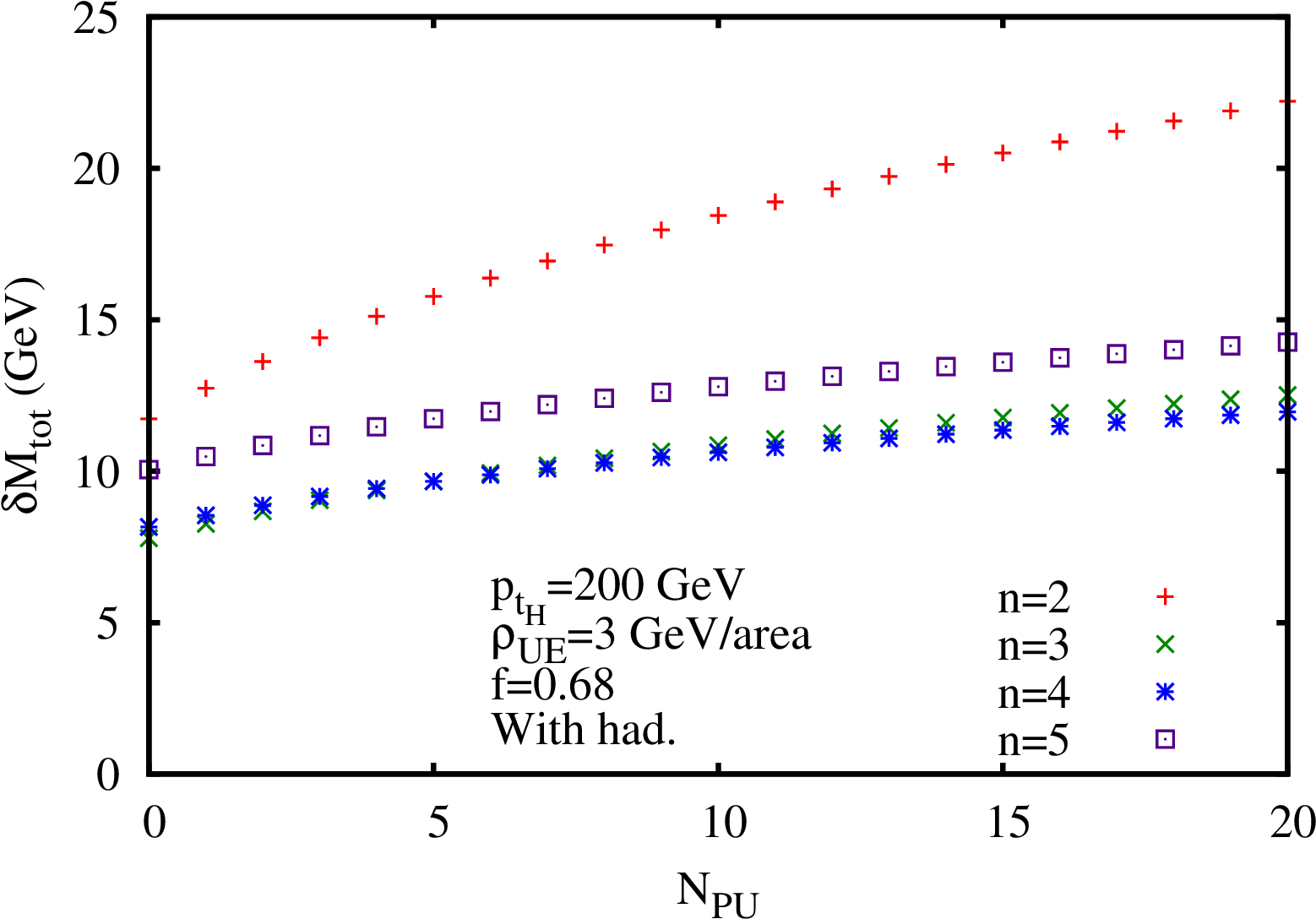}
    \label{dMtot_wrt_NPU_whad}
  }
  \caption{$\delta M_{tot}$ including hadronisation corrections computed at $\eta = \eta_{opt}$ as a function of \subref{dMtot_wrt_ptH_whad} $p_{t_H}$ and \subref{dMtot_wrt_NPU_whad} $N_{PU}$ for different values of $n$.}
  \label{dMtot_with_had}
\end{figure}

It is difficult to calculate what happens during the process of hadronisation, though some analytical results can be found concerning jet studies for instance \cite{Dasgupta:2007wa,Korchemsky:1994is,Dasgupta:2009tm}. In particular, it was shown in \cite{Dasgupta:2007wa} that such non-perturbative corrections lead to a $p_t$ shift for QCD jets equals on average $\sim -\Lambda/\filt{R}C_i$ where $\Lambda = 0.4$ GeV and $C_i$ $=$ $C_F$ or $C_A$ depending on whether it is a quark jet or a gluon jet. This can be translated in our study by the following averaged $p_t$ shift for the filtered jet:
\begin{align}
  \langle \delta p_t \rangle_{had} & = -\left(2C_F+(n-2)C_A\right)\frac{\Lambda}{\filt{R}}\,,\nonumber\\
  & \simeq -\frac{(n-1)N_c\Lambda}{\filt{R}}\,,
\end{align}
where the second equality holds in the large $N_c$ limit. Unfortunately, there is no result concerning the dispersion of the $p_t$ distribution, which is the relevant quantity to compute in our case. Therefore, we are going to assume that the spread is of the same order of magnitude as the shift. This is in principle a crude approximation, but the only aim here is to illustrate the consequences of including hadronisation corrections in order to emphasize the fact that taking $n$ too large is certainly not a good choice. Therefore, we use eq.~(\ref{average_delta_M_UE}) to estimate very roughly the hadronisation corrections to the reconstructed Higgs mass peak width:
\begin{equation}
  \delta M_{had} \sim \frac{(n-1)N_c\Lambda}{\filt{R}}\frac{M_H}{p_{t_H}} = \frac{(n-1)N_c\Lambda}{2\eta}\,,\label{dM_had}
\end{equation}
when $z=1/2$. As before, one should know how to combine perturbative radiation with UE/PU and hadronisation corrections in order to minimize the resulting combined width. However, we simply choose to minimize the quantity
\begin{equation}
  \delta M_{tot} = \sqrt{\delta M_{PT}^2+\delta M_{UE}^2+\delta M_{had}^2}\,,
\end{equation}
with respect to $\eta$ and plot the resulting minimal $\delta M_{tot}$ for different values of $n$ (fig.~\ref{dMtot_with_had}).

The first thing one can notice on these plots is that increasing $n$ also increases the hadronisation corrections. For $n=5$ they become so important that it is now clearly not an optimal filtering parameter contrary to what could be deduced from figure~\ref{solution_for_n_opt}. The relevant $p_t$ region in our study is roughly $200-400$ GeV where we find the major part of the Higgs cross-section above $200$ GeV and where our results are more reliable (see section~\ref{sec:higgs-width}). In this region $n=3$ gives the best result. And at high PU, $n=3$ and $n=4$ both seem optimal, whereas $n=2$ is far from being as good.

To conclude, our estimates seem to indicate within the accuracy of our calculations that $n=2$ is not a good choice, nor is $n\ge 5$. Taking $n=3$ or $n=4$ gives equally good quality to the mass peak. Increasing the hadronisation effects with respect to eq.~(\ref{dM_had}), would lead to $n_{opt}=3$, whereas if we lower them, we would find $n_{opt}=4$. The only thing we can say is that $n=3$ and $n=4$ both seem to work rather well.

One way to go beyond these results would be to use event generators like Herwig \cite{Corcella:2000bw,Corcella:2002jc} or Pythia \cite{Sjostrand:2006za}, to compute directly the Higgs width in presence of UE/PU, perturbative radiation, ISR and hadronisation, and to find for which value of the couple $(n,\eta)$ the reconstructed Higgs mass peak width $\delta M_H$ becomes minimal (that would still depend on $p_{t_H}$ and the level of UE/PU). But our study here aimed to understand as much as possible the physical aspects behind such an optimisation, the price to pay being larger uncertainties on the result because of the necessary simplifications that were made. 

\section{Conclusions}
\label{sec:conclusions}

This work has investigated the effect of QCD radiation on the
reconstruction of hadronically decaying boosted heavy particles,
motivated in part by the proposal of \cite{MyFirstPaper} to use a
boosted search channel for the $H\to b\bar b$ decay. Though this
article took the Higgs boson as an example, all the results presented
here can be applied to the $W$ and $Z$ bosons, as well as any new 
colorless resonance decaying hadronically that might be observed
at the LHC.
The main effect of the QCD radiation is to distort and spread out the
boosted heavy resonance shape well beyond the intrinsic width of the
resonance.
The aim therefore is to calculate the resulting resonance lineshape.
This is a function of the parameters of the reconstruction method,
notably of the ``filtering'' procedure, which aims to limit
contamination from underlying event and pile-up, but which causes more
perturbative radiation to be lost than would otherwise be the case.

Calculations were performed in a leading (single) logarithmic and
leading colour approximation, which is the state of the art for this
kind of problem.
Analytic results were provided up to $\as^2 \ln^2 \frac{M_H}{\Delta M}$ for
$n=2$, and all-orders analytic results for the cases $n=2$ and $n=3$
were given for the terms that dominate in the small $\eta$
limit.
Numerical fixed-order results up to $\as^5 \ln^5 \frac{M_H}{\Delta M}$ and all-orders resummed results were also given and are treated in more details in appendix~\ref{app:convergence_of_the_non-global_series} for a range of $n$ and $\eta$.
For the $n=2$ and $n=3$ cases there is quite acceptable
agreement between the small-$\eta$ analytic results and the full
numerical results, even for values of $\eta\simeq 0.5$.

One unexpected feature that was observed was the behaviour of
order-by-order expansion as compared to the resummed result: indeed
there are indications that the series in $\as \ln \frac{M}{\Delta M}$ has a radius of convergence that is zero (but in a way that is unrelated to the renormalon divergence of the perturbative QCD series).
This seems to be a general feature of the non-global logarithm series.
Its practical impact seems to be greater for large $\eta$, or
equivalently when the coefficients of the ``primary'' logarithms are
small.

With these results in hand, it was then possible to examine how the
perturbative width of the resonance peak depends on the parameters of
the filtering.
Though this was accessible only numerically for the full range of
filtering parameters, figure~\ref{Higgs_PT_width_t} lends itself to a
simple parametrisation for practically interesting parameter-ranges. 

This parametrisation was then used in section~\ref{sec:higgs-width}
together with a parametrisation for the effect of UE and PU, so as
to examine how to minimize the overall resonance width as a function
of the filtering parameters and of the physical parameters of the
problem such as the resonance $p_t$ and the level of UE/PU.
The approximations used might be described as overly simple, yet they
do suggest interesting relations between optimal choices of
the filtering parameters and the physical parameters of the problem.
Though it is beyond the scope of this article to test these relations
in full Monte Carlo simulation, we believe that investigation of their
applicability in realistic conditions would be an interesting subject
for future work. It should also be noticed that the methods used in
this paper may be adapted to other reconstruction procedures like 
jet pruning \cite{Ellis:2009me} and jet trimming \cite{Krohn:2009th}
as well as filtering as applied to jets without explicit substructure
\cite{Cacciari:2008gd}.

\section*{Acknowledgements}
\label{sec:acknowledgements}

I am very grateful to Gavin Salam for suggesting this work and for helpful
discussions while it was being carried out. I also wish to thank him as well as Sebastian Sapeta for comments on the manuscript.
This work was supported in part by the French ANR under contract
ANR-09-BLAN-0060.

\appendix
\appendixpage
\addappheadtotoc

\section{Analytical considerations on the non-global structure of the perturbative expansion \label{app:analytical_considerations}}

\begin{figure}[htb]
  \begin{center}
    \includegraphics[scale=0.4]{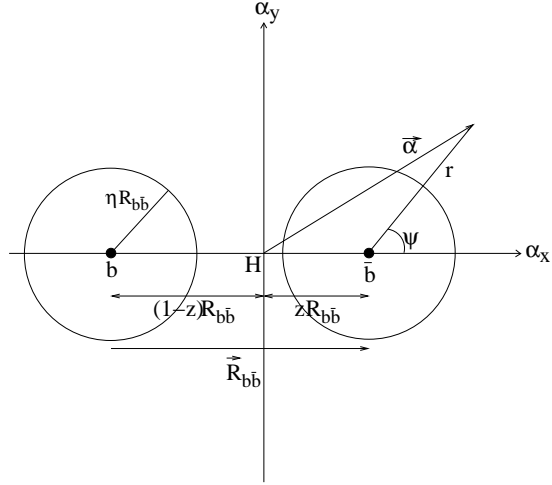}
  \end{center}
  \caption{The $\vec{\alpha}$ plane, with various variables used in the calculation and all along this study. In this figure, the $b$ quark is supposed to carry a fraction $z$ of the Higgs energy, and the center of the frame coincides with the direction of the Higgs boson momentum.}
  \label{variables}
\end{figure}

We derive here all the results presented in section~\ref{NG_structure_analytical_insights}. The Higgs boson is taken to move along the $x$ direction and the angular coordinates $(\theta,\phi)$ are defined with respect to the Higgs direction (so that $\theta = 0$ corresponds to the Higgs direction and $\phi = 0$ to the $y$ axis for instance). As it is very boosted, the $b$ and $\bb$ resulting from its decay should be close to the Higgs, i.e. with $\theta_{b,\bb}\ll 1$. Due to angular ordering, so will be the major part of the perturbative radiation from $b\bb$. To take this property into account, the angular coordinates $(\theta,\phi)$ are slightly changed into a two-dimensional vector $\vec{\alpha}$:
\begin{align}
  \vec{\alpha} & = \theta(\cos\phi,\sin\phi)\,, \nonumber\\
  d^2\vec{\alpha} & = \theta d\theta d\phi\,.
\end{align}
This vector is useful because of the property that, at small $\theta$, one can express the angle $\theta_{ij}$ between $2$ particles $i$ and $j$ as:
\begin{equation}
  \theta_{ij} = |\vec{\alpha}_i-\vec{\alpha}_j|\,.
\end{equation}
$\vec{R}_{b\bb}$ is defined to be the vector from $b$ to $\bb$ in the $\vec{\alpha}$ plane (see fig.~\ref{variables}). Even if $|\vec{\alpha}|<{\cal O}(1)$, the integrations over $\alpha_x$ and $\alpha_y$ are extended to $\infty$. By doing so, the error made on a ${\cal O}(1)$ result is of order ${\cal O}\left(R_{bb}^2\right)$.

\subsection{Primary coefficients}

Let us go back to the integral of eq.~(\ref{PrimaryIntegral}) and write it in the $\vec{\alpha}$ plane:
\begin{align}
  I(\Delta M) & = -\int_{\vec{k}\notin J_{b\bb}}\frac{d^3\vec{k}}{(2\pi)^32|\vec{k}|}|M(k)|^2\Theta\left(\Delta M(k) - \Delta M)\right)\,,\nonumber \\
  & = -\frac{\alpha_sC_F}{2\pi^2}\int_0^{p_{t_H}}\frac{dk_t}{k_t}\int_{\vec{\alpha}\notin J_{b\bb}}d^2\vec{\alpha}\frac{2R_{bb}^2}{|\vec{\alpha} - z\vec{R}_{bb}|^2|\vec{\alpha}+(1-z)\vec{R}_{bb}|^2}\Theta\left(\Delta M(k)-\Delta M\right)\,.
\end{align}
For $\Delta M(k) = M_H - M$ one easily finds
\begin{equation}
  \Delta M(k) = k_t\frac{M_H}{p_{t_H}}A\left(\vec{\alpha}\right)\,,  
\end{equation}
with
\begin{align}
  A\left(\vec{\alpha}\right) & = \frac{|\vec{\alpha}|^2p_{t_H}^2}{2M_H^2}+\frac12\,,\\
  & = \frac{|\vec{\alpha}|^2}{2z(1-z)R_{bb}^2}+\frac12\,,
\end{align}
where we used eq.~(\ref{Rbb_boosted_limit}). With this expression, the integration over $k_t$ is straightforward:
\begin{align}
  \int_0^{p_{t_H}}\frac{dk_t}{k_t}\,\Theta\left(\Delta M(k)-\Delta M\right) & = \int_{\frac{p_{t_H}\Delta M}{M_HA(\vec{\alpha})}}^{p_{t_H}}\frac{dk_t}{k_t}\,,\nonumber\\
  & = \ln\frac{M_H}{\Delta M} + \ln A(\vec{\alpha})\,.
\end{align}
Notice that $|\vec{\alpha}|\sim (1-z)R_{bb}$ or $|\vec{\alpha}|\sim zR_{bb}$, depending on whether the perturbative gluon emission is relatively close to $b$ or $\bb$ (due to the collinear divergence of QCD). Thus, given that a Higgs splitting is most of the time roughly symmetric ($z\sim 1/2$), this leads to $A\left(\vec{\alpha}\right) = {\cal O}(1)$. Therefore:
\begin{equation}
   \ln A(\vec{\alpha}) \ll \ln\frac{M_H}{\Delta M}\,,
\end{equation}
and the $\ln A(\vec{\alpha})$ term can be neglected in a leading-log calculation. One thus obtains:
\begin{equation}
  I(\Delta M) =  -\frac{\alpha_sC_F}{\pi}\ln\left(\frac{M_H}{\Delta M}\right)J(\eta)\,,
\end{equation}
with $J(\eta)$ the remaining angular integral. Introducing the coordinates $(r,\psi)$ defined in fig.~\ref{variables}, $J(\eta)$ can be rewritten in the following form:\footnote{A simple shift of the $\vec{\alpha}$ coordinates gets rid of the $z$ dependence, for instance $\vec{\alpha}'=\vec{\alpha}-\left(z-\frac12\right)\vec{R}_{b\bb}$}
\begin{align}
  J(\eta) & = J_0(\eta) = \frac{1}{\pi}\int_{\eta R_{bb}}^{+\infty}\frac{dr}{r}\int_{-\arccos(-\frac{R_{bb}}{2r})}^{\arccos(-\frac{R_{bb}}{2r})}d\psi \frac{2R_{bb}^2}{r^2+2rR_{bb}\cos\psi+R_{bb}^2} \quad \mbox{if }\frac{1}{2}<\eta<1\,,\nonumber \\
  & = J_0\left(\frac{1}{2}\right)+\frac{1}{\pi}\int_{\eta R_{bb}}^{\frac{R_{bb}}{2}}\frac{dr}{r}\int_{-\pi}^{\pi}d\psi\frac{2R_{bb}^2}{r^2+2rR_{bb}\cos\psi+R_{bb}^2} \hspace{1.7cm} \mbox{if } \eta < \frac{1}{2}\,.
\end{align}
Performing the $\psi$ integration and the $r$ one (for $\eta<1/2$), one arrives at the formulae~(\ref{primary_emission_result_eta_small},\ref{primary_emission_result_eta_large}), where all the $R_{bb}$ dependence is cancelled.

One remark: if $\eta > 1$, the $b$ and $\bb$ quarks cluster together, and the result then depends on the $z$ fraction of the splitting. For instance, if $\eta>2$, $J(\eta)$ can now be written:\footnote{the intermediate case $1<\eta<2$ has a more complicated phase space integration and is not presented here.}
\begin{equation}
  J(\eta,z) = \ln\left(\frac{(\eta^2+z(1-z))^2}{(\eta^2-z^2)(\eta^2-(1-z)^2)}\right)\,.
\end{equation}

\subsection{Non-Global coefficients}

\begin{figure}[htb]
  \begin{center}
    \subfigure[$S_{tot}$]{\epsfig{figure=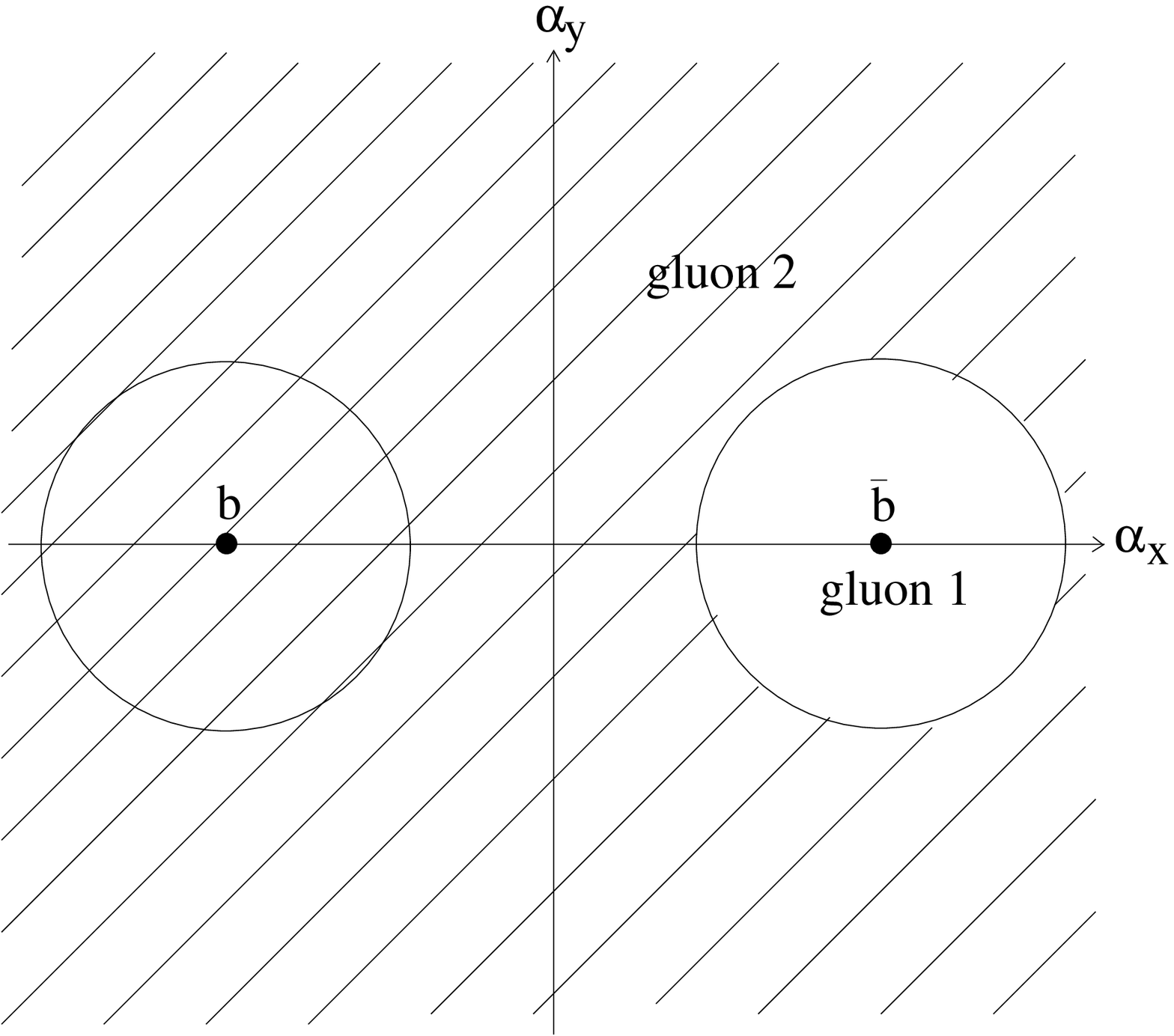,scale=0.33}}\qquad
    \subfigure[$S_{int}$]{\epsfig{figure=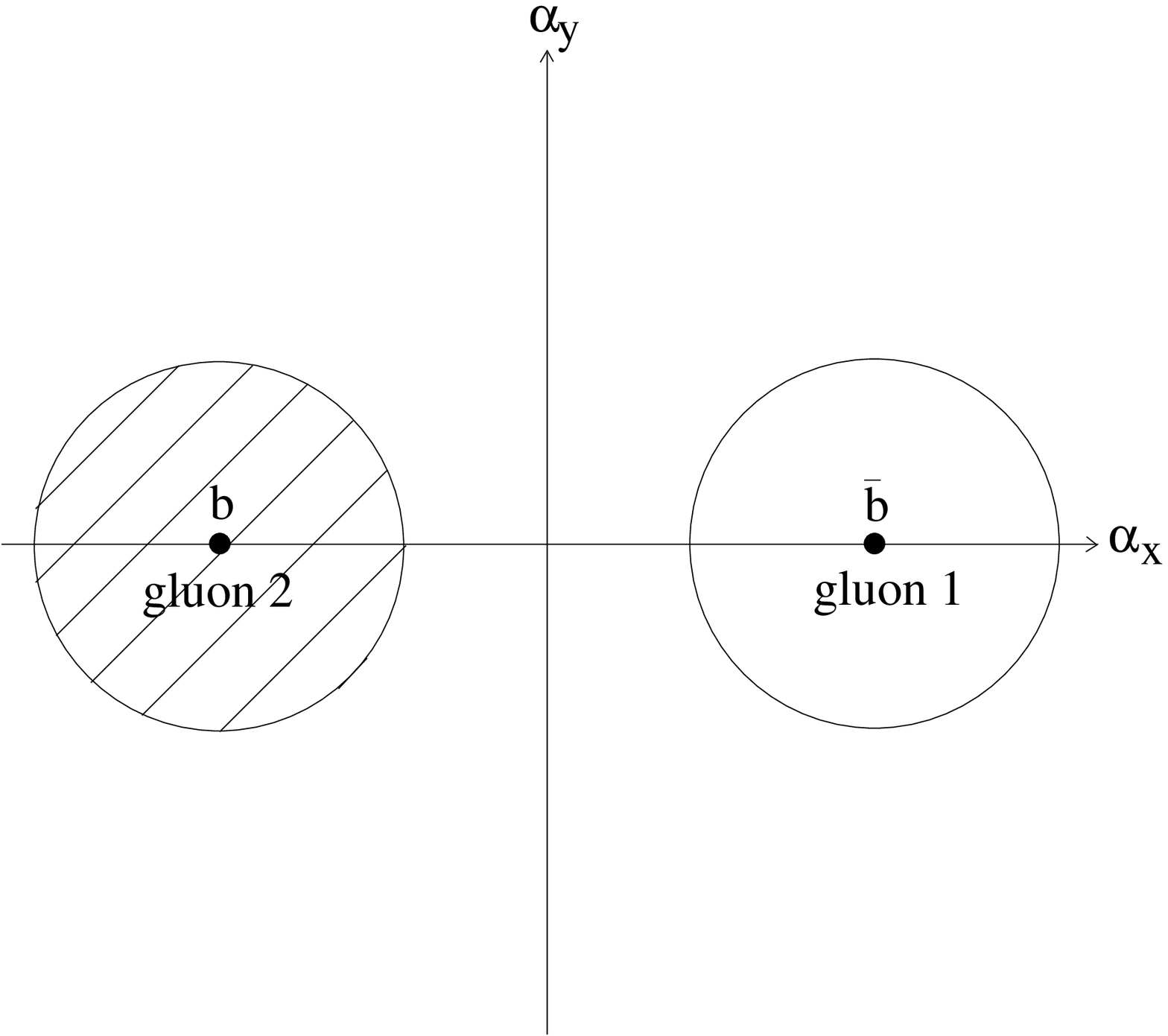,scale=0.33}}
  \end{center}
  \caption{regions of integration (dashed region) for gluon $2$ when computing $S_{tot}$ (left) and $S_{int}$ (right).}
\label{graphical_S_tot_and_S_int}
\end{figure}

The starting point here is eq.~(\ref{DefinitionOfI_2}) for $I_2^{NG}$:
\begin{equation*}
   I_2^{NG}(\Delta M) = - \int dk_1dk_2 (4\pi\alpha_s)^2\Theta(E_1-E_2)\Theta(k_1\in J_{b\bb})\Theta(k_2\notin J_{b\bb})W_2\Theta(\Delta M(k_2) - \Delta M)\,.
\end{equation*}
In the same way as the primary case, using the $\vec{\alpha}$ plane and integrating over the energies of the $2$ gluons, one arrives at:
\begin{equation}
  I_2^{NG}(\Delta M) = \frac{1}{2}C_FC_A\left(\frac{\alpha_s}{\pi}\log\left(\frac{M_H}{\Delta M}\right)\right)^2S_2(\eta)\,,
\end{equation}
with
\begin{align}
S_2(\eta) & = -\frac{R_{bb}^2}{2\pi^2}\int_{\vec{\alpha}_1\in J_{b\bb}} d^2\vec{\alpha}_1\int_{\vec{\alpha}_2\notin J_{b\bb}} d^2\vec{\alpha}_2\Big( \nonumber\\
& \qquad \qquad {} - \frac{R_{bb}^2}{|\vec{\alpha}_1-z\vec{R}_{bb}|^2||\vec{\alpha}_1+(1-z)\vec{R}_{bb}|^2 |\vec{\alpha}_2-z\vec{R}_{bb}|^2||\vec{\alpha}_2+(1-z)\vec{R}_{bb}|^2} \nonumber \\
& \qquad \qquad {} + \frac{1}{|\vec{\alpha}_1+(1-z)\vec{R}_{bb}|^2|\vec{\alpha}_1-\vec{\alpha}_2|^2|\vec{\alpha}_2-z\vec{R}_{bb}|^2}\nonumber\\
& \qquad \qquad {} +\frac{1}{|\vec{\alpha}_1-z\vec{R}_{bb}|^2|\vec{\alpha}_1-\vec{\alpha}_2|^2|\vec{\alpha}_2+(1-z)\vec{R}_{bb}|^2}\Big)\,.
\end{align}
In all this part, $\eta<1/2$ is assumed. To deal with this integral, the frame is centered around $\bb$ for instance, and $2$ quantities are computed: $S_{tot}$ where gluon $1$ is in the jet region $J_{\bb}$ around $\bb$ and gluon $2$ covers the whole phase space except $J_{\bb}$, from which is subtracted $S_{int}$ where gluon $2$ covers $J_b$, the jet region around $b$ (fig.~\ref{graphical_S_tot_and_S_int}).
Therefore:
\begin{equation}
  S_2(\eta) = 2(S_{tot}(\eta)-S_{int}(\eta))\,,
  \label{Final_S_2}
\end{equation}
where the factor $2$ is for the symmetric case (gluon $1$ in $J_b$).

\subsubsection{Calculation of $S_{tot}$}

Using the variables $\left(u=\frac{r}{\eta R_{bb}},\psi\right)$ (fig.~\ref{variables}), $S_{tot}$ can be written
\begin{align}
  S_{tot}(\eta)  = & -\frac{1}{2\pi^2}\int_0^1u_1du_1\int_1^{+\infty}u_2du_2\int_0^{2\pi}d\psi_1\int_0^{2\pi}d\psi_2\Big(\nonumber\\
  & {}\quad - \frac{1}{u_1^2u_2^2(1+2\eta u_1\cos\psi_1+\eta^2u_1^2)(1+2\eta u_2\cos\psi_2+\eta^2u_2^2)} \nonumber \\
  & {}\quad + {} \frac{1}{u_1^2(1+2\eta u_2\cos\psi_2+\eta^2u_2^2)(u_1^2-2u_1u_2\cos(\psi_1-\psi_2)+u_2^2)} \nonumber \\
  & {}\quad + {} \frac{1}{u_2^2(1+2\eta u_1\cos\psi_1+\eta^2u_1^2)(u_1^2-2u_1u_2\cos(\psi_1-\psi_2)+u_2^2)}\Big)\,.
\end{align}
Doing the angular integrations, one arrives at:
\begin{equation}
  S_{tot}(\eta) = -4\int_0^1du_1\frac{u_1}{1-\eta^2u_1^2}\int_1^{\frac{1}{\eta}}\frac{du_2}{u_2}\frac{1}{u_2^2-u_1^2}\,.
\end{equation}
$2$ remarks about this result:
\begin{enumerate}
  \item The first and second terms have divergences in $u_1=0$ and $u_2=\frac{1}{\eta}$, so respectively when gluon $1$ is collinear to $\bb$ and gluon $2$ is collinear to $b$, but they cancel when adding these terms.
  \item The part of the integral corresponding to $u_2>\frac{1}{\eta}$, i.e. $r_2>R_{bb}$, is null, which can be simply interpreted as a manifestation of the angular ordering.
\end{enumerate}
Performing this integration with Maple for instance gives the following result:
\begin{align}
  S_{tot}(\eta) & = \frac{\pi^2}{2}-\frac{3}{2}\ln^2(1-\eta)+\ln(2\eta)\ln(1-\eta)-\ln 2\ln(1+\eta) - 2\Re\left(\mbox{dilog}\left(\frac{\eta}{\eta - 1}\right)\right) -\nonumber\\
  & {} -2\Li_2(1-\eta)-\Li_2\left(\frac{1}{1+\eta}\right)+\Li_2\left(-\eta\right)+\Li_2\left(\frac{1-\eta}{1+\eta}\right)-\Li_2\left(\frac{-2\eta}{1-\eta}\right)\,,
\end{align}
where 
\begin{align}
  \mbox{dilog}(x) & = \int_1^xdt\frac{\ln t}{1-t}\,,\\
  \Li_2(x) & = \int_x^0dt\frac{\ln(1-t)}{t}\,.
\end{align}
This is rather complicated, but that expression can be greatly simplified using the relations:
\begin{align}
  \forall x>0\mbox{ , } & \mbox{dilog}(x) = \Li_2(1-x)\,,\\
  & \mbox{dilog}(-x) = \frac{\pi^2}{3}-\frac{1}{2}\ln^2(1+x)-\Li_2\left(\frac{1}{1+x}\right)-i\pi\ln(1+x)\,,\\
  \forall x \mbox{ with } 0<x<1\mbox{ , } & \Li_2(x)+\Li_2(1-x) = \frac{\pi^2}{6}-\ln(x)\ln(1-x)\,,\\
  & \Li_2(1-x)+\Li_2(1-1/x) = -\frac{1}{2}\ln^2x\,.
\end{align}
The final answer is then:
\begin{equation}
  S_{tot}(\eta) = -\frac{\pi^2}{6}\,.
\end{equation}
The remarkable point to notice is of course that $S_{tot}$ does not depend on $\eta$. But no simple explanation was found to interpret this result.

\subsubsection{Calculation of $S_{int}$}

$S_{int}$ must now be subtracted from $S_{tot}$. The computation is similar to the previous one and is not detailed here. However, contrary to $S_{tot}$, a simple analytical result was not obtained, only the following expansion:
\begin{align}
  S_{int}(\eta) & = -4\int_0^1\frac{du_1}{u_1}\int_0^1\frac{du_2}{u_2}\left(\frac{1}{\sqrt{(1-\eta^2(u_1^2+u_2^2))^2-4\eta^4u_1^2u_2^2}}-\frac{1}{1-\eta^2(u_1^2+u_2^2)}\right)\,, \\
  & = -2\eta^4-6\eta^6-\frac{31}{2}\eta^8-40\eta^{10}-\frac{1921}{18}\eta^{12}-\frac{889}{3}\eta^{14}-\frac{20589}{24}\eta^{16}-\frac{7643}{3}\eta^{18}+{\cal O}(\eta^{20})\,.
\end{align}
Using eq.~(\ref{Final_S_2}) with the $2$ previous results, one arrives at the expression eq.~(\ref{S_2_computed}).

\section{Analytical considerations on the dependence of the results on $z$~and~$f$ \label{app:some_analytical_results_for_the_dependence_on_z_and_f}}

\subsection{Dependence on $z$}

It is interesting to understand analytically how $\eta_{opt}$ evolves when the decay of the Higgs boson occurs with a $z$ fraction different from $1/2$. In fact, we obtain the same result as eqs.~(\ref{solution_for_L_opt},\ref{eta_approximate_solution}) up to a modification of the constant $C_{UE}$ (more generally, we have to modify the coefficients $c_{\sigma}$, $c_{\delta\rho}$ and $c_{\Sigma}$, see eqs.~(\ref{exact_equation_for_L}-\ref{c_Sigma_with_PU})). The starting point is eq.~(\ref{average_DeltaM_for_any_z_and_n_2}) generalized to any value of $n$:
\begin{equation}
  \Delta M \simeq \frac{M_H}{2p_{t_H}}\sum_{i=1}^na_i(z)\rho A_i\,,\label{Delta_M_for_every_z_and_n}
\end{equation}
where $a_i(z)$ is either $\frac1z$ or $\frac{1}{1-z}$ depending on whether subjet $i$ is in the $J_1$ region (around the $b$ quark) or in the $J_2$ region (around the $\bb$ quark). This is because the UE/PU particles tend to cluster around the perturbative radiation, which itself is emitted close to $b$ and $\bb$. We call a ``configuration'' the set of all the coefficients $a_i(z)$.

The result on the fluctuations depends on which ones are considered. Let us start with the fluctuations originating from the $\sigma$ and $\Sigma$ terms. In this case the $\rho A_i$ terms vary independently, thus leading to a contribution to $\delta M_{UE}$ similar to that of eq.~(\ref{z_dependence_n_2_for_s_and_drho}) for $n=2$:
\begin{equation}
  \delta M_{UE,\sigma,\Sigma,\{a_i\}}^2 = 4\left(\frac{M_H\rho_{UE}}{2p_{t_H}}\right)^2\sum_{i=1}^na_i(z)^2\delta_{i,\sigma,\Sigma}^2\,, \label{delta_M_UE_sigma_Sigma_a_i}
\end{equation}
with (see eq.~(\ref{delta_sigma_Sigma})):
\begin{align}
  \delta_{i,\sigma,\Sigma}^2 & = c_{\sigma}^2e^{-2L}+c_{\Sigma}^2e^{-4L}\,,\nonumber\\
  & \equiv \delta_{\sigma,\Sigma}^2\,.
\end{align}
The coefficients $c_{\sigma}$ and $c_{\Sigma}$ are still computed for $n=1$ in this formula. But eq.~(\ref{delta_M_UE_sigma_Sigma_a_i}) is only valid for a given configuration $\{a_i\}$. We thus have to average over all the $2^{n-2}$ possible configurations (the $b$ and $\bb$ subjets are fixed). As the perturbative radiation pattern does not depend on $z$ (for $z$ not too small), each configuration arises with the same probability. Therefore, if $k$ is the number of subjets in the $J_1$ region and $n-2-k$ the number of subjets in the $J_2$ region apart from the $b$ and $\bb$ subjets, we obtain:
\begin{align}
  \delta M_{UE,\sigma,\Sigma}^2 & = 4\left(\frac{M_H\rho_{UE}}{2p_{t_H}}\right)^2\frac{1}{2^{n-2}}\sum_{k=0}^{n-2}\binom{n-2}{k}\left(\frac{k+1}{z^2}+\frac{n-1-k}{(1-z)^2}\right)\delta_{\sigma,\Sigma}^2\,,\\
  & = 4\left(\frac{M_H\rho_{UE}}{p_{t_H}}\right)^2\frac{n}{8}\frac{1-2z(1-z)}{z^2(1-z)^2}\delta_{\sigma,\Sigma}^2\,.
\end{align}
We can follow a similar reasoning for the $\delta\rho$ fluctuations, except that the $\rho A_i$ terms in eq.~(\ref{Delta_M_for_every_z_and_n}) vary the same way from one event to the next. Therefore, for a given configuration $\{a_i\}$, one can deduce the following contribution to $\delta M_{UE}$:
\begin{equation}
  \delta M_{UE,\delta\rho,\{a_i\}}^2 = 4\left(\frac{M_H\rho_{UE}}{2p_{t_H}}\right)^2\left(\sum_{i=1}^na_i(z)\right)^2\delta_{\delta\rho}^2\,,
\end{equation}
with $\delta_{\delta\rho}$ given by eq.~(\ref{delta_delta_rho}). As before, we have to average this result over all the $2^{n-2}$ possible configurations, leading to:
\begin{align}
  \delta M_{UE,\delta\rho}^2 & = 4\left(\frac{M_H\rho_{UE}}{2p_{t_H}}\right)^2\frac{1}{2^{n-2}}\sum_{k=0}^{n-2}\binom{n-2}{k}\left(\frac{k+1}{z}+\frac{n-1-k}{1-z}\right)^2\delta_{\delta\rho}^2\,,\\
  & = 4\left(\frac{M_H\rho_{UE}}{p_{t_H}}\right)^2\frac{n^2+(n-2)(1-2z)^2}{16z^2(1-z)^2}\delta_{\delta\rho}^2\,.
\end{align}
One can absorb all the dependence of the resulting $\delta M_{UE}^2$ in $n$ and $z$ into the coefficients $c_{\sigma}$, $c_{\delta\rho}$ and $c_{\Sigma}$, and define new coefficients $c'$ such that
\begin{equation}
  \delta M_{UE} = 2\sqrt{c_{\sigma}'^2\eta^2+c_{\delta\rho}'^2\eta^4+c_{\Sigma}'^2\eta^4}\frac{M_H\rho_{UE}}{p_{t_H}}\,,
\end{equation}
with
\begin{align}
  c_{\sigma}'(n,N_{PU},R_{bb},z) & \simeq 0.6\sqrt{\pi}\sqrt{n}\frac{\sqrt{1-2z(1-z)}}{2\sqrt{2}z(1-z)}R_{bb}\sqrt{1+\frac{N_{PU}}{4}}\,,\\
  c_{\delta\rho}'(n,N_{PU},R_{bb},z) & \simeq 0.8\pi \frac{\sqrt{n^2+(n-2)(1-2z)^2}}{4z(1-z)}R_{bb}^2\sqrt{1+\frac{N_{PU}}{4}}\,,\\
  c_{\Sigma}'(n,N_{PU},R_{bb},z) & \simeq 0.26\pi\sqrt{n}\frac{\sqrt{1-2z(1-z)}}{2\sqrt{2}z(1-z)}R_{bb}^2\left(1+\frac{N_{PU}}{4}\right)\,.
\end{align}
With these results in hand, we can easily generalize eq.~(\ref{solution_for_L_opt}) for any value $z$ of the Higgs splitting. One only has to modify the value of $C_{UE}$. For instance, the curves in fig.~\ref{eta_optimal_ptH_z0.2} were obtained using eqs.~(\ref{value_of_B_PT}$-$\ref{eta_approximate_solution}) with:
\begin{equation}
  C_{UE}(n,N_{PU},z) = 1.6\pi\frac{\sqrt{n^2+(n-2)(1-2z)^2}}{4z(1-z)}\sqrt{1+\frac{N_{PU}}{4}}\,,
\end{equation}
where we just use the dominant $c_{\delta\rho}'$ term.

There is also another source of fluctuations that we haven't accounted for so far. Indeed, even if the values of $\rho$ and of the jets area were constant, we should consider the fact that the filtered subjets can be either in the $J_1$ or in the $J_2$ region. The calculation of its effect is similar to that of the fluctuations in $\sigma$, $\delta\rho$ and $\Sigma$. Its contribution on $\delta M_{UE}^2$ can be written:
\begin{equation}
  \delta M_{UE,\mbox{\scriptsize other}}^2 = 4\left(\frac{M_H\rho_{UE}}{p_{t_H}}\right)^2c_{\mbox{\scriptsize other}}'^2\eta^4\,,
\end{equation}
with
\begin{equation}
  c_{\mbox{\scriptsize other}}'(n,N_{PU},R_{bb},z) = \pi \frac{\sqrt{n-2}|1-2z|}{4z(1-z)} R_{bb}^2 \left(1+\frac{N_{PU}}{4}\right)\,.
\end{equation}
But we checked that its effect on $\delta M_{UE}^2$ is negligible compared to that from the dominant $\delta\rho$ fluctuation,\footnote{This effect is strictly null when $z=1/2$. When $z=0.2$, we obtain:$$\delta M_{UE,\delta\rho}^2 \simeq 16.6\,\delta M_{UE,\mbox{\scriptsize other}}^2\,,$$ for $p_{t_H}=200$ GeV and $N_{PU}=0$.} and therefore we did not include it.

\subsection{Comments on the uncertainty due to the choice of $f$}

Let us return to the observation of section~\ref{variations_of_the_results_with_z_and_f} that $\eta_{sat}(n,f)$ is almost independent of $f$ for $n=3$. To understand why, we have to estimate $t_{sat}(f)$ and $C_{PT}(3,f)$ (cf eq.~(\ref{eta_sat})). $t_{sat}$ can be deduced from the equation:
\begin{equation}
  \Sigma^{(n)}(\eta = 1, t_{sat}) = f\,,
\end{equation}
for any value of $n$, as $t_{sat}$ only depends on $f$ (see for instance figs.~(\ref{Higgs_PT_width_GeV},\ref{Higgs_PT_width_t}). To be simple, we can use the function $\Sigma^{(2)}$ which was widely studied in this paper. Unfortunately we cannot take the primary emission result eq.~(\ref{exponentiated_primary_result}) because the non-global part becomes important when $\eta=1$. However, as an approximation, we can numerically compute the second order coefficient $a_2\simeq -3$ of $\Sigma^{(2)}(\eta=1,t)$ and solve:
\begin{equation}
  1-J(1)N_ct_{sat}+a_2t_{sat}^2 \simeq f\,,
\end{equation}
which simply leads to
\begin{equation}
  t_{sat}(f) \simeq \frac{J(1)N_c-\sqrt{J(1)^2N_c^2-4a_2(1-f)}}{2a_2}\,.
\end{equation}
This expression can be shown numerically to give $t_{sat}(f)$ with a precision better than $1\%$ for $f\in [0.5,0.8]$. $C_{PT}(3,f)$ is harder to evaluate. One must solve 
\begin{equation}
  \Sigma^{(3)}(L,t) = f\,, \label{width_equation_for_nf3}
\end{equation}
in the limit of large $L$. Using eq.~(\ref{Sigma3LargeLLimit}), which is valid in this limit, and defining the function $h$ such that $h(4N_cLt)=\Sigma^{(3)}(L,t)-f$, we can use Newton's method with only one iteration on $h$, evaluated numerically for $4N_cLt=1$,\footnote{Starting from $4N_cLt\sim 1$ quickly converges to the true solution.} to solve approximately eq.~(\ref{width_equation_for_nf3}):
\begin{align}
  4N_cLt & \simeq 1-\frac{h(1)}{h'(1)}\,,\nonumber\\
  & \simeq 2.543-2.334f\,,\nonumber\\
  & \equiv 4N_cC_{PT}(3,f)\,,
\end{align}
which gives $C_{PT}(3,f)$ within a few $\%$. Therefore, eq.~(\ref{eta_sat}) leads to
\begin{align}
  \eta_{sat}(3,f) & = e^{-\frac{C_{PT}(3,f)}{t_{sat}}}\,,\nonumber\\
  & \simeq e^{-0.58}\left(1+0.044\left(f-\frac12\right)+{\cal O}\left(\left(f-\frac12\right)^2\right)\right)\,,
\end{align}
which is eq.~(\ref{analytical_eta_sat_for_nf3}).

To be complete, the same analysis can be done when $n=2$, using eq.~(\ref{C_PT_2_f}) for $C_{PT}(2,f)$, to obtain $\eta_{sat}(2,f)$. We will just mention that 
\begin{equation}
  \lim_{f\rightarrow 1}\eta_{sat}(2,f) = e^{-\frac{J(1)}{4}}\simeq 0.85\,.
\end{equation}

\section{Convergence of the non-global series \label{app:convergence_of_the_non-global_series}}

In this appendix, we go back to the study of the convergence of the non-global series that was already briefly examined in section~\ref{comparison_with_fixed_order_results}, trying to understand a little more what may be behind the observed behaviour.

\subsection {Case  $\filt{n}=2$}

\begin{figure}[hbt]
  \centering
  \subfigure[]{
    \includegraphics[scale=0.275]{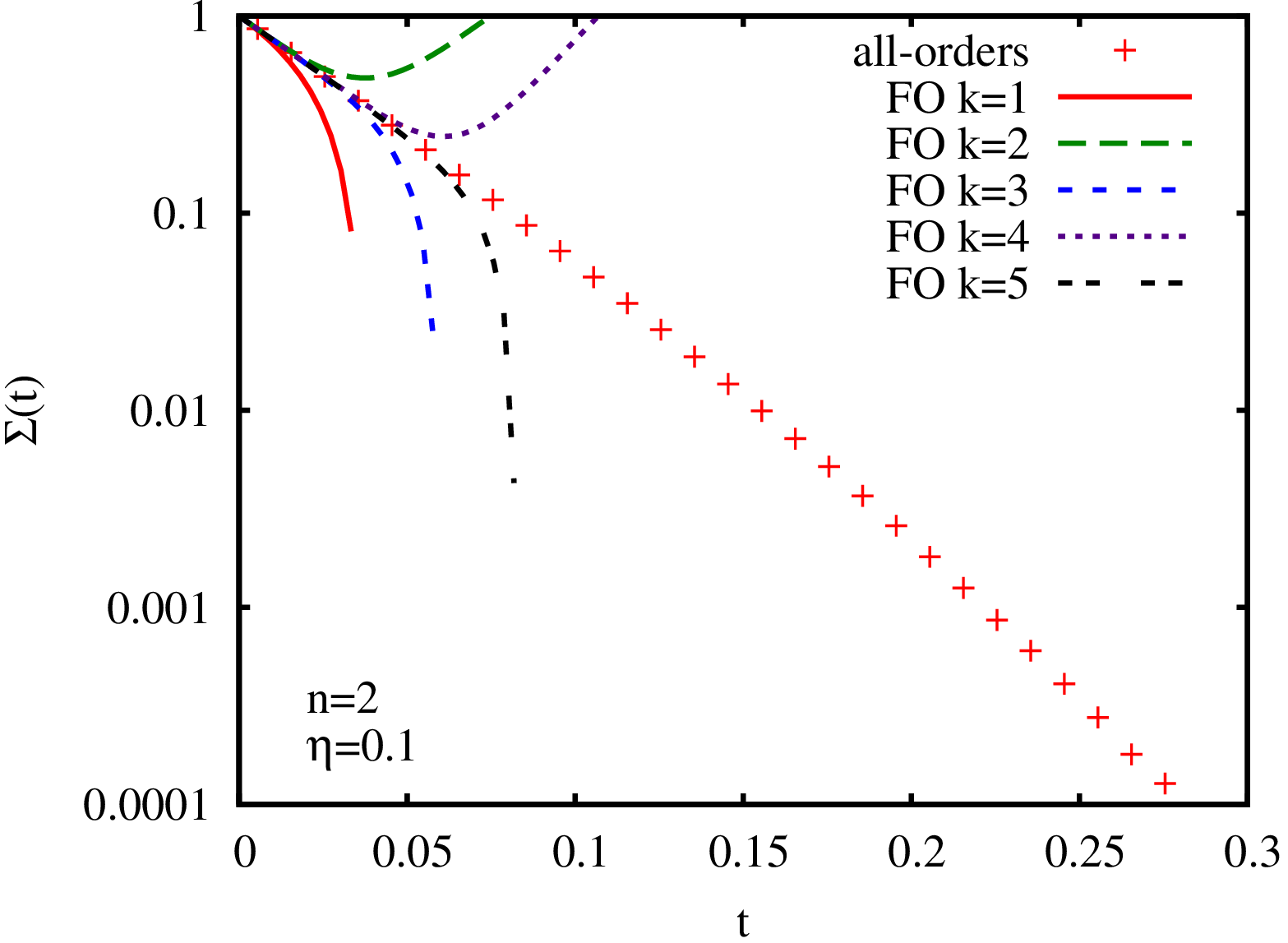}
    \label{ao_vs_fo_nf2_eta0.1}
  }~
  \subfigure[]{
    \includegraphics[scale=0.275]{figures/ao_vs_fo_nf2_eta0.3.ps}
    \label{ao_vs_fo_nf2_eta0.3_app}
  }
  \caption{Comparison between fixed-order (FO) expansion and all-orders result when $n=2$ for \subref{ao_vs_fo_nf2_eta0.1} $\eta=0.1$ and \subref{ao_vs_fo_nf2_eta0.3_app} $\eta=0.3$.}
  \label{ao_vs_fo_nf2_eta_small}
\end{figure}

First, we start by studying the convergence of the non-global perturbative series when $n=2$ for $\eta$ small, where the primary coefficients are known to be enhanced with respect to the purely non-global ones because of the presence of large collinear logarithms (cf section~\ref{some_results_for_nf2}). Figure~\ref{ao_vs_fo_nf2_eta_small} compares the fixed-order results to the all-orders one up to $\alpha_s^5$.
On these $2$ plots one can notice that the series seems to converge, as was already shown in section~\ref{comparison_with_fixed_order_results}. In other respects, the convergence looks better when $\eta$ is larger. This may be understood using the following simple explanation: if we make an expansion up to order $k$, then the series starts to diverge from the exact result when the term of order $k+1$ becomes roughly of the same size as the function itself. In the $\Sigma^{(2)}(L,t)$ case, using the analytic expression (eq.~(\ref{Sigma2LargeLLimit})), this can be translated to
\begin{equation}
  \frac{\left(4N_cLt\right)^{k+1}}{(k+1)!} \sim e^{-4N_cLt}\,,
\end{equation}
with $L = \ln\frac{1}{\eta}$. At large $k$,\footnote{the derivation was done at large $k$, but the result seems to be reasonable even for $k=2$.} the solution gives
\begin{equation}
  t \sim \frac{k+1}{4N_cL}\,.
\end{equation}
So the convergence is better when $k$ and $\eta$ increase.
\begin{figure}[hbt]
  \centering
  \subfigure[]{
    \includegraphics[scale=0.275]{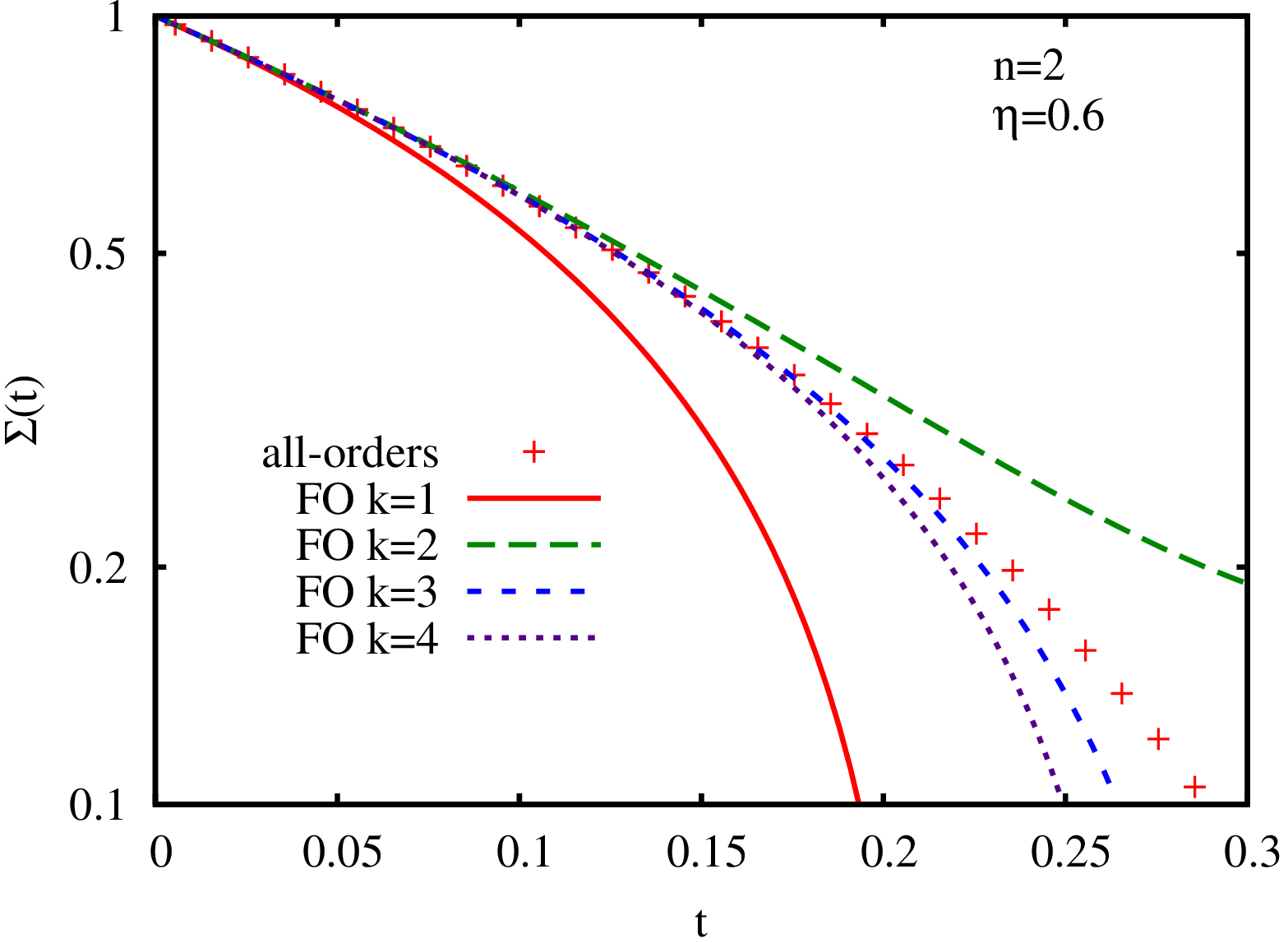}
    \label{ao_vs_fo_nf2_eta0.6}
  }~
  \subfigure[]{
    \includegraphics[scale=0.275]{figures/ao_vs_fo_nf2_eta0.9.ps}
    \label{ao_vs_fo_nf2_eta0.9_app}
  }
  \caption{Comparison between fixed-order (FO) expansion and all-orders result when $n=2$ for \subref{ao_vs_fo_nf2_eta0.6} $\eta=0.6$ and \subref{ao_vs_fo_nf2_eta0.9_app} $\eta=0.9$.}
  \label{ao_vs_fo_nf2_eta_large}
\end{figure}

Let us go to larger $\eta$, where there is no large collinear logarithm anymore, and check what happens. This is done for $\eta = 0.6$ and $\eta = 0.9$ on figure~\ref{ao_vs_fo_nf2_eta_large}. One striking feature of these plots is that the convergence seems acceptable up to the third order, but the $4^{th}$ order does not give as good a result, and the situation becomes worse as $\eta$ increases. Said another way, the perturbative expansion can be trusted until the third order, but then it starts to diverge. The fact that the convergence looks better for small $\eta$ may come from the dominant behavior of the primary series, which converges very well to a nice exponential function. However, if one could go to sufficiently high orders, it might be possible to observe the same divergence as in figure~\ref{ao_vs_fo_nf2_eta_large}, when the purely non-global coefficients become of the same order of magnitude as the primary ones.

To get an idea of these coefficients, the series of the plots are explicitly written below:
\begin{align}
  \Sigma^{(2)}(\eta=0.1,t) & = 1-27.57t+368.8t^2-3195t^3+20200t^4-99300t^5+{\cal O}\left(t^6\right)\,,\label{expansion_eta_0.1}\\
  \Sigma^{(2)}(\eta=0.3,t) & = 1-13.88t+87.49t^2-334t^3+860t^4-1500t^5+{\cal O}\left(t^6\right)\,,\\
  \Sigma^{(2)}(\eta=0.6,t) & = 1-4.656t+6.53t^2-6.7t^3-10t^4+{\cal O}\left(t^5\right)\,,\\
  \Sigma^{(2)}(\eta=0.9,t) & = 1-2.320t-2.27t^2+1.9t^3-12t^4+{\cal O}\left(t^5\right)\,.
\end{align}
For $\eta$ small, the growth of these coefficients comes from the powers of the large collinear logarithm $L$. For $\eta$ near $1$, the small growth that one can start to observe at the $4^{th}$ order essentially comes from the purely non-global part. Indeed, at the $4^{th}$ order, the primary coefficient is positive, and it is not the case for $\eta>0.6$. This non-global growth will be confirmed at higher orders for the slice observable in section~\ref{the_slice_case}.

\begin{figure}[htbp]
  \centering
  \subfigure[]{
    \includegraphics[scale=0.275]{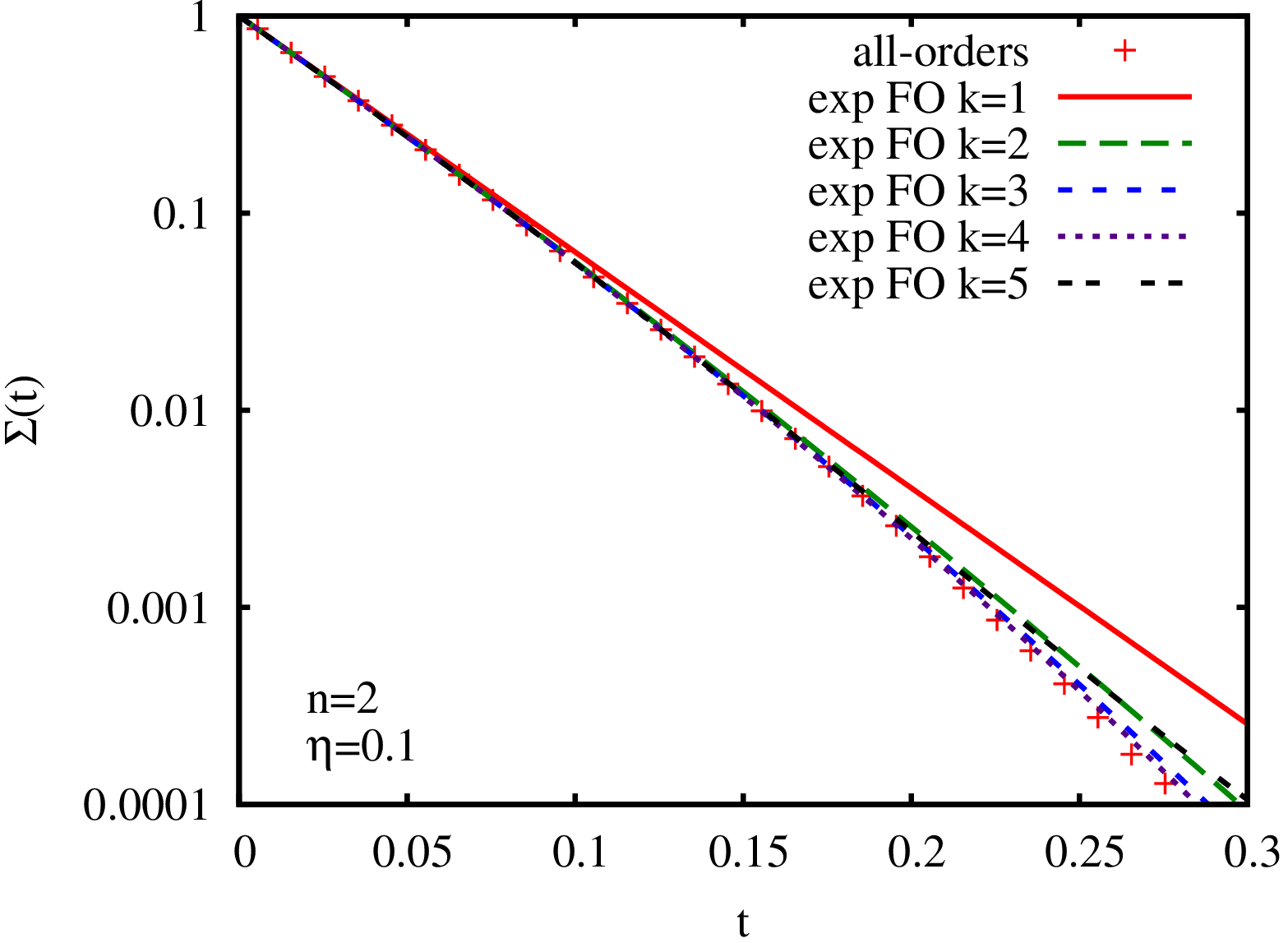}
    \label{ao_vs_fo_nf2_eta0.1_exp}
  }~
  \subfigure[]{
    \includegraphics[scale=0.275]{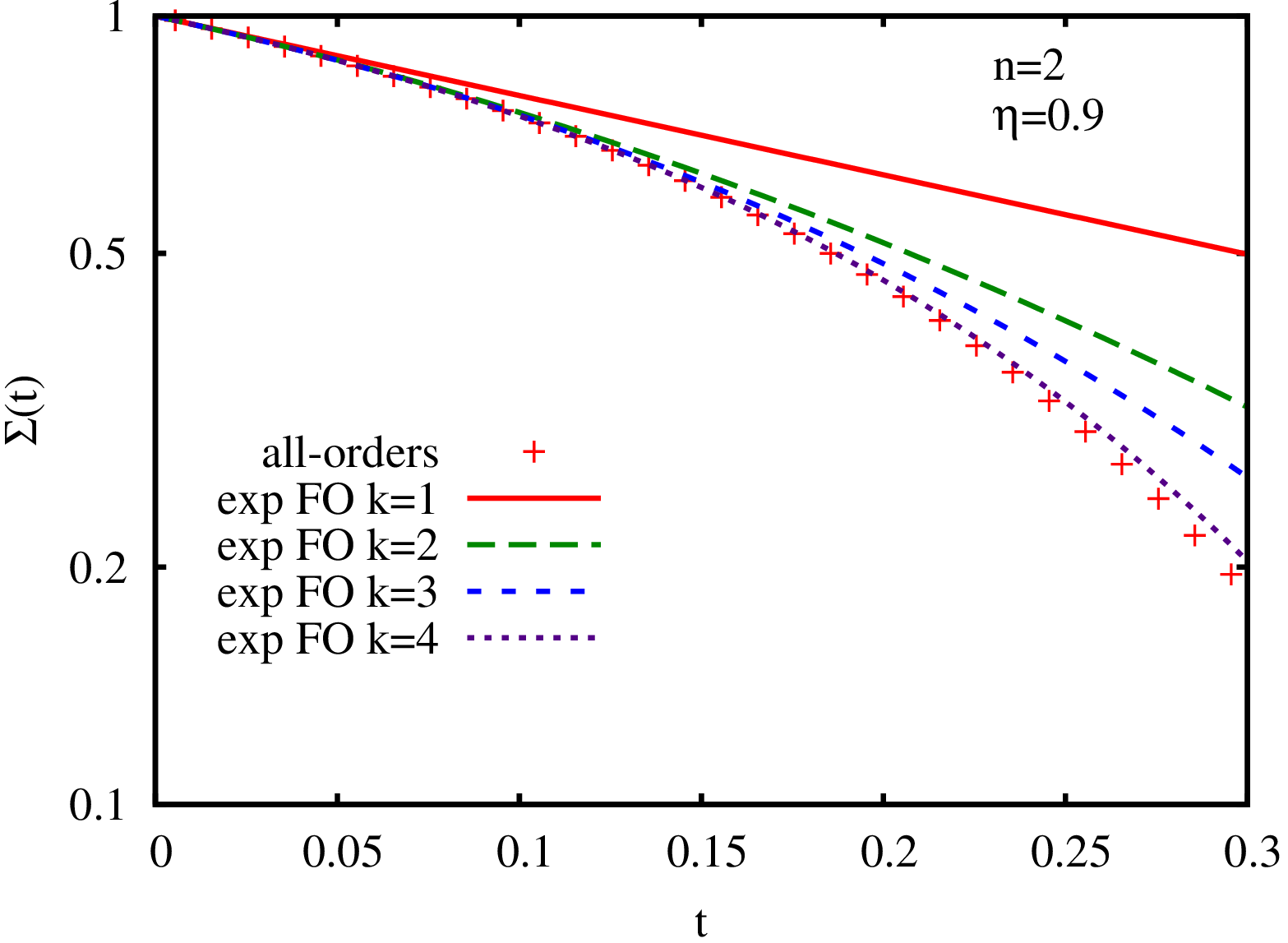}
    \label{ao_vs_fo_nf2_eta0.9_exp}
  }
  \caption{Comparison between exponentiated fixed-order (exp FO) expansion and all-orders result when $n=2$ for \subref{ao_vs_fo_nf2_eta0.1_exp}  $\eta=0.1$ and \subref{ao_vs_fo_nf2_eta0.9_exp} $\eta=0.9$.}
  \label{ao_vs_fo_nf2_exp}
\end{figure}

Let us now see what happens if the perturbative series is exponentiated. This means that instead of plotting $g(t) = 1+\sum\limits_{i=1}^ka_it^i$, which is the perturbative series, one plots $e^{f(t)}$ where
\begin{equation}
f(t)=\sum_{i=1}^kc_it^i \mbox{ and } e^{f(t)}=g(t)+{\cal O}\left(t^{k+1}\right)\,,
\end{equation}
so that 
\begin{align}
  c_1 & = a_1\,,\\
  c_2 & = a_2-\frac{a_1^2}{2}\,,\\
  & \mathellipsis\nonumber
\end{align}
 Notice that the exponentiated first order corresponds to the analytical estimate for $n=2$. One can observe on figure~\ref{ao_vs_fo_nf2_exp} a nice convergence for $\eta=0.1$ until the $4^{th}$ order. Concerning the order $5$, it seems that it diverges a little, but if the coefficients of the series are varied within their respective errors, then it can coincide with the all-orders curve. However, one can guess that it should not converge at the end because the all-orders function is not strictly a simple exponential. This will be confirmed later with the slice. When $\eta=0.9$, the exponentiated $4^{th}$ order surprisingly almost fits the all-orders result. Even more surprising, if the $5^{th}$ order coefficient (not shown here) is varied within its error band, it seems that the exponentiated fit is still improved. Is it accidental? Here again, the slice example and good sense lead us to answer yes but we cannot be completely sure.

\subsection{Case $\filt{n}=3$}

\begin{figure}[htbp]
  \centering
  \subfigure[]{
    \includegraphics[scale=0.275]{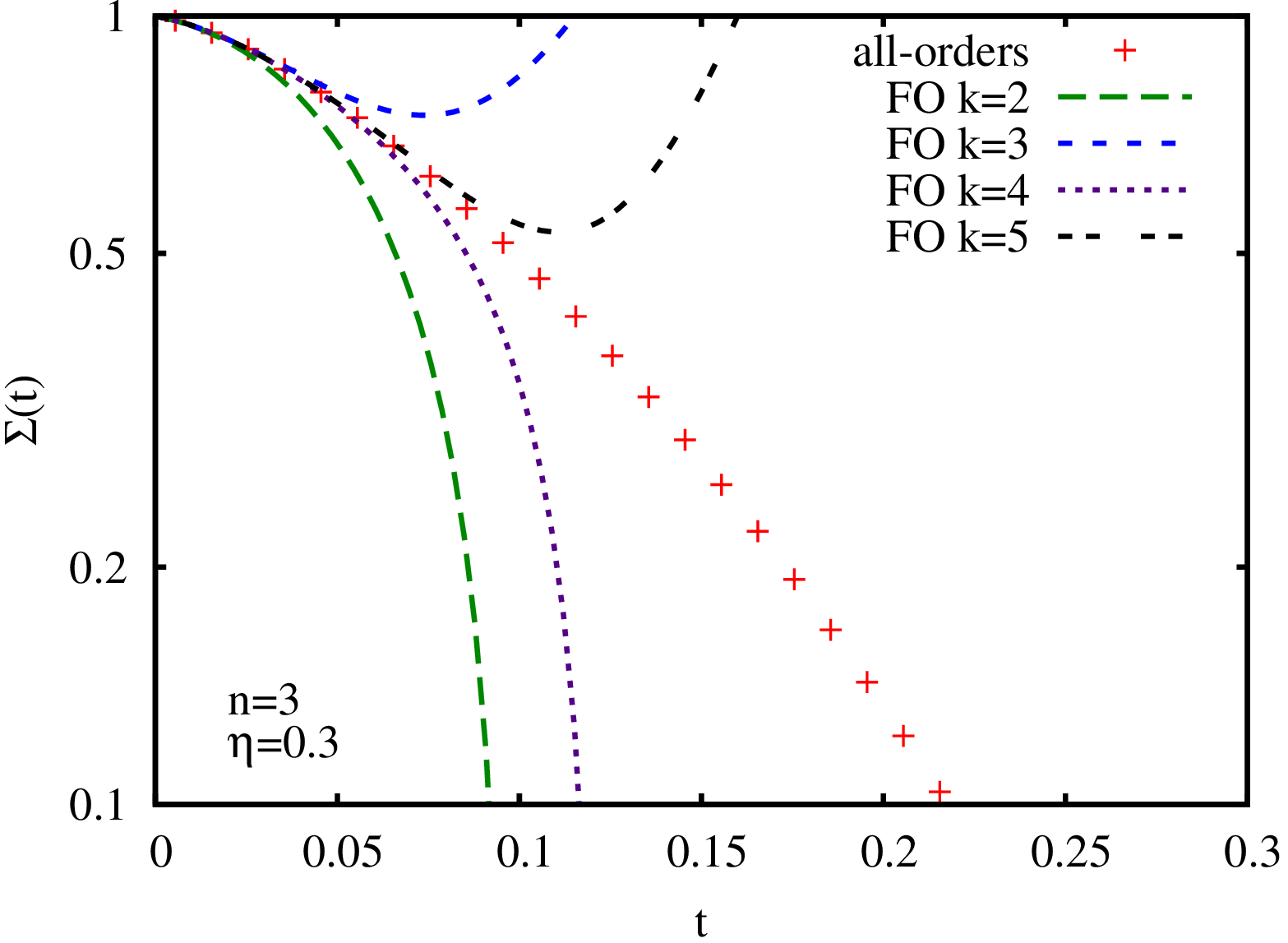}
    \label{ao_vs_fo_nf3_eta0.3}
  }~
  \subfigure[]{
    \includegraphics[scale=0.275]{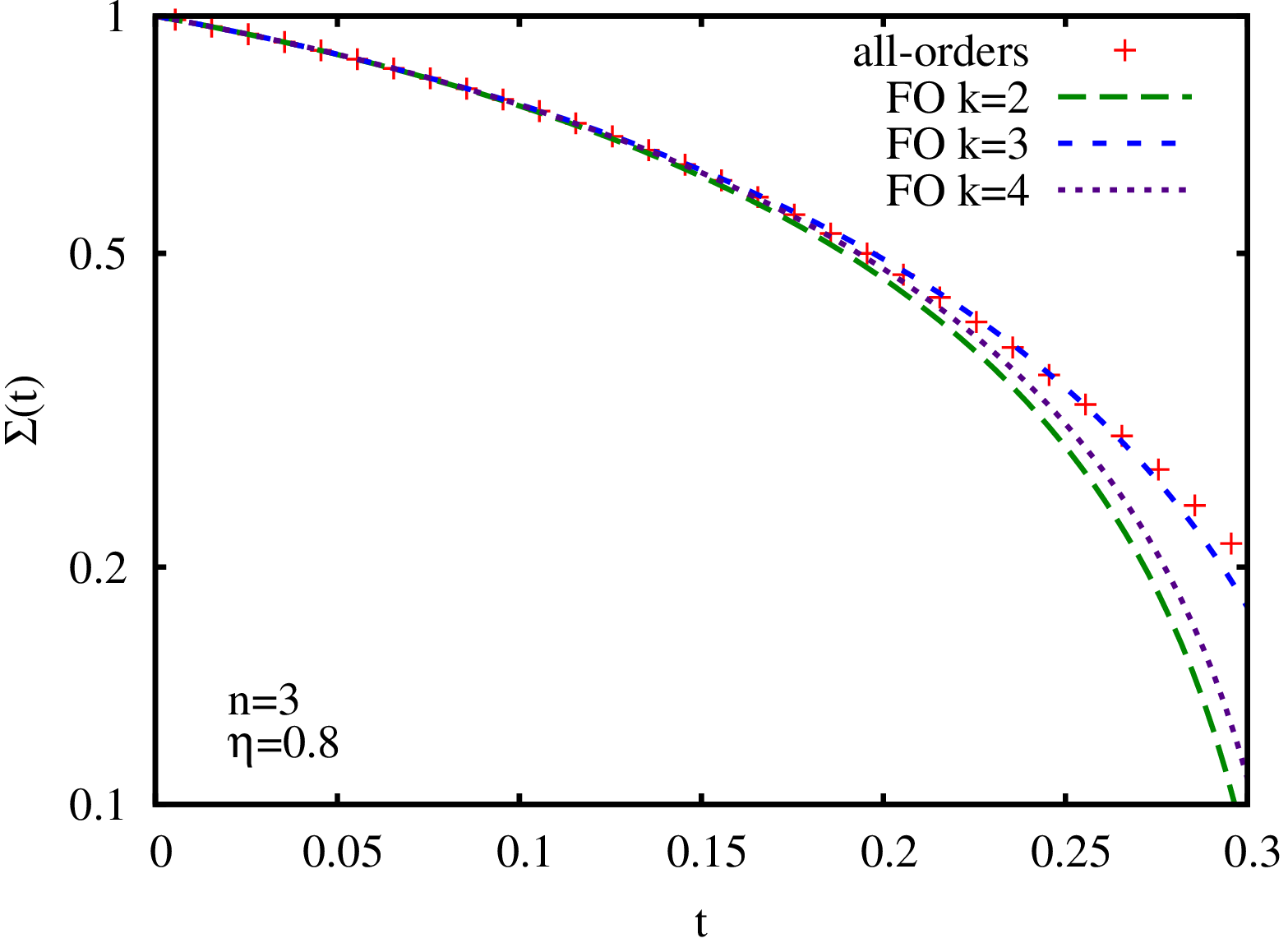}
    \label{ao_vs_fo_nf3_eta0.8}
  }
  \caption{Comparison between fixed-order expansion and all-orders result when $n=3$ for \subref{ao_vs_fo_nf3_eta0.3} $\eta=0.3$ and \subref{ao_vs_fo_nf3_eta0.8} $\eta=0.8$.}
  \label{ao_vs_fo_nf3}
\end{figure}
For small enough $\eta$, it was found analytically in section~\ref{Some_results_for_nfilt_3} that the all-orders result for the Higgs $\Delta M$ distribution can be expressed as a simple function of $t$ (eq.~(\ref{Sigma3LargeLLimit})), which is of the form $g(4N_cLt)$ with:
\begin{equation}
  g(t) = e^{-t}\left(1+\int_0^tdt'\frac{1-e^{t'}}{t'}\right)\,.
\end{equation}
The radius of convergence of the Taylor series for $g$ is infinite as the coefficients $a_k$ resulting from its expansion can be shown to be bounded for large $k$ by:
\begin{equation}
  |a_k| < \frac{1}{k!}(2^k+{\cal O}(1))\,.
\end{equation}
 As the expansion of this function converges, one would expect the same to occur for the curves obtained numerically. However, the expansion does not converge as fast as for the exponential, so that the $t$ window for which there is a convergence should be smaller than what was obtained for $n=2$. This is indeed what is observed in figure~\ref{ao_vs_fo_nf3_eta0.3} for $\eta=0.3$, if it is compared for instance with the plot~\ref{ao_vs_fo_nf2_eta0.3_app}.

In other respects, the curves on the plot~\ref{ao_vs_fo_nf3_eta0.8} for $\eta$ near $1$ behave similarly as for $n=2$, i.e. a perturbative series that converges until the $3^{rd}$ order, and that starts to diverge from the $4^{th}$ order. Notice that the same behaviour is observed for $n>3$.

\subsection{The Slice case \label{the_slice_case}}

\begin{figure}[htbp]
  \begin{center}
    \includegraphics[scale=0.45]{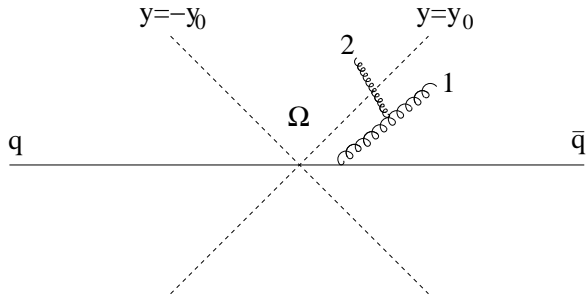}
  \end{center}
  \caption{The region $\Omega$ between the $2$ dashed lines, and the gluon's configuration leading to the appearance of non-global logarithms.}
\label{Slice_observable}
\end{figure}

\begin{figure}[p]
  \centering
  \subfigure[]{
    \includegraphics[scale=0.275]{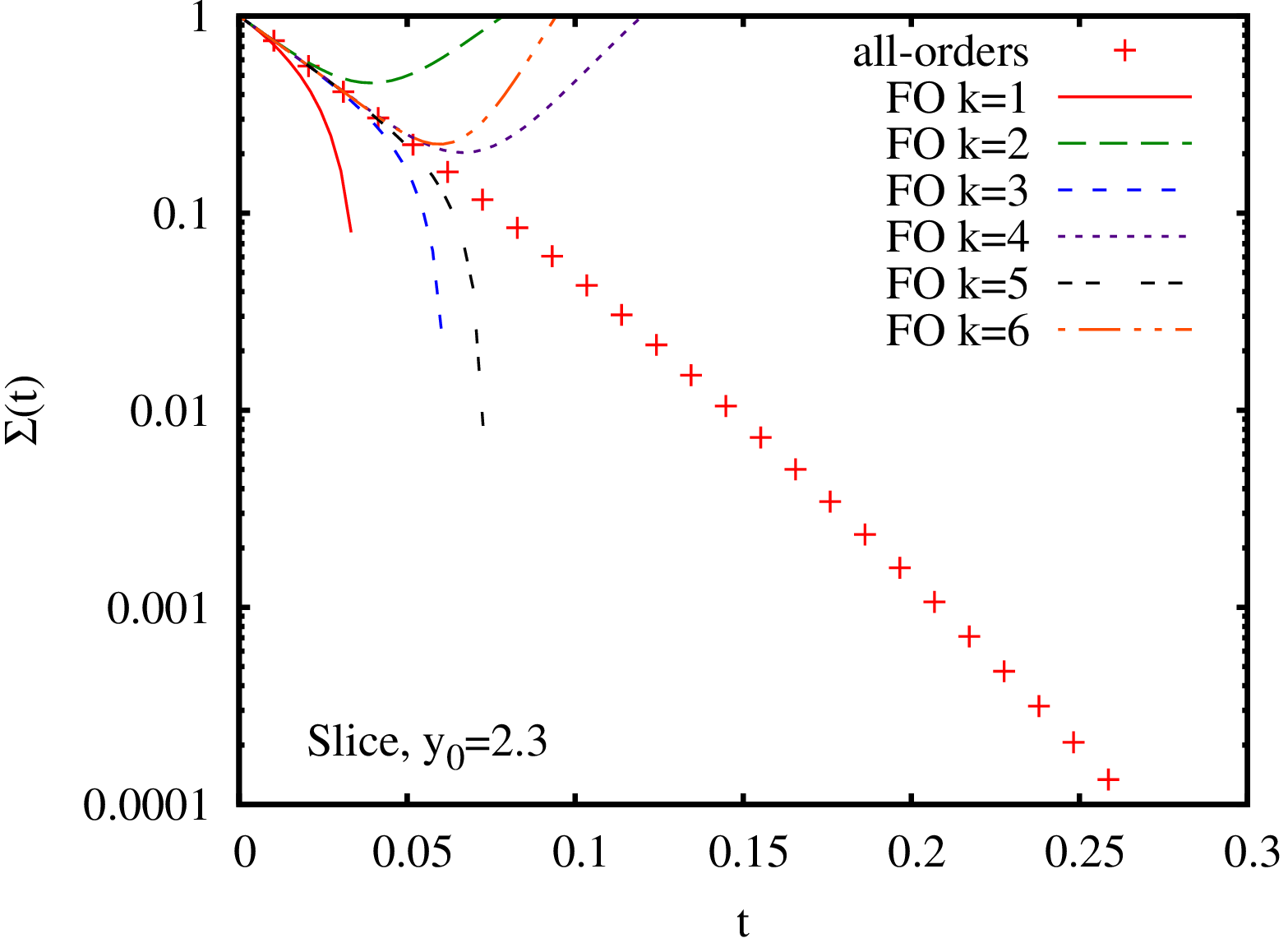}
    \label{Slice_ao_vs_fo_eta2.3}
  }~
  \subfigure[]{
    \includegraphics[scale=0.275]{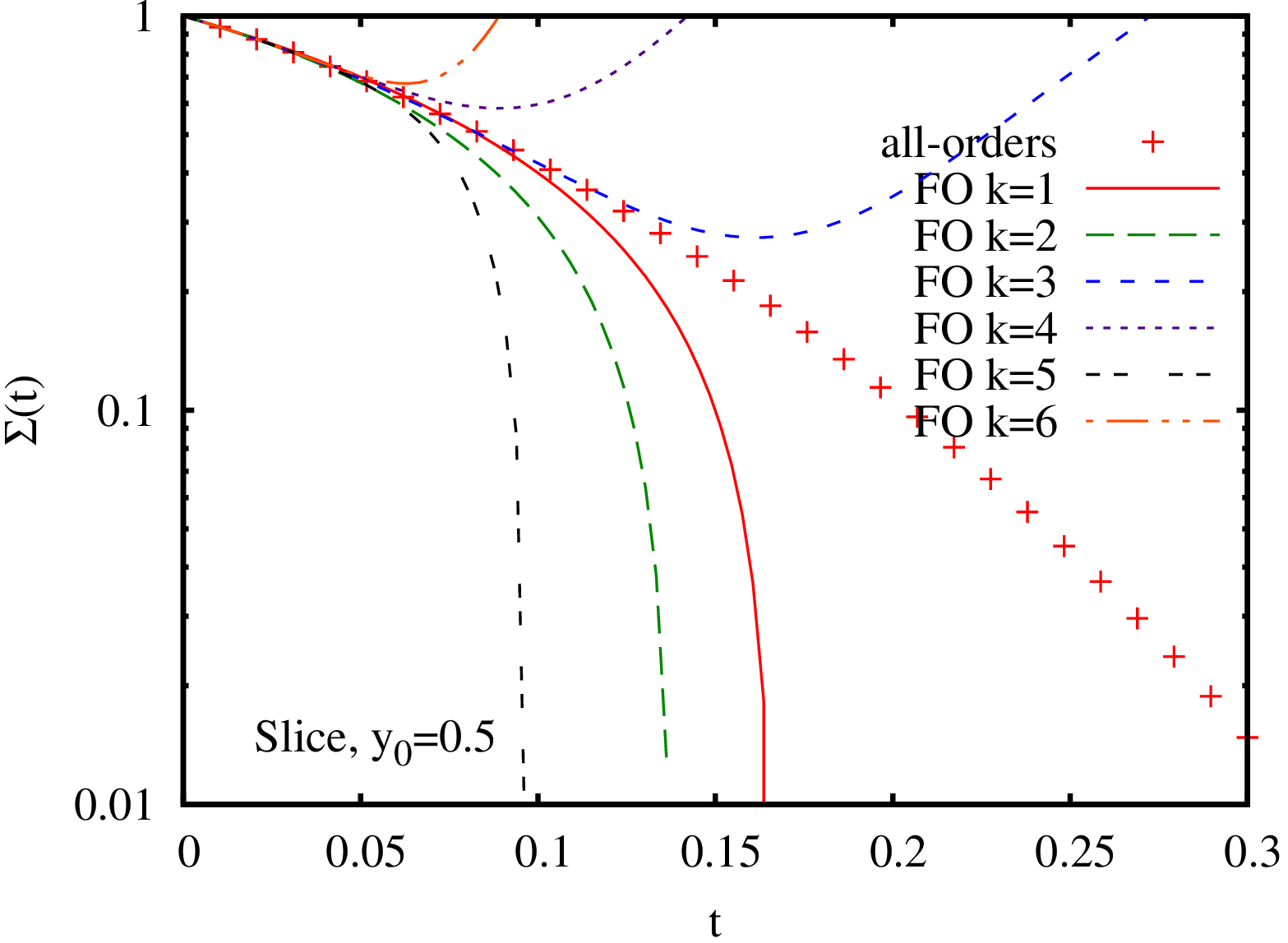}
    \label{Slice_ao_vs_fo_eta0.5}
  }
  \caption{Comparison between all-orders and fixed-order results for the Slice when \subref{Slice_ao_vs_fo_eta2.3} $y_0=2.3$ and \subref{Slice_ao_vs_fo_eta0.5} $y_0=0.5$.}
  \label{Slice_ao_vs_fo}
\end{figure}

The Slice observable, studied for instance in~\cite{Dasgupta:2002bw}, gives an interesting example of the strange behaviour of the non-global leading-log series. It is simply defined by the sum of all the particles' energy flowing into a region $\Omega$ of the phase space corresponding to a rapidity $y \in [-y_0,y_0]$, with $y_0$ a parameter of the observable (figure~\ref{Slice_observable}). Here, we work in the $q\bar{q}$ center of mass frame and the quarks are assumed to move along the $z$ axis.
This observable is non-global as shown in fig.~\ref{Slice_observable}, and it is interesting in $2$ ways:
\begin{itemize}
  \item A simple change of frame\footnote{or more simply comparison between primary results, eq.~(2.5) from \cite{Dasgupta:2002bw} and eq.~(\ref{exponentiated_primary_result}) from this paper.} shows that it is more or less equivalent in the leading-log approximation to the $\Delta M$ observable in the filtering analysis using anti-$k_t$ with $n=2$ and $\eta = e^{-y_0}$ when $\eta\ll 1$, while being faster to compute numerically due to the absence of clustering.
  \item When $e^{-y_0}\sim 1$, which should very approximately correspond to $\eta = {\cal O}(1)$, the strange behaviour of the non-global series, observed in section~\ref{NG_structure_numerical_results}, is clearly confirmed with the addition of the $5^{th}$ and $6^{th}$ orders.
\end{itemize}
The series are represented in figure~\ref{Slice_ao_vs_fo} for $2$ different values of $y_0$. The plot for $y_0=2.3$ is here for a comparison with figure~\ref{ao_vs_fo_nf2_eta0.1}, i.e. when $n=2$ and $\eta=0.1$. One can notice that until order $5$ the behaviours of the $2$ series are very similar. The slight difference comes from the fact that the $C/A$ algorithm was used there instead of anti-$k_t$. This can also be seen on the expansions eqs.~(\ref{expansion_eta_0.1},\ref{expansion_y0_2.3}). However, a remarkable effect is the $6^{th}$ order curve which does not improve the fit with the all-orders result anymore (it even makes it worse). When $y_0=0.5$ (fig.~\ref{Slice_ao_vs_fo_eta0.5}), this effect is enhanced: indeed, one notices that the $3^{rd}$ order gives the best result, with the $2^{nd}$ order even worse than the $1^{st}$ one. And here again, adding more orders shifts point of disagreement to smaller $t$.
\begin{figure}[p]
  \centering
  \subfigure[]{
    \includegraphics[scale=0.275]{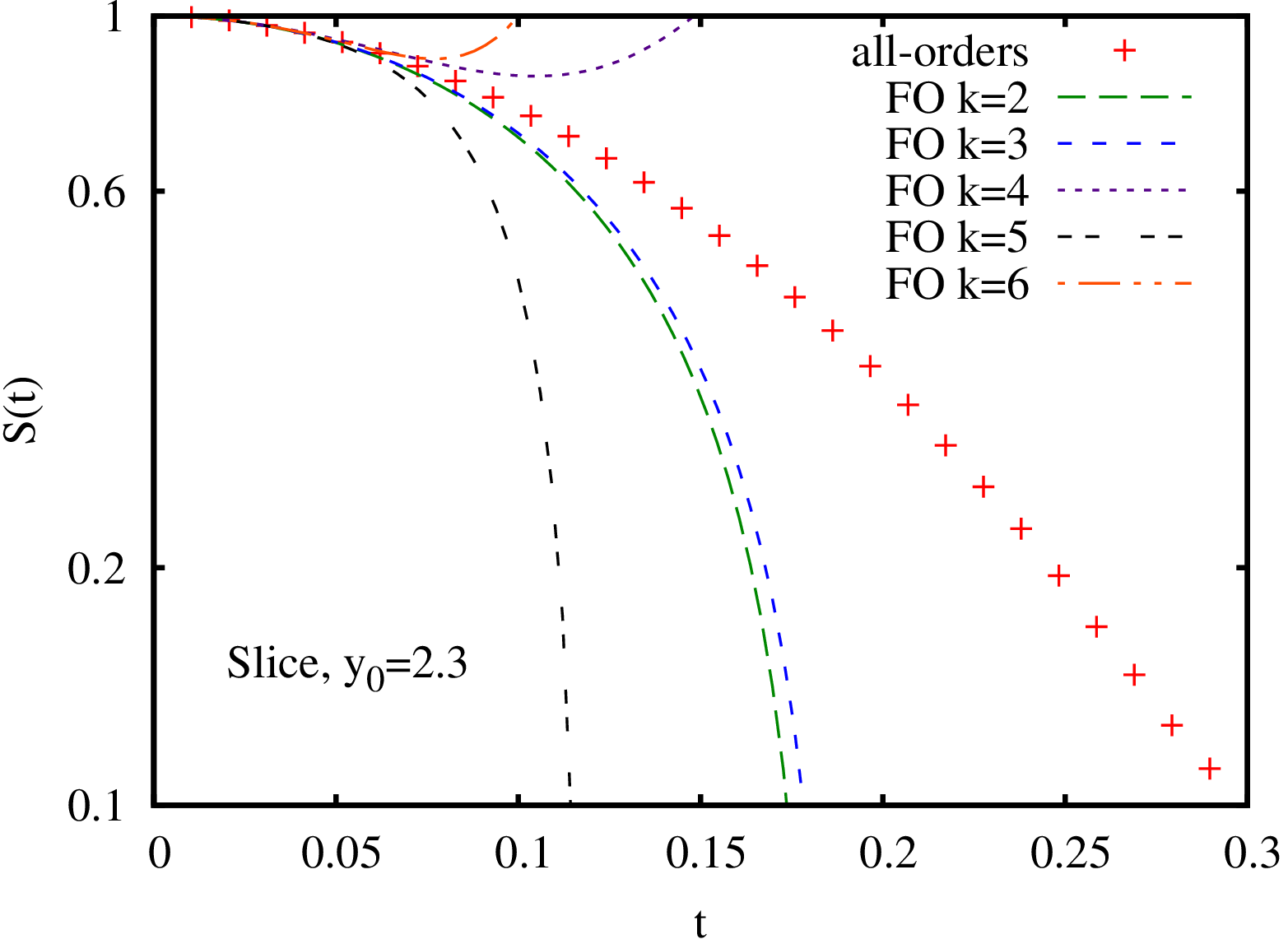}
    \label{St_eta2.3}
  }~
  \subfigure[]{
    \includegraphics[scale=0.275]{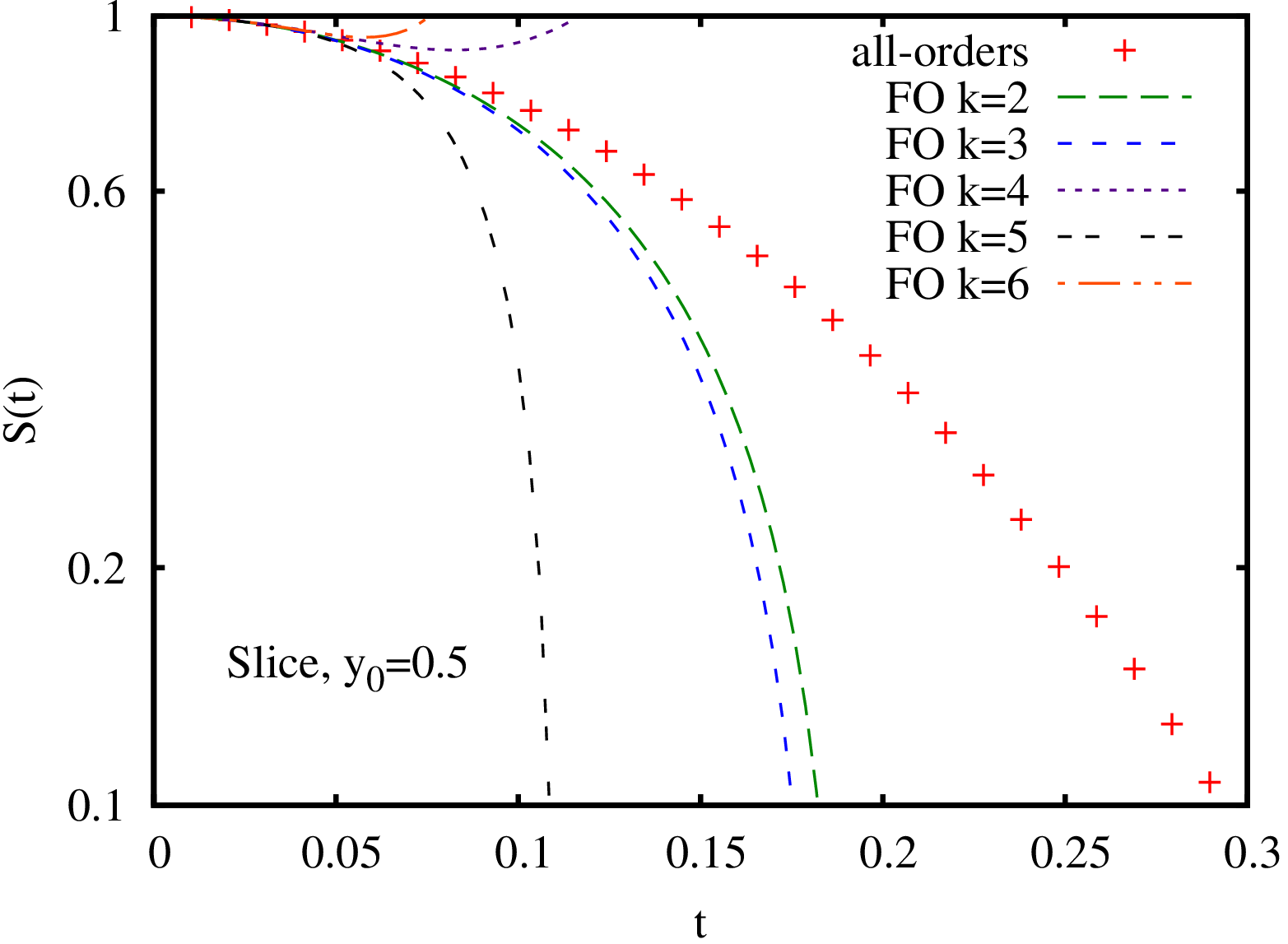}
    \label{St_eta0.5}
  }
  \caption{Comparison between all-orders and fixed-order results for the purely non-global part $S(t)$ of the Slice when \subref{St_eta2.3} $y_0=2.3$ and \subref{St_eta0.5} $y_0=0.5$.}
  \label{Slice_ao_vs_fo_for_St}
\end{figure}

To get an idea of the coefficients in this case, the series of the plots are given below:
\begin{align}
  \Sigma(y_0=2.3,t) & = 1-27.6t+351.1t^2-2673t^3+13900t^4-99500t^5+2\, 10^6t^6 + {\cal O}(t^7)\,,\label{expansion_y0_2.3}\\
  \Sigma(y_0=0.5,t) & = 1-6t-9.108t^2+114t^3+1740t^4-68400t^5+1.6\, 10^6t^6 + {\cal O}(t^7)\,.
\end{align}
The growth of the coefficients for $y_0=2.3$ until the $5^{th}$ order essentially comes from the powers of $4N_cy_0$ when expanding the primary result, which is the expression of the collinear divergence near $q$ and $\bar{q}$, whereas, for $y_0=0.5$, it essentially takes its origin from the purely non-global part. There is no large enhancement due to collinear divergence which can explain it. As an example, we also show the function $S(y_0,t)$, which is defined as in \cite{Dasgupta:2001sh,Dasgupta:2002bw} to contain the purely non-global part of the result:
\begin{equation}
  S(y_0,t) \equiv \frac{\Sigma(y_0,t)}{\Sigma^{(P)}(y_0,t)}\,,
\end{equation}
where, for the slice, the primary contribution $\Sigma^{(P)}$ can be written as:
\begin{equation}
  \Sigma^{(P)}(y_0,t) = e^{-4N_cy_0t}\,,
\end{equation}
in the large-$N_c$ limit. The plots for $S(y_0,t)$ are shown in fig.~\ref{Slice_ao_vs_fo_for_St}. One observes the saturation property noticed in \cite{Dasgupta:2002bw} which leads to very similar plots for $y_0=0.5$ and $y_0=2.3$. There is clearly no convergence of the perturbative series.

Therefore, the non-global leading-log large-$N_c$ series seems to behave badly at high orders. Does this mean that it is an asymptotic series like the Standard Model is known to be \cite{Beneke:1998ui}?\footnote{though the origin would be different if it were the case.} This study cannot answer such a question but, at least, one should be aware of the strange behaviour of the non-global series.

\begin{figure}[hbtp]
  \centering
  \subfigure[]{
    \includegraphics[scale=0.275]{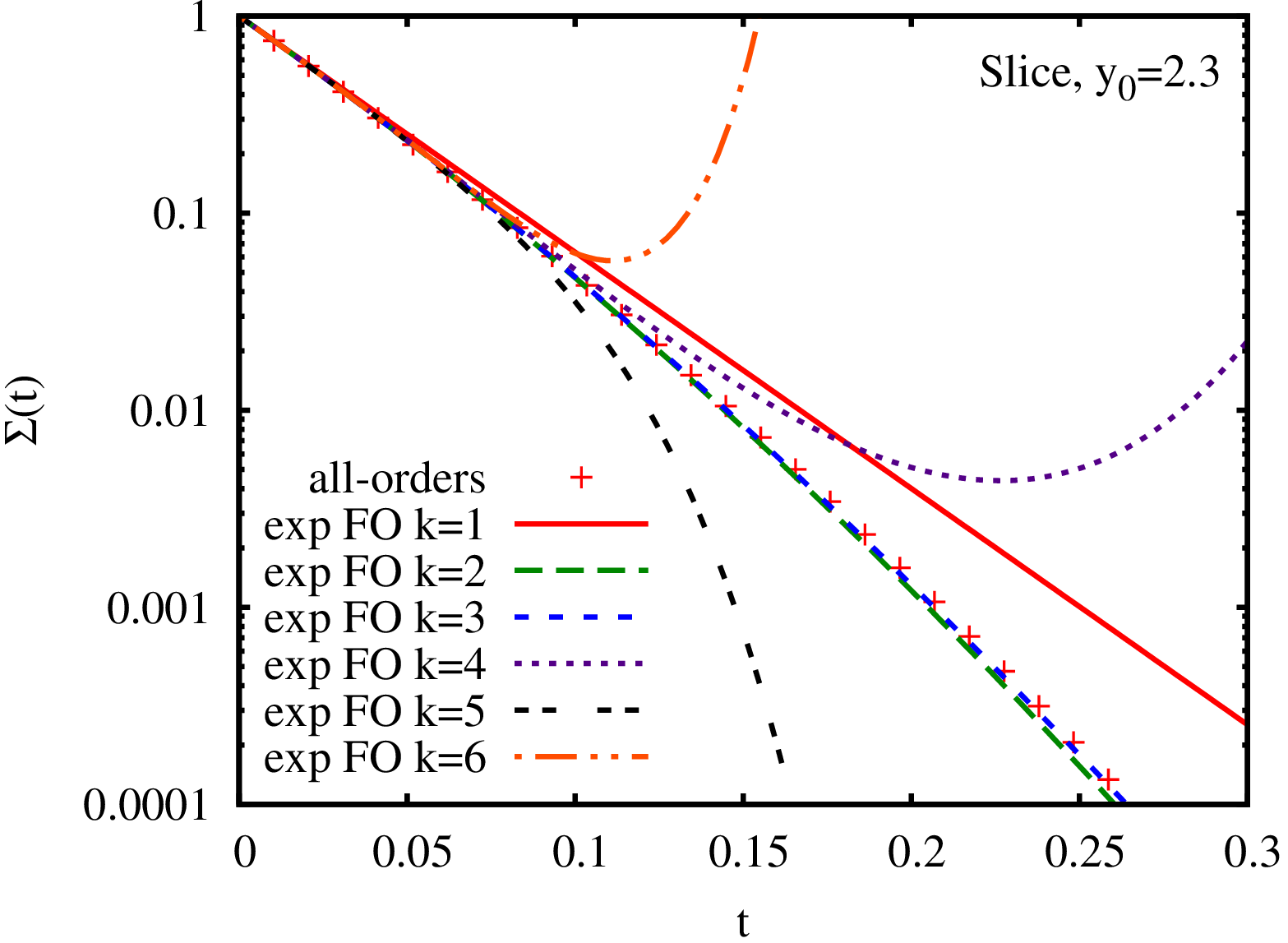}
    \label{Slice_ao_vs_fo_eta2.3_exp}
  }~
  \subfigure[]{
    \includegraphics[scale=0.275]{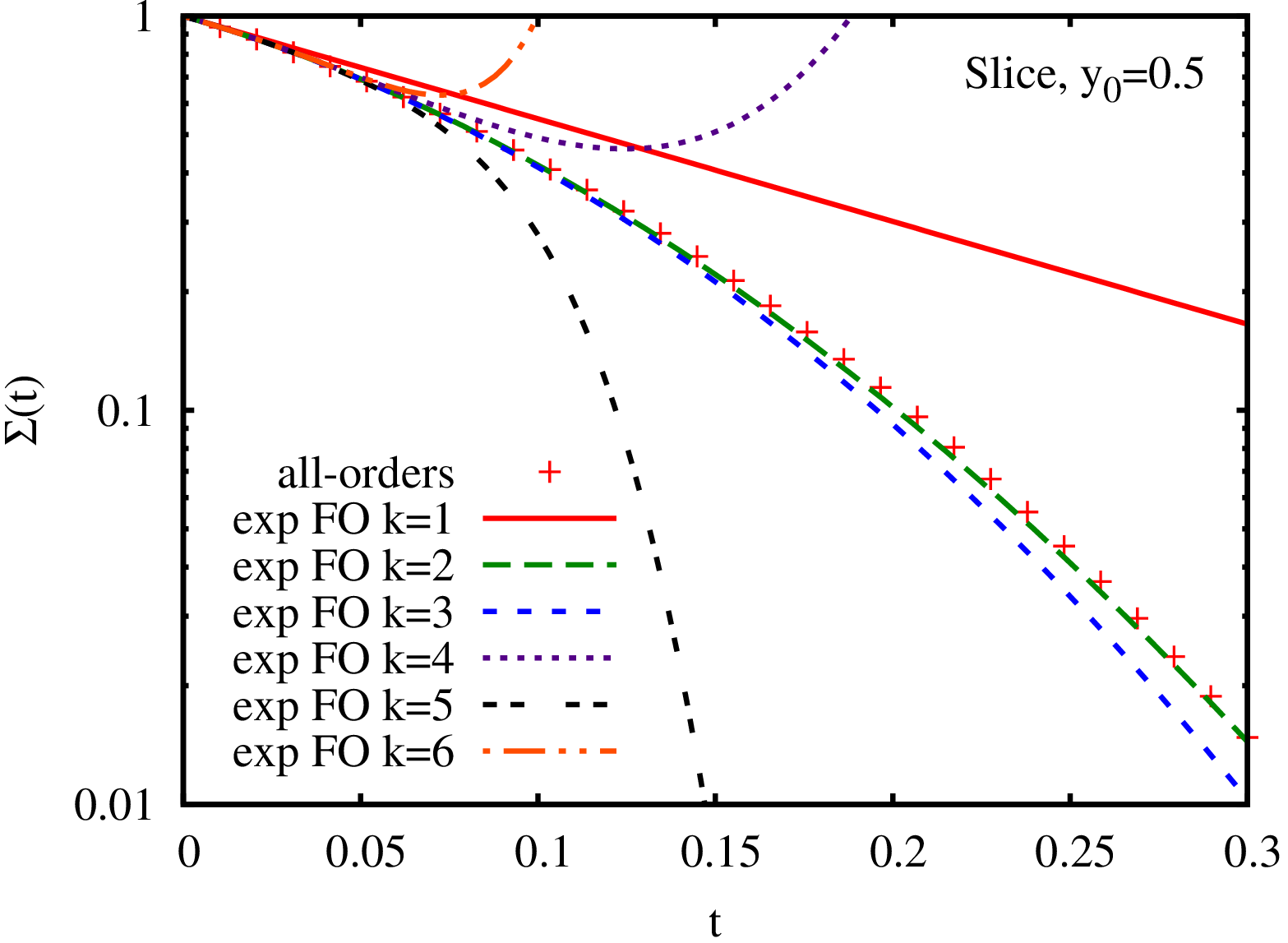}
    \label{Slice_ao_vs_fo_eta0.5_exp}
  }
  \caption{Comparison between all-orders and exponentiated fixed-order results for the Slice when \subref{Slice_ao_vs_fo_eta2.3} $y_0=2.3$ and \subref{Slice_ao_vs_fo_eta0.5} $y_0=0.5$.}
  \label{Slice_ao_vs_fo_exp}
\end{figure}
To finish, let us mention the exponentiated results represented in figure~\ref{Slice_ao_vs_fo_exp}. They show that the convergence observed in the case $n=2$ is clearly only illusory. One point to notice is that the case $y_0=2.3$ is slightly different from its $\eta=0.1$ counterpart as the divergence starts to be visible at order $4$ instead of $5$. This may be because using the $C/A$ algorithm in the filtered Higgs mass observable reduces the impact of the non-global logarithms \cite{Delenda:2006nf,Appleby:2002ke} and the primaries still dominate at order $4$, whereas it is not the case anymore for the slice. Notice also the very nice fit given by the exponentiated $2^{nd}$ order curve when $y_0=0.5$.


\bibliographystyle{JHEP}

\bibliography{paper}

\providecommand{\href}[2]{#2}\begingroup\raggedright\begin{thebibliography}{10}

\bibitem{Seymour:1993mx}
M.~H. Seymour, {\it {Searches for new particles using cone and cluster jet
  algorithms: A Comparative study}},  {\em Z. Phys.} {\bf C62} (1994) 127--138.

\bibitem{Butterworth:2002tt}
J.~M. Butterworth, B.~E. Cox, and J.~R. Forshaw, {\it {$W W$ scattering at the
  CERN LHC}},  {\em Phys. Rev.} {\bf D65} (2002) 096014,
  [\href{http://xxx.lanl.gov/abs/hep-ph/0201098}{{\tt hep-ph/0201098}}].

\bibitem{Skiba:2007fw}
W.~Skiba and D.~Tucker-Smith, {\it {Using jet mass to discover vector quarks at
  the LHC}},  {\em Phys. Rev.} {\bf D75} (2007) 115010,
  [\href{http://xxx.lanl.gov/abs/hep-ph/0701247}{{\tt hep-ph/0701247}}].

\bibitem{Holdom:2007nw}
B.~Holdom, {\it {t' at the LHC: The physics of discovery}},  {\em JHEP} {\bf
  03} (2007) 063, [\href{http://xxx.lanl.gov/abs/hep-ph/0702037}{{\tt
  hep-ph/0702037}}].

\bibitem{Almeida:2008tp}
L.~G. Almeida, S.~J. Lee, G.~Perez, I.~Sung, and J.~Virzi, {\it {Top Jets at
  the LHC}},  {\em Phys. Rev.} {\bf D79} (2009) 074012,
  [\href{http://xxx.lanl.gov/abs/0810.0934}{{\tt arXiv:0810.0934}}].

\bibitem{Kaplan:2008ie}
D.~E. Kaplan, K.~Rehermann, M.~D. Schwartz, and B.~Tweedie, {\it {Top Tagging:
  A Method for Identifying Boosted Hadronically Decaying Top Quarks}},  {\em
  Phys. Rev. Lett.} {\bf 101} (2008) 142001,
  [\href{http://xxx.lanl.gov/abs/0806.0848}{{\tt arXiv:0806.0848}}].

\bibitem{Thaler:2008ju}
J.~Thaler and L.-T. Wang, {\it {Strategies to Identify Boosted Tops}},  {\em
  JHEP} {\bf 07} (2008) 092, [\href{http://xxx.lanl.gov/abs/0806.0023}{{\tt
  arXiv:0806.0023}}].

\bibitem{Plehn:2009rk}
T.~Plehn, G.~P. Salam, and M.~Spannowsky, {\it {Fat Jets for a Light Higgs}},
  \href{http://xxx.lanl.gov/abs/0910.5472}{{\tt arXiv:0910.5472}}.

\bibitem{Butterworth:2007ke}
J.~M. Butterworth, J.~R. Ellis, and A.~R. Raklev, {\it {Reconstructing
  sparticle mass spectra using hadronic decays}},  {\em JHEP} {\bf 05} (2007)
  033, [\href{http://xxx.lanl.gov/abs/hep-ph/0702150}{{\tt hep-ph/0702150}}].

\bibitem{Butterworth:2009qa}
J.~M. Butterworth, J.~R. Ellis, A.~R. Raklev, and G.~P. Salam, {\it
  {Discovering baryon-number violating neutralino decays at the LHC}},  {\em
  Phys. Rev. Lett.} {\bf 103} (2009) 241803,
  [\href{http://xxx.lanl.gov/abs/0906.0728}{{\tt arXiv:0906.0728}}].

\bibitem{Baur:2008uv}
U.~Baur and L.~H. Orr, {\it {Searching for $t \bar{t}$ Resonances at the Large
  Hadron Collider}},  {\em Phys. Rev.} {\bf D77} (2008) 114001,
  [\href{http://xxx.lanl.gov/abs/0803.1160}{{\tt arXiv:0803.1160}}].

\bibitem{FileviezPerez:2008ib}
P.~Fileviez~Perez, R.~Gavin, T.~McElmurry, and F.~Petriello, {\it {Grand
  Unification and Light Color-Octet Scalars at the LHC}},  {\em Phys. Rev.}
  {\bf D78} (2008) 115017, [\href{http://xxx.lanl.gov/abs/0809.2106}{{\tt
  arXiv:0809.2106}}].

\bibitem{Bai:2008sk}
Y.~Bai and Z.~Han, {\it {Top-antitop and Top-top Resonances in the Dilepton
  Channel at the CERN LHC}},  {\em JHEP} {\bf 04} (2009) 056,
  [\href{http://xxx.lanl.gov/abs/0809.4487}{{\tt arXiv:0809.4487}}].

\bibitem{Ellis:2009me}
S.~D. Ellis, C.~K. Vermilion, and J.~R. Walsh, {\it {Recombination Algorithms
  and Jet Substructure: Pruning as a Tool for Heavy Particle Searches}},
  \href{http://xxx.lanl.gov/abs/0912.0033}{{\tt arXiv:0912.0033}}.

\bibitem{MyFirstPaper}
J.~M. Butterworth, A.~R. Davison, M.~Rubin, and G.~P. Salam, {\it {Jet
  substructure as a new Higgs search channel at the LHC}},  {\em Phys. Rev.
  Lett.} {\bf 100} (2008) 242001,
  [\href{http://xxx.lanl.gov/abs/0802.2470}{{\tt arXiv:0802.2470}}].

\bibitem{ATL-PHYS-PUB-2009-088}
{\it Atlas sensitivity to the standard model higgs in the hw and hz channels at
  high transverse momenta},  Tech. Rep. ATL-PHYS-PUB-2009-088.
  ATL-COM-PHYS-2009-345, CERN, Geneva, Aug, 2009.

\bibitem{Kribs:2009yh}
G.~D. Kribs, A.~Martin, T.~S. Roy, and M.~Spannowsky, {\it {Discovering the
  Higgs Boson in New Physics Events using Jet Substructure}},
  \href{http://xxx.lanl.gov/abs/0912.4731}{{\tt arXiv:0912.4731}}.

\bibitem{Cacciari:2008gd}
M.~Cacciari, J.~Rojo, G.~P. Salam, and G.~Soyez, {\it {Quantifying the
  performance of jet definitions for kinematic reconstruction at the LHC}},
  {\em JHEP} {\bf 12} (2008) 032,
  [\href{http://xxx.lanl.gov/abs/0810.1304}{{\tt arXiv:0810.1304}}].

\bibitem{Krohn:2009th}
D.~Krohn, J.~Thaler, and L.-T. Wang, {\it {Jet Trimming}},
  \href{http://xxx.lanl.gov/abs/0912.1342}{{\tt arXiv:0912.1342}}.

\bibitem{Fadin:1983aw}
V.~S. Fadin, {\it {Double logarithmic asymptotics of the cross-sections of $e^+e^-$ annihilation into quarks and gluons. (In Russian)}}, {\em Yad. Fiz.} {\bf
  37} (1983) 408--423.

\bibitem{Ermolaev:1981cm}
B.~I. Ermolaev and V.~S. Fadin, {\it {Log - Log Asymptotic Form of Exclusive
  Cross-Sections in Quantum Chromodynamics}},  {\em JETP Lett.} {\bf 33} (1981)
  269--272.

\bibitem{Mueller:1981ex}
A.~H. Mueller, {\it {On the Multiplicity of Hadrons in QCD Jets}},  {\em Phys.
  Lett.} {\bf B104} (1981) 161--164.

\bibitem{Dokshitzer:1982xr}
Y.~L. Dokshitzer, V.~S. Fadin, and V.~A. Khoze, {\it {Double Logs of
  Perturbative QCD for Parton Jets and Soft Hadron Spectra}},  {\em Zeit.
  Phys.} {\bf C15} (1982) 325.

\bibitem{Bassetto:1984ik}
A.~Bassetto, M.~Ciafaloni, and G.~Marchesini, {\it {Jet Structure and Infrared
  Sensitive Quantities in Perturbative QCD}},  {\em Phys. Rept.} {\bf 100}
  (1983) 201--272.

\bibitem{Catani:1991kz}
S.~Catani, G.~Turnock, B.~R. Webber, and L.~Trentadue, {\it {Thrust
  distribution in e+ e- annihilation}},  {\em Phys. Lett.} {\bf B263} (1991)
  491--497.

\bibitem{Catani:1991pm}
S.~Catani, Y.~L. Dokshitzer, F.~Fiorani, and B.~R. Webber, {\it {Average number
  of jets in e+ e- annihilation}},  {\em Nucl. Phys.} {\bf B377} (1992)
  445--460.

\bibitem{Catani:1992ua}
S.~Catani, L.~Trentadue, G.~Turnock, and B.~R. Webber, {\it {Resummation of
  large logarithms in e+ e- event shape distributions}},  {\em Nucl. Phys.}
  {\bf B407} (1993) 3--42.

\bibitem{Catani:1992jc}
S.~Catani, G.~Turnock, and B.~R. Webber, {\it {Jet broadening measures in
  $e^{+} e^{-}$ annihilation}},  {\em Phys. Lett.} {\bf B295} (1992) 269--276.

\bibitem{Catani:1998sf}
S.~Catani and B.~R. Webber, {\it {Resummed C-parameter distribution in e+ e-
  annihilation}},  {\em Phys. Lett.} {\bf B427} (1998) 377--384,
  [\href{http://xxx.lanl.gov/abs/hep-ph/9801350}{{\tt hep-ph/9801350}}].

\bibitem{Dokshitzer:1998kz}
Y.~L. Dokshitzer, A.~Lucenti, G.~Marchesini, and G.~P. Salam, {\it {On the
  {QCD} analysis of jet broadening}},  {\em JHEP} {\bf 01} (1998) 011,
  [\href{http://xxx.lanl.gov/abs/hep-ph/9801324}{{\tt hep-ph/9801324}}].

\bibitem{Antonelli:1999kx}
V.~Antonelli, M.~Dasgupta, and G.~P. Salam, {\it {Resummation of thrust
  distributions in DIS}},  {\em JHEP} {\bf 02} (2000) 001,
  [\href{http://xxx.lanl.gov/abs/hep-ph/9912488}{{\tt hep-ph/9912488}}].

\bibitem{Burby:1999yb}
S.~J. Burby, {\it {The four-jet rate in e+ e- annihilation}},  {\em Phys.
  Lett.} {\bf B453} (1999) 54--58,
  [\href{http://xxx.lanl.gov/abs/hep-ph/9902305}{{\tt hep-ph/9902305}}].

\bibitem{Burby:2001uz}
S.~J. Burby and E.~W.~N. Glover, {\it {Resumming the Light Hemisphere Mass and
  Narrow Jet Broadening distributions in $e^+e^-$ annihilation}},  {\em JHEP}
  {\bf 04} (2001) 029, [\href{http://xxx.lanl.gov/abs/hep-ph/0101226}{{\tt
  hep-ph/0101226}}].

\bibitem{Banfi:2000si}
A.~Banfi, G.~Marchesini, Y.~L. Dokshitzer, and G.~Zanderighi, {\it {QCD
  analysis of near-to-planar 3-jet events}},  {\em JHEP} {\bf 07} (2000) 002,
  [\href{http://xxx.lanl.gov/abs/hep-ph/0004027}{{\tt hep-ph/0004027}}].

\bibitem{Dasgupta:2001sh}
M.~Dasgupta and G.~P. Salam, {\it {Resummation of non-global QCD observables}},
   {\em Phys. Lett.} {\bf B512} (2001) 323--330,
  [\href{http://xxx.lanl.gov/abs/hep-ph/0104277}{{\tt hep-ph/0104277}}].

\bibitem{Banfi:2003jj}
A.~Banfi and M.~Dasgupta, {\it {Dijet rates with symmetric E(t) cuts}},  {\em
  JHEP} {\bf 01} (2004) 027,
  [\href{http://xxx.lanl.gov/abs/hep-ph/0312108}{{\tt hep-ph/0312108}}].

\bibitem{Banfi:2004nk}
A.~Banfi, G.~P. Salam, and G.~Zanderighi, {\it {Resummed event shapes at hadron
  - hadron colliders}},  {\em JHEP} {\bf 08} (2004) 062,
  [\href{http://xxx.lanl.gov/abs/hep-ph/0407287}{{\tt hep-ph/0407287}}].

\bibitem{Banfi:2004yd}
A.~Banfi, G.~P. Salam, and G.~Zanderighi, {\it {Principles of general
  final-state resummation and automated implementation}},  {\em JHEP} {\bf 03}
  (2005) 073, [\href{http://xxx.lanl.gov/abs/hep-ph/0407286}{{\tt
  hep-ph/0407286}}].

\bibitem{Becher:2008cf}
T.~Becher and M.~D. Schwartz, {\it {A Precise determination of $\alpha_s$ from
  LEP thrust data using effective field theory}},  {\em JHEP} {\bf 07} (2008)
  034, [\href{http://xxx.lanl.gov/abs/0803.0342}{{\tt arXiv:0803.0342}}].

\bibitem{Dissertori:2009ik}
G.~Dissertori {\em et.~al.}, {\it {Determination of the strong coupling
  constant using matched NNLO+NLLA predictions for hadronic event shapes in
  e+e- annihilations}},  {\em JHEP} {\bf 08} (2009) 036,
  [\href{http://xxx.lanl.gov/abs/0906.3436}{{\tt arXiv:0906.3436}}].

\bibitem{Dasgupta:2002bw}
M.~Dasgupta and G.~P. Salam, {\it {Accounting for coherence in interjet E(t)
  flow: A case study}},  {\em JHEP} {\bf 03} (2002) 017,
  [\href{http://xxx.lanl.gov/abs/hep-ph/0203009}{{\tt hep-ph/0203009}}].

\bibitem{Delenda:2006nf}
Y.~Delenda, R.~Appleby, M.~Dasgupta, and A.~Banfi, {\it {On QCD resummation
  with k(t) clustering}},  {\em JHEP} {\bf 12} (2006) 044,
  [\href{http://xxx.lanl.gov/abs/hep-ph/0610242}{{\tt hep-ph/0610242}}].

\bibitem{Appleby:2002ke}
R.~B. Appleby and M.~H. Seymour, {\it {Non-global logarithms in inter-jet
  energy flow with kt clustering requirement}},  {\em JHEP} {\bf 12} (2002)
  063, [\href{http://xxx.lanl.gov/abs/hep-ph/0211426}{{\tt hep-ph/0211426}}].

\bibitem{Banfi:2002hw}
A.~Banfi, G.~Marchesini, and G.~Smye, {\it {Away-from-jet energy flow}},  {\em
  JHEP} {\bf 08} (2002) 006,
  [\href{http://xxx.lanl.gov/abs/hep-ph/0206076}{{\tt hep-ph/0206076}}].

\bibitem{Hatta:2009nd}
Y.~Hatta and T.~Ueda, {\it {Jet energy flow at the LHC}},  {\em Phys. Rev.}
  {\bf D80} (2009) 074018, [\href{http://xxx.lanl.gov/abs/0909.0056}{{\tt
  arXiv:0909.0056}}].

\bibitem{Weigert:2003mm}
H.~Weigert, {\it {Non-global jet evolution at finite N(c)}},  {\em Nucl. Phys.}
  {\bf B685} (2004) 321--350,
  [\href{http://xxx.lanl.gov/abs/hep-ph/0312050}{{\tt hep-ph/0312050}}].

\bibitem{Forshaw:2006fk}
J.~R. Forshaw, A.~Kyrieleis, and M.~H. Seymour, {\it {Super-leading logarithms
  in non-global observables in QCD}},  {\em JHEP} {\bf 08} (2006) 059,
  [\href{http://xxx.lanl.gov/abs/hep-ph/0604094}{{\tt hep-ph/0604094}}].

\bibitem{Cacciari:2008gp}
M.~Cacciari, G.~P. Salam, and G.~Soyez, {\it {The anti-$k_t$ jet clustering
  algorithm}},  {\em JHEP} {\bf 04} (2008) 063,
  [\href{http://xxx.lanl.gov/abs/0802.1189}{{\tt arXiv:0802.1189}}].

\bibitem{Dokshitzer:1997in}
Y.~L. Dokshitzer, G.~D. Leder, S.~Moretti, and B.~R. Webber, {\it {Better Jet
  Clustering Algorithms}},  {\em JHEP} {\bf 08} (1997) 001,
  [\href{http://xxx.lanl.gov/abs/hep-ph/9707323}{{\tt hep-ph/9707323}}].

\bibitem{Wobisch:1998wt}
M.~Wobisch and T.~Wengler, {\it {Hadronization corrections to jet cross
  sections in deep- inelastic scattering}},
  \href{http://xxx.lanl.gov/abs/hep-ph/9907280}{{\tt hep-ph/9907280}}.

\bibitem{Fiorani:1988by}
F.~Fiorani, G.~Marchesini, and L.~Reina, {\it {SOFT GLUON FACTORIZATION AND
  MULTI - GLUON AMPLITUDE}},  {\em Nucl. Phys.} {\bf B309} (1988) 439.

\bibitem{Dokshitzer:1992ip}
Y.~L. Dokshitzer, G.~Marchesini, and G.~Oriani, {\it {Measuring color flows in
  hard processes: Beyond leading order}},  {\em Nucl. Phys.} {\bf B387} (1992)
  675--714.

\bibitem{Dokshitzer_et_al}
Y.~Dokshitzer, V.~Khoze, A.~Mueller, and S.~Troyan, {\em {Basics of
  Perturbative QCD}}.
\newblock Editions Fronti\`eres, Gif-sur-Yvette, France, 1991.

\bibitem{Cacciari:2005hq}
M.~Cacciari and G.~P. Salam, {\it {Dispelling the $N^{3}$ myth for the $k_t$
  jet-finder}},  {\em Phys. Lett.} {\bf B641} (2006) 57--61,
  [\href{http://xxx.lanl.gov/abs/hep-ph/0512210}{{\tt hep-ph/0512210}}].

\bibitem{Buttar:2008jx}
C.~Buttar {\em et.~al.}, {\it {Standard Model Handles and Candles Working
  Group: Tools and Jets Summary Report}},
  \href{http://xxx.lanl.gov/abs/0803.0678}{{\tt arXiv:0803.0678}}.

\bibitem{Dasgupta:2007wa}
M.~Dasgupta, L.~Magnea, and G.~P. Salam, {\it {Non-perturbative QCD effects in
  jets at hadron colliders}},  {\em JHEP} {\bf 02} (2008) 055,
  [\href{http://xxx.lanl.gov/abs/0712.3014}{{\tt arXiv:0712.3014}}].

\bibitem{Cacciari:2008gn}
M.~Cacciari, G.~P. Salam, and G.~Soyez, {\it {The Catchment Area of Jets}},
  {\em JHEP} {\bf 04} (2008) 005,
  [\href{http://xxx.lanl.gov/abs/0802.1188}{{\tt arXiv:0802.1188}}].

\bibitem{Cacciari:2009dp}
M.~Cacciari, G.~P. Salam, and S.~Sapeta, {\it {On the characterisation of the
  underlying event}},  \href{http://xxx.lanl.gov/abs/0912.4926}{{\tt
  arXiv:0912.4926}}.

\bibitem{Cacciari:2007fd}
M.~Cacciari and G.~P. Salam, {\it {Pileup subtraction using jet areas}},  {\em
  Phys. Lett.} {\bf B659} (2008) 119--126,
  [\href{http://xxx.lanl.gov/abs/0707.1378}{{\tt arXiv:0707.1378}}].

\bibitem{Sjostrand:2000wi}
T.~Sjostrand {\em et.~al.}, {\it {High-energy physics event generation with
  PYTHIA 6.1}},  {\em Comput. Phys. Commun.} {\bf 135} (2001) 238--259,
  [\href{http://xxx.lanl.gov/abs/hep-ph/0010017}{{\tt hep-ph/0010017}}].

\bibitem{Sjostrand:2003wg}
T.~Sjostrand, L.~Lonnblad, S.~Mrenna, and P.~Z. Skands, {\it {Pythia 6.3
  physics and manual}},  \href{http://xxx.lanl.gov/abs/hep-ph/0308153}{{\tt
  hep-ph/0308153}}.

\bibitem{Rubin:2009ft}
M.~Rubin, {\it {Light Higgs searches at the LHC using jet substructure}},
  \href{http://xxx.lanl.gov/abs/0905.2124}{{\tt arXiv:0905.2124}}.

\bibitem{Korchemsky:1994is}
G.~P. Korchemsky and G.~Sterman, {\it {Nonperturbative corrections in resummed
  cross-sections}},  {\em Nucl. Phys.} {\bf B437} (1995) 415--432,
  [\href{http://xxx.lanl.gov/abs/hep-ph/9411211}{{\tt hep-ph/9411211}}].

\bibitem{Dasgupta:2009tm}
M.~Dasgupta and Y.~Delenda, {\it {On the universality of hadronisation
  corrections to QCD jets}},  {\em JHEP} {\bf 07} (2009) 004,
  [\href{http://xxx.lanl.gov/abs/0903.2187}{{\tt arXiv:0903.2187}}].

\bibitem{Corcella:2000bw}
G.~Corcella {\em et.~al.}, {\it {HERWIG 6.5: an event generator for Hadron
  Emission Reactions With Interfering Gluons (including supersymmetric
  processes)}},  {\em JHEP} {\bf 01} (2001) 010,
  [\href{http://xxx.lanl.gov/abs/hep-ph/0011363}{{\tt hep-ph/0011363}}].

\bibitem{Corcella:2002jc}
G.~Corcella {\em et.~al.}, {\it {HERWIG 6.5 release note}},
  \href{http://xxx.lanl.gov/abs/hep-ph/0210213}{{\tt hep-ph/0210213}}.

\bibitem{Sjostrand:2006za}
T.~Sjostrand, S.~Mrenna, and P.~Z. Skands, {\it {PYTHIA 6.4 Physics and
  Manual}},  {\em JHEP} {\bf 05} (2006) 026,
  [\href{http://xxx.lanl.gov/abs/hep-ph/0603175}{{\tt hep-ph/0603175}}].

\bibitem{Beneke:1998ui}
M.~Beneke, {\it {Renormalons}},  {\em Phys. Rept.} {\bf 317} (1999) 1--142,
  [\href{http://xxx.lanl.gov/abs/hep-ph/9807443}{{\tt hep-ph/9807443}}].

\end{thebibliography}\endgroup

\end{document}